\documentclass[12pt,a4paper]{article}
\usepackage{eurosym}
\usepackage{amssymb}
\usepackage{graphicx}
\usepackage{amsmath}
\usepackage[toc,page]{appendix}

\begin{document}

\title{Field theoretical formulation of the asymptotic relaxation states of
two-dimensional ideal fluids}
\author{F. Spineanu and M. Vlad \\
National Institute of Laser, Plasma and Radiation Physics\\
Magurele, Bucharest 077125, Romania}
\maketitle

\begin{abstract}
The ideal incompressible fluid in two dimensions (Euler fluid) evolves at
relaxation from turbulent states to highly coherent states of flow. For the
case of double spatial periodicity and zero total vorticity it is known that
the streamfunction verifies the \emph{sinh}-Poisson equation. These
exceptional states can only be identified in a description based on the
extremum of an action functional. Starting from the discrete model of interacting point-like vortices
it was possible to write a Lagrangian in terms of a matter
function and a gauge potential. They provide a dual representation of the same physical object, the vorticity. This classical field theory identifies the stationary, coherent, states of the $%
2D $ Euler fluid as derived from the self-duality. We first provide a more detailed analysis of this model, including a comparison with the approach based on the statistical physics of point-like vortices. The second main objective is the study of the dynamics in close proximity of the stationary self-dual state, \emph{i.e.} before the system has reached the absolute extremum of the action functional. Finally, limitations and possible extensions of
this field theoretical model for the $2D$ fluids model are discussed and some possible applications are mentioned.
\end{abstract}

\tableofcontents

\bigskip


\section{Introduction}

The ideal (non-dissipative) incompressible fluid in two - dimensions,
which we will shortly call $2D$ Euler fluid, can be described by three
related functions $\left( \psi ,\mathbf{v,\omega }\right) $. The
streamfunction is a scalar field $\psi \left( x,y,t\right) $ from which the
velocity vector field $\mathbf{v}\left( x,y,t\right) $ is derived: from the
incompressibility $\mathbf{\nabla \cdot v}=0$, one can write $\mathbf{v}=%
\mathbf{\nabla }\chi -\mathbf{\nabla }\psi \times \widehat{\mathbf{e}}_{z}$
where $\chi $ is a harmonic function, $\Delta \chi =0$, $\widehat{\mathbf{e}}%
_{z}$ is the versor perpendicular on the plane of the motion and the
operators $\mathbf{\nabla }$ and $\Delta $ are restricted to $2D$. Applying
the rotational operator one obtains the vorticity $\omega \widehat{\mathbf{e}%
}_{z}=\mathbf{\nabla \times v=}\Delta \psi \widehat{\mathbf{e}}_{z}$ and the
Euler equation is%
\begin{equation}
\frac{d\omega }{dt}=\frac{\partial }{\partial t}\Delta \psi +\left[ \left( -%
\mathbf{\nabla }\psi \times \widehat{\mathbf{e}}_{z}\right) \cdot \mathbf{%
\nabla }\right] \Delta \psi =0.  \label{eq1}
\end{equation}%
The velocity vector field $\mathbf{v}\left( x,y,t\right) $ has the
fundamental quality that it can be measured in physical fluids, offering a
direct connection with the experiments and observations. It is then natural
that almost all studies on the fluid dynamics are expressed in terms of
these three functions and any more abstract description must finally return
to them.

It is known that in $2D$ there is inverse cascade, \emph{i.e.} there is flow
of energy in the spectrum from small spatial scales towards the large
spatial scales. The numerical simulation of the $2D$ Euler fluid in a box
with doubly periodic boundary conditions fully confirms this behavior.
Adding just a small viscosity and starting from a state of turbulence, the
fluid evolves to a state of highly ordered flow: the positive and negative
vorticities contained in the initial flow are separated and collected into
two large scale vortical flows of opposite sign. Fully convincing pictures
of the asymptotic states are shown in Ref. \cite{Montgomery1991} and \cite%
{Montgomery1992}. The motion is stationary for a long time, being finally
dissipated by the friction associated to the small viscosity. It has been
found that the streamfunction $\psi $ in these states reached asymptotically
at relaxation from turbulence verify the \emph{sinh}-Poisson equation%
\begin{equation}
\Delta \psi +\lambda \sinh \psi =0  \label{eq2}
\end{equation}%
where $\lambda >0$ is a parameter. The significance of this fact is very
deep and can be appreciated by the following considerations. If we want to
find the stationary solution of Eq.(\ref{eq1}) we take $\partial \psi
/\partial t=0$ and look for the solutions of%
\begin{equation}
\left[ \left( -\mathbf{\nabla }\psi \times \widehat{\mathbf{e}}_{z}\right)
\cdot \mathbf{\nabla }\right] \Delta \psi =0  \label{st}
\end{equation}%
It is obvious (and widely adopted) that we can solve this equation by taking
the vorticity to be an arbitrary function $F$ of the streamfunction: $\omega
=\Delta \psi =F\left( \psi \right) $. Equivalently this is
a recognition of the fact that Eq.(\ref{st}) has an indefinitely large space
of solutions. However the nature does not confirm this: the fluid left to
evolve from a turbulent initial state will end up by taking one of the
functions $\psi \left( x,y\right) $ that verify Eq.(\ref{eq2}), \emph{i.e.}
it goes precisely towards a tiny subset within the whole function space that
seemingly was at its disposal. This dramatically underlines the contrast:
while $\omega =F\left( \psi \right) $ with arbitrary $F$ is a result of the
conservation law $d\omega /dt=0$, the strict evolution towards solutions $%
\psi $ of Eq.(\ref{eq2}) suggests that there are exceptional states and they
should be chosen by some variational principle that is expected to apply to
this system.

The equation (\ref{eq2}) is exactly integrable \cite{TingChenLee}. Since in
general the coherent structures and the integrability are connected with
self-duality \cite{MasonWoodhouse}, one may be interested to identify the
analytical framework where the coherent structures of the stationary $2D$
Euler fluid flow appear as a consequence of the self-duality (SD). We note
that, at least at first sight, the classical formulation in terms of $\left(
\psi ,\mathbf{v},\mathbf{\omega }\right) $ does not appear to be adequate to
express the property of self-duality.

Although the accumulation of results on the dynamics of the $2D$ Euler fluid
is immense, there is an obstacle if we want to exploit it in order to
construct a formulation that exhibits the connection \textquotedblleft
coherent flow\textquotedblright\ - \textquotedblleft self
duality\textquotedblright . The classical formulation uses the conservation
laws as dynamical equations. The zero-divergence of the velocity field is
equivalent to the continuity equation, \emph{i.e.} the conservation of the
fluid mass. The conservation of the momentum is the zero-dissipation version
of the Navier-Stokes equation, which, after applying the operator $\mathbf{%
\nabla \times }$ , becomes Eq.(\ref{eq1}). Further, commonly used are the
conservation of the energy, of angular momentum, etc. If there is a change
of one of the variables of which the state of the system depends, the
conservation laws show how the other variables must change such that certain
quantities (mass, momentum, energy, etc.) remain invariant. The conservation
laws cannot identify exceptional states. For this we need a functional of
the state of the fluid and a variational principle able to identify the
evolutions toward particular, exceptional states, like those given by Eq.(%
\ref{eq2}). In other words, we need a description of the fluid motion in
terms of the density of a Lagrangian, whose integral over space-time is an
action functional. The dynamical equations would then be derived as
Euler-Lagrange variational equations, by extremizing the action.
Summarizing, we currently use the conservation laws as dynamical equations,
which is formally not correct: the \emph{dynamical} equations are by
definition the Euler-Lagrange equations obtained from functional variation
of an action functional. The difficult problem is, of course, to find the
Lagrangian. The Lagrangian must be inferred from basic physical facts about
the system, and it is not satisfactory to simply find a functional (like a
minimizer) or Lyapunov-type.

Finding the adequate Lagrangian for the two-dimensional Euler fluid is
however possible. The reason is the existence of a model consisting of a
discrete version of the physical dynamics expressed by Eq.(\ref{eq1}): a set
of point-like vortices interacting in plane by a potential generated by
themselves. The interaction is long - range (Coulombian) and the equations
of motion are a discrete version of the advection of the elementary vortices
by the velocity field produced by themselves. It is well established (and
will be reminded below) that the set of discrete, point-like, vortices can
be treated as a statistical ensemble with the result that at maximum entropy
the Eq.(\ref{eq2}) is obtained. Several other applications of the discrete
model have led to interesting results but in general the model is difficult
to be used directly. From the point of view of what we are looking for, 
\emph{i.e.} a Lagrangian for the Euler fluid, the discrete model is however
extremely suggestive \cite{FlorinMadi2003}. Instead of $\left( \psi ,\mathbf{%
v},\mathbf{\omega }\right) $ it uses \emph{matter} (density of point-like
vortices), \emph{field} (corresponding to the potential generated by the
discrete vortices) and \emph{interaction}. This means that returning to the
continuum limit but preserving this structure, we can formulate a classical
field theory. This shift is a conceptual change and some inferrence is still
needed in order to write the Lagrangian functional. Following the suggestion
of the point-like vortex model two fields are involved, a field $\phi \left(
x,y,t\right) $ representing the \emph{matter} and a vector field $A^{\mu
}\left( x,y,t\right) $ with $\mu =0,x,y$, representing the \emph{gauge
potential}. The vortical nature of the elementary objects can be reproduced
by a classical spin-like quantity. It is convenient to represent the negative vortices
as positive vortices having backward time propagation, \emph{i.e.} the
positive and negative vortices behave as particles and antiparticles. The
matter $\phi $ will be represented by a mixed spinor of the type $x^{\alpha 
\overset{\cdot }{\beta }}$, a $2\times 2$ matrix with complex entries, with
distinct spinorial transformations on its two indices (this is the reason of
the dot on the index $\beta $). Accordingly the potential is a complex $%
2\times 2$ matrix, an element of $sl\left( 2,\mathbf{C}\right) $. The
Lorentz-type motion of the elementary vortices is represented by the
Chern-Simons term in the Lagrangian. A nonlinear self-interaction of the
matter field cancels, via Gauss constraint, the part of the kinetic energy which is due to the  interaction between the
rotational of the potential (the magnetic field) and the matter density. The
extremum of the action corresponds to self-duality and the states are
stationary with the streamfunction obeying Eq.(\ref{eq2}). This shows that
the coherent flows reached \ by the Euler fluid at relaxation belong to the
same exceptional family of soliton or instanton-like solutions, a purely
nonlinear effect. This represents also an analytical derivation of the Eq.(%
\ref{eq2}), alternative to the statistical analysis.

A full framework for the description of the $2D$ Euler fluid is built in
this way, using the powerful field theoretical (FT) formalism and ready to
benefit from its achievements in the physics of vortices (Bose-Einstein
condensate, superconductivity, topological field theories like $O\left(
n\right) $, cosmic strings, etc.). Naturally there are limitations too: one
still has to include dissipation and the change of topology of flows by
breaking and reconnection of streamlines, study the isotopological dynamical
aspects (\emph{i.e.} between reconnection events), and adapt the formalism
to various boundary conditions, etc. In the present work we focus on the $2D$
Euler fluid evolving in a box with boundary conditions leading to double
periodicity. This is known to evolve asymptotically to solutions of Eq.(\ref%
{eq2}) and, in the FT, exhibit the property of self-duality. We attach most
importance to this fact since it has become more and more clear that all
known coherent structures are connected with self-duality \cite%
{MasonWoodhouse}.

The states identified by the FT as extrema of the action functional are
characterized by: (1) stationarity; (2) double periodicity, \emph{i.e.} the
function $\psi \left( x,y\right) $ must only be determined on a
\textquotedblleft fundamental\textquotedblright\ square in plane; (3) the
total vorticity is zero; (4) the states verify Eq.(\ref{eq2}). The self-dual
states are the absolute minimum of the energy but in order them to be attained
the system must have access to the class of configurations defined by these
symmetry conditions : zero total vorticity in the field and double spatial
periodicity. In the non-dissipative fluid these conditions are fixed at the
initial state and the SD state cannot be reached in general. This means that
here a large class of fluid asymptotic states will not be examined. The
relevance of all these, for the physics of fluids, is an interesting subject,
which we will not discuss here.

In Ref.\cite{FlorinMadi2003} we have presented the derivation of the \emph{sinh}-Poisson equation in a field theoretical model for the $2D$  Euler fluid. The objective of the present work is the study of the time evolution in
close proximity of the SD state, for a system that asymptotically reaches
the SD state. We derive the specific form taken by the equations of motion
in this regime, the current of \textquotedblleft matter\textquotedblright\
and the equations for the magnitudes of the positive and negative parts of
the matter field that combines into a single physical variable, the
vorticity. These equations are similar but not identical to equations of
continuity and generalize the equations of the Abelian model \cite{JackiwPi}.

The SD state depends on the equality between two parameters that enter the
expression of the Lagrangian. Since we are interested in the states that are
close, - but not exactly at SD, we suggest (in a qualitative discussion) that it may be possible to include situations
where these two parameters are not equal but evolve slowly toward equality. 
It arises a possible reflection in the theory of the events of dissipative
reconnection of streamlines and increase of the topological order of the
flow, toward SD. Now there is attraction between \emph{mesoscopic}
vortices. (We use this name for the few vortices remaining in the late
phase, which have already concentrated a large part of the initial
vorticity; they move slowly in plane and their encounters and mergings is
the last phase of the evolution toward the final, fully organized, state).
The FT suggests the interpretation that an excess of \textquotedblleft
helicity\textquotedblright\ is removed at each reconnection until identity
is reached between two different contributions to the energy: the FT energy
is exactly zero at SD since the energy is only due to the motion of the centers of the \emph{%
mesoscopic} vortices, which stop at SD, while the motion of the fluid on streamlines has zero
energy. We suggest that a FT formalism similar with the \emph{baryogenesis}
but in reversed direction, \emph{i.e.} decrease of Chern-Simons topological
number, may provide an analytical description. Since the term of the
Lagrangian that is so decreased becomes at SD the square vorticity, it seems
that there is compatibility with the known decay of the enstrophy during
vorticity self-organization in weakly dissipative fluids.

\bigskip

\section{The model of interacting point-like vortices}

The physical quantities describing the two-dimensional fluid dynamics are $%
\psi \equiv $streamfunction, $\mathbf{v}\mathbf{\equiv }$velocity, $\omega 
\widehat{\mathbf{e}}_{z}=$vorticity, which are related by 
\begin{equation}
\mathbf{v}=-\mathbf{\nabla }\psi \times \widehat{\mathbf{e}}_{z}\ \ ,\ \
\omega =\Delta \psi  \label{vom}
\end{equation}%
and are solutions of the Euler equation (\ref{eq1}). The discretized form of
this equation has been extensively studied \cite{KraichnanMontgomery}, \cite%
{RobertSommeria}, \cite{RobertSommeria1992}, \cite{Chavanis1}. The continuum
limit of the discretization is matematically equivalent with the fluid
dynamics. We just review few elements of this theory, for further reference.

Consider the discretization of the vorticity field $\omega \left( x,y\right) 
$ in a set of $2N$ point-like vortices $\omega _{i}$ each carrying the
elementary quantity $\omega _{0}$ (= const $>0$) of vorticity which can be
positive or negative $\omega _{i}=\pm \omega _{0}$. There are $N$ vortices
with the vorticity $+\omega _{0}\ \ $and $N$ vortices with the vorticity $%
-\omega _{0}$. The current position of a point-like vortex is $\left(
x_{i},y_{i}\right) $ at the moment $t$. The vorticity is expressed as%
\begin{equation}
\omega \left( x,y\right) =\sum\limits_{i=1}^{2N}\omega _{i}a^{2}\delta
\left( x-x_{i}\right) \delta \left( y-y_{i}\right)   \label{omeg}
\end{equation}%
where $a$ is the radius of an effective support of a smooth representation
of the Dirac $\delta $ functions approximating the product of the two $%
\delta $ functions \cite{KraichnanMontgomery}. Instead of $\omega _{i}a^{2}$
we can use the \emph{circulation} $\gamma _{i}$ which is the integral of the
vorticity over a small area around the point $\left( x_{i},y_{i}\right) $: $%
\gamma _{i}=\int d^{2}x\omega _{i}$ \cite{Chavanis1}. The formal solution of
the equation $\Delta \psi =\omega $, connecting the vorticity and the
streamfunction, can be obtained using the Green function for the Laplace
operator%
\begin{equation}
\Delta _{x,y}G\left( x,y;x^{\prime },y^{\prime }\right) =\delta (x-x^{\prime
})\delta \left( y-y^{\prime }\right)   \label{greendef}
\end{equation}%
where $\left( x^{\prime },y^{\prime }\right) $ is a reference point in the
plane. As shown in Ref.\cite{KraichnanMontgomery} $G\left( \mathbf{r};%
\mathbf{r}^{\prime }\right) $ can be approximated for $a$ small compared to
the space extension of the fluid, $L$, $a\ll L$, as the Green function of
the Laplacian 
\begin{equation}
G\left( \mathbf{r};\mathbf{r}^{\prime }\right) \approx \frac{1}{2\pi }\ln
\left( \frac{\left\vert \mathbf{r}-\mathbf{r}^{\prime }\right\vert }{L}%
\right)   \label{Green}
\end{equation}%
where $L$ is the length of the side of the square domain. The solution of
the equation $\Delta \psi =\omega $ is obtained using the Green function,
using the circulation $\gamma _{i}=\omega _{i}a^{2}$,%
\begin{equation}
\psi \left( \mathbf{r}\right) =\sum\limits_{i=1}^{2N}\gamma _{i}\frac{1}{%
2\pi }\ln \left( \frac{\left\vert \mathbf{r}-\mathbf{r}_{i}\right\vert }{L}%
\right)   \label{psilog}
\end{equation}%
The velocity of the $k$-th point-vortex is $\mathbf{v}_{k}=-\left. \mathbf{%
\nabla }\psi \right\vert _{\mathbf{r=r}_{k}}\times \widehat{\mathbf{e}}_{z}$
and the equations of motion are%
\begin{eqnarray}
\frac{dx_{k}}{dt} &=&v_{x}^{\left( k\right) }=-\sum\limits_{i=1,i\neq k}^{2N}\gamma
_{i}\frac{1}{2\pi }\frac{y_{k}-y_{i}}{\left\vert \mathbf{r}_{k}-\mathbf{r}%
_{i}\right\vert ^{2}}  \label{statiseqs} \\
\frac{dy_{k}}{dt} &=&v_{y}^{\left( k\right) }=\sum\limits_{i=1,i\neq k}^{2N}\gamma
_{i}\frac{1}{2\pi }\frac{x_{k}-x_{i}}{\left\vert \mathbf{r}_{k}-\mathbf{r}%
_{i}\right\vert ^{2}}\   \nonumber
\end{eqnarray}

The equations can be derived from a Hamiltonian%
\begin{equation}
H=\frac{1}{2\pi }\underset{i<j}{\sum\limits_{i=1}^{2N}\sum\limits_{j=1}^{2N}}%
\gamma _{i}\ln \left( \frac{\left\vert \mathbf{r}_{i}-\mathbf{r}%
_{j}\right\vert }{L}\right) \gamma _{j}  \label{hamilt}
\end{equation}

The standard way of describing the discrete model is within a statistical
approach \cite{Onsager}, \cite{KraichnanMontgomery}, \cite{RobertSommeria}, \cite{MajdaWang}, \cite{DritschelLuciaPoje}.
The elementary vortices are seen as elements of a system of interacting
particles (like a gas) that explore an ensemble of microscopic states
leading to the macroscopic manifestation that is the fluid flow. The number
of positive vortices in the state $i$ is $N_{i}^{+}$ and the number of
negative vortices in the state $i$ is $N_{i}^{-}$. The total numbers of
positive and respectively negative vortices are equal: $N^{+}=\sum%
\limits_{i}N_{i}^{+}=\sum\limits_{i}N_{i}^{-}=N^{-}$. This system has a
statistical temperature that is negative when the energy is zero or positive 
\cite{EdwardsTaylor}. The energy of the discrete system of point-like
vortices is $\mathcal{E}=\frac{1}{2}\sum\limits_{ij}\omega \left( \mathbf{r}%
_{i}\right) G\left( \mathbf{r}_{i},\mathbf{r}_{j}\right) \omega \left( 
\mathbf{r}_{j}\right) $ where $\omega \left( \mathbf{r}_{i}\right) =-\left(
N_{i}^{+}-N_{i}^{-}\right) $ is the vorticity. The probability of a state is
calculated as a combinatorial expression%
\begin{equation}
\mathcal{W}=\left\{ \frac{N^{+}!}{\prod\limits_{i}N_{i}^{+}!}\right\}
\left\{ \frac{N^{-}!}{\prod\limits_{i}N_{i}^{-}!}\right\}   \label{W}
\end{equation}%
The \emph{entropy} is the logarithm of this expression and by extremization
one finds 
\begin{equation}
\ln N_{i}^{\pm }+\alpha ^{\pm }\pm \beta \sum\limits_{j}G\left( \mathbf{r}%
_{i},\mathbf{r}_{j}\right) \left( N_{j}^{+}-N_{j}^{-}\right) =0
\end{equation}%
for $i=1,N$, where $\alpha ^{\pm }$ and $\beta $ are Lagrange multipliers introduced to
ensure $\sum N_{i}^{+}=\sum N_{i}^{-}=N=$ const and conservation of the
Energy $\mathcal{E}$. The solutions are written in terms of a continuous
function $\psi \left( x,y\right) $%
\begin{equation*}
N_{i}^{\pm }=\exp \left[ -\alpha ^{\pm }\mp \beta \psi \left( x,y\right) %
\right] 
\end{equation*}%
implying $N_{i}^{+}N_{i}^{-}=$ const, and this leads to the \emph{sinh}%
-Poisson equation (\ref{eq2}). The statistical approach has had to face
particular problems: the system has finite phase space; there is no
thermodynamic limit; there is no ergodicity; the temperature is negative;
the entropy extremum is counter-intuitive, leading to maximum order; the
final state of the system is not a \emph{statistical} equilibrium but
consists of non-fluctuating positions of the elementary vortices,
composing a solution of (\ref{eq2}). However the statistical approach
succeeds to derive Eq.(\ref{eq2}), is fully confirmed and generates
successful exploration of similar problems. Since the field theoretical
approach is different in an essential way it appears that the statistical
approach has identified, in its specific way, the self-duality. Few aspects
of the statistical approach will be discussed below in connection with FT
formulation.

\section{Field theoretical formulation of the continuum limit of the
point-like vortex model}

The physical vortical flow is represented by the Lorentz-type motion of the
discrete set of point-like, massless, vortices. We note however that nowhere
in the formulation (\ref{statiseqs}) is made explicit the fact that we are
dealing with \emph{vortices}. The same equations describe a system of guiding-centers  \cite{JoyceMontgomery}, 
point-like \emph{charges} \cite{JackiwPi} or
currents \cite{Taylor1993}. The information that it is question of vortices, 
\emph{i.e.} objects that have the nature of vectors, must be supplemented to
the system (\ref{statiseqs}). We then also note that the third axis $\left(
z\right) $ although irrelevant for the plane motion, is implicitely present
in the model.

In the basic model (Kraichnan and Montgomery \cite{KraichnanMontgomery},
which will be taken as the reference model) it is assumed that the
elementary vortices have equal magnitudes of vorticity $\omega _{0}$ and,
for periodicity, the numbers of positive and of negative vortices are equal, 
$N$. This $N$ is invariant, \emph{i.e.} there are no flip and/or
annihilations. Physical vorticity $\omega $ in a point $\left( x,y\right) $
is obtained by placing together $n$ elementary vortices, $\omega \approx
n\omega _{0}$ in an infinitesimal area around $\left( x,y\right) $. The
model does not allow building up higher similar objects \emph{i.e. }$\pm
2\omega _{0}$, $\pm 3\omega _{0}$, etc. are not allowed as independent
objects. In this representation the physical vorticity comes from the
density of elementary vortices, \emph{i.e.} like-sign vortices are not
superposed one to the other, similar to the Pauli exclusion principle for
fermion particles.

Therefore we have two types of elementary objects, carrying $+\omega _{0}$
and respectively $-\omega _{0}$ vorticity. The elementary vortices are
similar to massless particles carrying half-integer spin but with fixed,
unchangeable, projection along the transversal axis. The interaction between
the two types of elementary vortices only affects their positions in plane,
without changing their spin and projection.

\bigskip

Taking a fixed vorticity $\pm \omega _{0}$ for an elementary vortex there is
no needs of an assumption on how this vorticity has been created, for
example there is no need to imagine the presence of a fluid between the
vortices. The model of point-like vortices fully replaces the model based on
the physical variables $\left( \psi ,\mathbf{v},\mathbf{\omega }\right) $.
However, for theoretical purposes we can imagine that around each elementary
vortex there is a fluid rotating such as to create the elementary vorticity $%
\omega _{0}$. Obviously there is no unique way of prescribing such a
velocity. The difference between the positive $\left( +\omega _{0}\right) $
and negative $\left( -\omega _{0}\right) $ vortices is the direction of the
associated vortical flow in the plane: we take \emph{anti-clockwise} for $%
+\omega _{0}$ and \emph{clockwise} for $-\omega _{0}$. We note a particular
property which is revealed by the representation based on the virtual-fluid
rotation: the negative vortex can be obtained from a positive vortex by
reversing the direction of time, since this leads to the reverse of the
sense of rotation of the virtual fluid. Moreover, this ensures the
invariance of the theory to time reversal transformation, if the total
numbers of positive and negative-vorticity elements are equal.

As explained by Kraichnan and Montgomery \cite{KraichnanMontgomery} the
elementary vortical structure in $3D$ is a vortex ring. A $3D$ ring of
infinitesimal cross section intersects a plane that contains the center of
the ring and is transversal on the plane of the ring, in two points. Close
to these points the $3D$ ring is approximately reduced to two elementary
linear vortices, perpendicular on the plane and with opposite vorticity. We
add to this picture the observation that an axial flow in a $3D$ ring vortex
becomes after reduction to $2D$ a flow perpendicular on the plane in
positive $z$ direction for one elementary vortex and in the negative $z$
direction for its pair with opposite vorticity. The particularity of the $2D$
Euler fluid, that has been transfered to the discrete system, is the
invariance against displacements on the $z$-axis, which can be defined
locally arbitrarily. We will consider, without restricting generality, that
the positive vortices $\left( +\omega _{0}\right) $ have a momentum $\mathbf{%
p}=p_{0}\widehat{\mathbf{e}}_{z}$ and in agreement with the picture that
represents a pair of opposite vortices as resulting from a $3D$ ring with
axial flow, the negative vortices have a motion $\mathbf{p}=-p_{0}\widehat{%
\mathbf{e}}_{z}$. The filaments can have a translation along the irrelevant
axis ($z$) with arbitrary momentum, $p_{0}$. Again we note that the time
inversion leaves invariant the system, with the positive vortices mapped
onto negative ones. This will make the negative vortices actually to be
defined as anti-vortices, similar to the anti-particles.

In the case of the point-like vortices for the Euler equation the positive
energy vortices propagating forward in time are the usual physical
point-like vortices. The time reflection vortices are propagating backward
in time but they can be considered physical vortices with opposite charge (%
\emph{i.e.} the vorticity $\omega _{0}\rightarrow -\omega _{0}$) and
propagates forward in time. They are simply physical point-like vortices
with opposite vorticity.

With relation to the chiral analogy, we have \textquotedblleft
right-handed\textquotedblright\ and respectively \textquotedblleft
left-handed\textquotedblright\ vortices.

\bigskip

The two elements of the flow are positive and negative elementary vortices
(point-like). The positive vortices: (1) rotate anti-clockwise in plane: $%
\omega \widehat{\mathbf{e}}_{z}\sim \mathbf{\sigma }$\ spin is up; (2) move
along the positive $z$ axis: $\mathbf{p=}\widehat{\mathbf{e}}_{z}p_{0}$; (3)
have positive chirality: $\chi =\frac{\mathbf{\sigma \cdot p}}{\left\vert 
\mathbf{p}\right\vert }$. The positive vortices can be represented as a
point that runs along a positive helix, upward. In projection from the above
the plane toward the plane we see a circle on which the point moves
anti-clockwise.

The \ negative vortices: (1) rotate clockwise in plane: $\left( -\omega
\right) \widehat{\mathbf{e}}_{z}\sim -\mathbf{\sigma }$ spin is down; (2)
move along the negative $z$ axis: $-\mathbf{p=}\widehat{\mathbf{e}}%
_{z}\left( -p_{0}\right) $, along $-z$; (3) have positive chirality: $\chi =%
\frac{\mathbf{\sigma \cdot p}}{\left\vert \mathbf{p}\right\vert }$. The
negative vortices can be represented as a point that runs along a positive
helix, the same as above, but runs downward. In projection from the above
the plane toward the plane we see a circle on which the point moves
clockwise.

The positive vortices and the negative vortices have the same \emph{%
chirality }and in a point where there is superposition of a positive and a
negative elementary vortices the \emph{chirality} is added. In particular,
the vacuum consists of paired positive and negative vortices, with no motion
of the fluid, which in physical variables means $\psi \equiv 0$, $\mathbf{%
v\equiv 0}$, $\omega \equiv 0$. In FT the vacuum consists of
superposition of positive and negative vortices, which means: (1) zero spin,
or zero \emph{vorticity}; (2) zero momentum $\mathbf{p}=\mathbf{0}$; (3) $%
2\times $chiral charge. The Euler fluid at equilibrium ($\psi =0$, $\mathbf{v=0}$%
, $\omega =0$) is in a vacuum with \emph{broken chiral invariance}.

We can now return to comment the result of the statistical analysis based on
the maximum entropy for the system of point-like vortices. The results was $%
N_{i}^{+}N_{i}^{-}=$const. This means that in order to reproduce a positive
physical vorticity $\left\vert \omega \right\vert $ in a point we cannot
simply take only positive elementary vortices $\left\vert \omega \right\vert
=N_{i}^{+}\omega _{0}$. We must also take a certain amount of negative
point-like vortices $\left( N_{i}^{-}\right) $ in the same differentially
small area of the discretization and obtain the physical vorticity $%
\left\vert \omega \right\vert $ as the difference between the two
contributions, $\left\vert \omega \right\vert =\left\vert
N_{i}^{+}-N_{i}^{-}\right\vert $. None of them can ever be exactly zero, $%
N_{i}^{\pm }\neq 0$. This was the first indication that the elementary
vortices are not as simple as pieces of vorticity. The zero vorticity does
not mean absence of $N_{i}^{\pm }$. Both these numbers \emph{must} remain
non-zero but they are now equal and implicitly there is mutual annihilation
of their virtual flows and of their $z$-momenta. This corresponds to \emph{%
pairing} of vortices with anti-vortices. In a fermionic picture of the
discrete system we have that the spin is zero but the chiral number is $2$.
The discrete system is an example, in the classical world, of the
spontaneous breaking of chiral symmetry.

\bigskip

The energy at the continuum limit of the model of discrete point-like
vortices is, according to Eq.(\ref{hamilt})%
\begin{equation}
E=\frac{1}{2\pi }\int d^{2}xd^{2}x^{\prime }\omega \left( \mathbf{x}%
\right) \ln \left( \frac{\left\vert \mathbf{x-x}^{\prime }\right\vert }{L}%
\right) \omega \left( \mathbf{x}^{\prime }\right)   \label{eomega}
\end{equation}%
We now have a problem that is similar to that mentioned above, about the
nature of point-like objects (vortices or charges or currents). This time
the problem arises because the same expression of energy can be written for
a Coulombian gas of charges of density $\rho \left( \mathbf{x}\right) $ in
plane, by replacing $\omega \left( \mathbf{x}\right) \rightarrow \rho \left( 
\mathbf{x}\right) $. In this case however the interaction leads to motion
that is along the line separating the charges. The direction of the relative
motion of two interacting point-like objects (charges) is given by the
gradient of the scalar function $\psi \left( \mathbf{x}\right) =\int
d^{2}x^{\prime }\ln \left( \left\vert \mathbf{x-x}^{\prime }\right\vert
/L\right) \rho \left( \mathbf{x}^{\prime }\right) $. The same scalar
potential is introduced for the point-like vortices (\ref{psilog}). But
there, two interacting vortices will move in directions that are
perpendicular on the line that connects them, \emph{i.e.} they tend to
rotate one around the other. Then, for the system of point-like vortices,
the Hamiltonian must be \emph{supplemented} with the prescription that the
velocity is like that of a charge in a magnetic field (or geostrophic) 
\begin{equation}
\mathbf{v=-\nabla }\psi \times \widehat{\mathbf{e}}_{z}  \label{vpsi}
\end{equation}%
or, equivalently, of the $E\times B$-type.

The preceding observations prove to be essential when we go to the FT
formulation: first, the fact that the equations of motion for the point-like
objects refer to \emph{vortices} (not charges, currents, etc.) imposes to
consider the non-Abelian representation of the objects, finally leading to
mixed spinors. Second, the fact that the Hamiltonian must be supplemented
with the prescription that the motion is purely kinematic (\emph{i.e.} we
derive directly the velocities from $\psi $ as in Eq.(\ref{vpsi}) ) and the
velocity is $E\times B$ - type requires to adopt the Chern-Simons (CS) term in the
Lagrangian of the system. The CS term supports the \emph{vortical} content
of the dynamics. To close the discussion about the contrast between the
interacting vortices $\left( \mathbf{v}\sim -\mathbf{\nabla }\psi \times 
\widehat{\mathbf{e}}_{z}\right) $ and the interacting charges in plane
(force $\sim \mathbf{\nabla }\psi $), we note that the Lagrangian for the
latter system does not include CS term and the asymptotic limit is the
Landau-Ginzburg equation \cite{Bethuel}. For the point-like vortices we must include CS
term and the asymptotic equation is \emph{sinh}-Poisson.

The FT model for the system of charges in plane moving according to the Eqs.(%
\ref{statiseqs}) was formulated by Jackiw and Pi \cite{JackiwPi} having in
view the application to the Fractional Quantum Hall Effect. The classical
part has identified the absolute extremum of the action as stationary self-dual
states, solutions of the Liouville equation. The non-Abelian extension of
this model has been introduced and discussed by Dunne \emph{et al.} for a
gauge algebra $su\left( N\right) $, with $N$ arbitrary \cite%
{DunneJackiwTrugenberg}, \cite{DunneBook}. The $sl(2,\mathbf{C})$
Non-Abelian structure is necessary due to the \emph{vortical} nature of the
elementary object. Due to the extension of the space of particles
(elementary vortices) with anti-particles (anti-vortices) requested by the
parity invariance, the vorticity matter will need to be represented by a 
\emph{mixed spinor}. By contrast, Jackiw and Pi obtain Liouville equation in
the model of \emph{scalar charges} evolving in plane.

The Lagrangian \cite{DunneBook}%
\begin{eqnarray}
\mathcal{L} &=&-\kappa \varepsilon ^{\mu \nu \rho }\mathrm{tr}\left( \left(
\partial _{\mu }A_{\nu }\right) A_{\rho }+\frac{2}{3}A_{\mu }A_{\nu }A_{\rho
}\right)  \label{Lagrange} \\
&&+i\mathrm{tr}\left( \phi ^{\dagger }\left( D_{0}\phi \right) \right) -%
\frac{1}{2m}\mathrm{tr}\left( \left( D_{k}\phi \right) ^{\dagger }\left(
D^{k}\phi \right) \right) \   \nonumber \\
&&-V\left( \left\vert \phi \right\vert \right) \   \nonumber
\end{eqnarray}%
where $D_{\mu }=\partial _{\mu }+\left[ A_{\mu },\right] $ and $\kappa $, $m$
are positive constants. The matter self-interaction potential is%
\begin{equation}
V\left( \left\vert \phi \right\vert \right) =-\frac{g}{2}\mathrm{tr}\left( %
\left[ \phi ^{\dagger },\phi \right] ^{2}\right)  \label{vpot}
\end{equation}%
The Euler - Lagrange equations for the action functional $\mathcal{S}=\int
dxdydt\mathcal{L}$ are the equations of motion 
\begin{eqnarray}
iD_{0}\phi &=&-\frac{1}{2m}D_{k}D^{k}\phi -g\left[ \left[ \phi ,\phi
^{\dagger }\right] ,\phi \right]  \label{eqmotion} \\
\kappa \varepsilon ^{\mu \nu \rho }F_{\nu \rho } &=&iJ^{\mu }\ 
\end{eqnarray}%
where the current 
\begin{eqnarray}
J^{0} &=&\left[ \phi ,\phi ^{\dagger }\right]  \label{current} \\
J^{k} &=&-\frac{i}{2m}\left( \left[ \phi ^{\dagger },\left( D^{k}\phi
\right) \right] -\left[ \left( D^{k}\phi \right) ^{\dagger },\phi \right]
\right) \ 
\end{eqnarray}%
is covariantly conserved $D_{\mu }J^{\mu }=0$. The energy density is 
\begin{equation}
E=\frac{1}{2m}\mathrm{tr}\left( \left( D_{k}\phi \right) ^{\dagger }\left(
D^{k}\phi \right) \right) -\frac{g}{2}\mathrm{tr}\left( \left[ \phi
^{\dagger },\phi \right] ^{2}\right)  \label{energia}
\end{equation}

The Gauss law is the zero component of the second equation of motion 
\begin{equation}
2\kappa F_{12}=iJ^{0}=i\left[ \phi ,\phi ^{\dagger }\right]  \label{ga1}
\end{equation}

In the following we will use convenient combinations of variables: $A_{\pm
}\equiv A_{x}\pm iA_{y}$, $\partial /\partial z=\frac{1}{2}\left( \partial
/\partial x-i\partial /\partial y\right) $, $\partial /\partial z^{\ast }=%
\frac{1}{2}\left( \partial /\partial x+i\partial /\partial y\right) $, and
similar. Writting 
\begin{eqnarray}
\mathrm{tr}\left( \left( D_{k}\phi \right) ^{\dagger }\left( D^{k}\phi
\right) \right) &=&\mathrm{tr}\left( \left( D_{-}\phi \right) ^{\dagger
}\left( D_{-}\phi \right) \right) -i\mathrm{tr}\left( \phi ^{\dagger }\left[
F_{12},\phi \right] \right)  \label{dkdk} \\
&&-m\varepsilon ^{ij}\partial _{i}\left[ \phi ^{\dagger }\left( D_{j}\phi
\right) -\left( D_{j}\phi \right) ^{\dagger }\phi \right] \   \nonumber
\end{eqnarray}%
we replace in the expression of the energy density and note that for smooth
fields we can ignore the last term, which is evaluated at the boundary 
\begin{equation}
E=\frac{1}{2m}\mathrm{tr}\left( \left( D_{-}\phi \right) ^{\dagger }\left(
D_{-}\phi \right) \right) +\left( -\frac{g}{2}+\frac{1}{4m\kappa }\right) 
\mathrm{tr}\left( \left[ \phi ^{\dagger },\phi \right] ^{2}\right)
\label{enersd}
\end{equation}%
For $\kappa =\left\vert \kappa \right\vert $ the choice of the constants 
\begin{equation}
g=\frac{1}{2m\kappa }>0  \label{match}
\end{equation}%
permits to obtain the absolute minimum of the action (the SD states) and it
will be adopted below. Later we will discuss the effect of not adopting Eq.(%
\ref{match}). The states are \emph{stationary} $\partial _{0}\phi =0$ and
minimise the energy ($E=0$). Adding the Gauss constraint we have a set of
two equations for stationary states corresponding to the absolute minimum of
the energy%
\begin{eqnarray}
D_{-}\phi &=&0  \label{eqmotsd} \\
F_{12} &=&\frac{i}{2\kappa }\left[ \phi ,\phi ^{\dagger }\right] \ 
\end{eqnarray}

From these equations the \emph{sinh}\textbf{-}Poisson equation is derived 
\cite{DunneJackiwTrugenberg}. The states correspond to zero curvature in a
formulation that involves the two dimensional reduction from a four
dimensional Self - Dual Yang Mills system, as shown in \cite%
{DunneJackiwTrugenberg}. Therefore we will denote this state as Self - Dual
(SD). The functions $\phi $ and $A_{\mu }$ are mixed spinors, elements of
the algebra $sl\left( 2,\mathbf{C}\right) $. Adopting the algebraic ansatz,%
\begin{equation}
\phi =\phi _{1}E_{+}+\phi _{2}E_{-}\ ,\ \phi ^{\dagger }=\phi _{1}^{\ast
}E_{-}+\phi _{2}^{\ast }E_{+}  \label{absatz1}
\end{equation}%
and%
\begin{equation}
A_{-}=aH\ ,\ A_{+}=-a^{\ast }H  \label{ansatz2}
\end{equation}%
which is based on the three generators $\left( E_{+},H,E_{-}\right) $ of the
Chevalley basis, the Gauss equation becomes

\begin{equation}
\frac{\partial a}{\partial x_{+}}+\frac{\partial a^{\ast }}{\partial x_{-}}=%
\frac{1}{k}\left( \rho _{1}-\rho _{2}\right)  \label{56}
\end{equation}

From the $E_{+}$ respectively the $E_{-}$ part of the first equation of
motion $D_{-}\phi =0$ we obtain 
\begin{equation}
\frac{\partial \phi _{1}}{\partial z}+a\phi _{1}=0  \label{64a}
\end{equation}%
\begin{equation}
\frac{\partial \phi _{2}}{\partial z}-a\phi _{2}=0  \label{64b}
\end{equation}

Using Eqs.(\ref{64a}) and its complex conjugate the left hand side of Eq.(%
\ref{56}) becomes%
\begin{equation*}
\frac{\partial a}{\partial x_{+}}+\frac{\partial a^{\ast }}{\partial x_{-}}%
=-2\frac{\partial ^{2}}{\partial z\partial z^{\ast }}\ln \left( \left\vert
\phi _{1}\right\vert ^{2}\right) =-\frac{1}{2}\Delta \ln \left( \left\vert
\phi _{1}\right\vert ^{2}\right) 
\end{equation*}%
The equation (\ref{56}) becomes 
\begin{equation}
-\frac{1}{2}\Delta \ln \rho _{1}=\frac{1}{\kappa }\left( \rho _{1}-\rho
_{2}\right)  \label{82}
\end{equation}

The Eq.(\ref{64b}) allows to express $a$ and $a^{\ast }$ in terms of $\phi
_{2}$. The left hand side of Eq.(\ref{56}) becomes 
\begin{equation*}
\frac{\partial a}{\partial x_{+}}+\frac{\partial a^{\ast }}{\partial x_{-}}=2%
\frac{\partial ^{2}}{\partial z\partial z^{\ast }}\ln \left( \left\vert \phi
_{2}\right\vert ^{2}\right) =\frac{1}{2}\Delta \ln \left( \left\vert \phi
_{2}\right\vert ^{2}\right) 
\end{equation*}%
The other form of Eq.(\ref{56}) is 
\begin{equation}
\frac{1}{2}\Delta \ln \rho _{2}=\frac{1}{\kappa }\left( \rho _{1}-\rho
_{2}\right)  \label{86}
\end{equation}%
The right hand side in Eqs.(\ref{82}) and (\ref{86}) is the same and if we
substract the equations we obtain 
\begin{equation}
\Delta \ln \rho _{1}+\Delta \ln \rho _{2} = 0  \label{87}
\end{equation}%
This means $\rho _{1}\rho _{2}=\exp \left( \sigma \right) $ where $\sigma $
is a \emph{harmonic function}, $\Delta \sigma =0$. We take $\sigma \equiv 0$%
, leading to $\rho _{1}=\rho _{2}^{-1}\equiv \rho $ and introduce a scalar
function $\psi $, defined by $\rho =\exp \left( \psi \right) $. Then the
Eqs.(\ref{82}) and (\ref{86}) take the unique form%
\begin{equation}
\Delta \ln \rho =-\frac{2}{\kappa }\left( \rho -\frac{1}{\rho }\right)
\label{eqbun}
\end{equation}%
which is the \emph{sinh}-Poisson equation (also known as the elliptic \emph{%
sinh}-Gordon equation)%
\begin{equation}
\Delta \psi +\frac{4}{\kappa }\sinh \psi =0  \label{sinhp}
\end{equation}

\bigskip

The model describes correctly the Self-Duality states and identifies the
asymptotic relaxation states of the fluid (known to be solutions of the 
\emph{sinh}-Poisson equation \cite{Montgomery1992}) with the \emph{self -
duality} states.

However we would like to examine the model when the system is not at
self-duality. Then the energy is not zero and $\rho _{1}\rho _{2}\neq 1$. It
is only at SD that we have the relationship $\rho _{1}=\rho _{2}^{-1}$ and
we can use a single $\psi $. We can however define $\omega $ on the basis of
the gauge field $A^{\mu }$, as $\omega \sim F_{12}\sim F_{+-}$. Before the
SD state is reached we see the gauge field as a velocity that carries the
matter $\phi $.

\bigskip

\section{Parallel between the field theoretical and the statistical
approaches}

One cannot establish a simple mapping from the notions and operations in the
fluid model ($\psi $,$\mathbf{v},\omega $), the point-like vortex model ($%
x_{i},y_{i}$) and the field-theoretical model ($\phi $, $A_{\mu }$). In the
following we note few suggestive connections.

\subsection{The condition of consistency}

For an arbitrary position $\left( x,y\right) $ in plane, the sum of the
contributions of all point-like vortices, propagated through $G\left(
x,y;x^{\prime },y^{\prime }\right) $ the Green functions of the Laplacian (%
\emph{i.e.} the right hand side of the Eq.(\ref{statiseqs}) ) gives the
velocity that would have a point-like vortex if it were placed in that point 
$\left( x,y\right) $. Knowing the local space variation of this velocity one
can calculate the vorticity in that particular point. On the other hand the
density of point-like vortices in that particular point (positive and
negative) also determines the vorticity. We then dispose of the vorticity in 
$\left( x,y\right) $ calculated in two ways: from the rotational of the
velocity derived from the contributions of all point-like vortices
(excepting the current point $\left( x,y\right) $ to avoid singularity),
and, on the other hand, from the density of positive/negative point-like
accumulations in a differential area around $\left( x,y\right) $. The
consistency imposes that these two values of vorticity are identical. For
the discrete model this remains an imaginary exercise but in FT this
compatibility is ensured by the Gauss law (or constraint) which is the
second of the equations of motion of the FT model, obtained after functional
variation to the time-like component of the gauge field $A_{0}\left(
x,y\right) $. It expresses the fact that $F_{12}$ , which is the magnetic
field $B$ or the rotational of the velocity, is equal to the zero-component,
(the \textquotedblleft charge\textquotedblright\ density) of the current,
which is the difference $\rho _{1}-\rho _{2}$ or the vorticity, at SD. A
similar conclusion is arrived at by Montgomery 1993: self-consistency means
that the\ \textquotedblleft most probable\textquotedblright\ state generates
the velocity field in which the vortices are convected.

The condition is $\frac{i}{2\kappa }J^{0}=F_{12}$ which must be read in this
order: the vorticity (the density of point-like positive/negative vortices,
more generally $J^{0}$) is equal with the rotational of the velocity, \emph{%
i.e.} the curvature of the connection $A_{\mu }$.

\subsection{None of the two kinds of point-like vortices in a point can be
zero}

In the discrete model the value of the vorticity in every cell is obtained
as an unbalance between the positive and negative vortices%
\begin{equation}
\omega _{i}=-\left( N_{i}^{+}-N_{i}^{-}\right)   \label{omnpnm}
\end{equation}%
Joyce and Montgomery \cite{JoyceMontgomery} find the relation%
\begin{equation}
N_{i}^{+}N_{i}^{-}=const  \label{npnmunu}
\end{equation}%
which means that in the same state $i$ the number of positve vortices is the
inverse of the number of negative vortices. The state $i$ is actually the
space position $\left( x,y\right) $. This excludes the situation that one of 
$N_{i}^{\pm }$ can be zero. The same relationship is derived in the FT model
[Eq.(\ref{87})]. This becomes at SD a property of invariance of the FT model
to the inversion: $\rho \rightarrow 1/\rho $.

\subsection{The energy}

The energy of the fluid is%
\begin{equation}
\mathit{E}=\frac{1}{2}\int d^{2}r\left\vert \mathbf{\nabla }\psi \right\vert
^{2}=-\frac{1}{2}\int d^{2}r\;\omega \psi  \label{efluid}
\end{equation}%
If we simply translate this expression in terms of FT variables at SD it
results%
\begin{eqnarray}
E^{FT} &=&\frac{1}{\kappa }\int d^{2}r\ \left( \rho -\frac{1}{\rho }\right)
\ln \rho  \label{efl} \\
&=&\frac{1}{\kappa }\int d^{2}r\left( \rho \ln \rho +\frac{1}{\rho }\ln 
\frac{1}{\rho }\right) \   \nonumber
\end{eqnarray}%
which is connected with the entropy $S=2\beta E$ of the discrete system but
expressed in terms of the variable $\rho $, 
\begin{equation}
S=\ln W=\sum\limits_{i}\left( N_{i}^{+}\ln N_{i}^{+}+N_{i}^{-}\ln
N_{i}^{-}\right)  \label{entr}
\end{equation}%
and suggests the identifications $N_{i}^{+}\rightarrow \rho $ and $%
N_{i}^{-}\rightarrow 1/\rho $ at SD.

\subsection{The helicity in the FT description}

The conventional helicity density is zero in $2D$: $\mathbf{v\cdot \omega }=0
$. However the Chern - Simons term in the Lagrangian carries a similar
significance (one easily recognizes that the CS term generalizes the product 
$\mathbf{A\cdot B}$ , \emph{i.e.} the helicity of a magnetic field
configuration). At stationarity, as is SD, the Chern - Simons term becomes%
\begin{eqnarray}
\hspace*{-1cm}
-\kappa \varepsilon ^{\mu \nu 0}\mathrm{tr}\left( \left( \partial _{\mu
}A_{\nu }\right) A_{0}+\frac{2}{3}A_{\mu }A_{\nu }A_{0}\right)  &=&-\kappa
\varepsilon ^{ij}\mathrm{tr}\left( A_{i}\overset{\cdot }{A}_{j}\right)
-\kappa \mathrm{tr}\left( A_{0}F_{12}\right)   \label{h1} \\
&=&-\kappa \mathrm{tr}\left( A_{0}F_{12}\right) \ 
\end{eqnarray}%
and from the Gauss constraint ($H$ is the Cartan generator)%
\begin{equation}
A_{0}=-\frac{i}{4m\kappa }\left[ \phi ,\phi ^{\dagger }\right] =-\frac{i}{%
4m\kappa }\left( \rho _{1}-\rho _{2}\right) H=\left( \frac{i}{8m}\omega
\right) H  \label{h2}
\end{equation}%
and $F_{12}\equiv F_{xy}=B=\left( -i\omega /4\right) H$. From (\ref{h2}) we
note that $A_{0}$ is purely imaginary. The field $B$ depends on the matter
functions $\rho _{1,2}$ via the Gauss constraint$\ \ $%
\begin{equation}
B=F_{12}=\frac{i}{2\kappa }\left[ \phi ,\phi ^{\dagger }\right] =\left( -%
\frac{i}{4}\omega \right) H  \label{h4}
\end{equation}%
with the last equality valid at SD. At stationarity 
\begin{equation}
\mathcal{L}_{CS}=-\kappa \mathrm{tr}\left( A_{0}F_{12}\right) =-\omega ^{2}%
\frac{\kappa }{16m}  \label{h5}
\end{equation}

This part of the action functional is related to the helicity of the field.
We note however that it has the same nature as the matter field
self-interaction (last term in the Lagrangian) which means that at SD the
physical vorticity is actually represented by two distinct functions: using
the matter field $\sim \left[ \phi ,\phi ^{\dagger }\right] $ and
respectively using the gauge field $F_{12}$.

\subsection{The Entropy}

The statistical approach (SA) to the discretized model uses the entropy of
the gas of point-like vortices and looks for its extremum under the
constraints of constant number of positive and negative vortices
(separately) and of constant energy. To draw a parallel between the
statistical approach and the FT model we write the \emph{partition function}
for the FT Lagrangian. Since the field theory is purely classical, a partition function has only a meaning if we have a statistical ensemble of realizations of the fields, due to either a random initialization or to an external random factor \cite{MartinSiggiaRose}, \cite{RVJensen}. Without an in-depth investigation, we just indicate below the possible mapping between the specific quantities in the two approaches%
\begin{eqnarray}
Z &=&\int D\left[ \phi \right] D\left[ \phi ^{\dagger }\right] D\left[
A^{\mu }\right] D\left[ A^{\mu \dagger }\right] \exp \left( i\int d^{2}xdt\ 
\mathcal{L}\right)  \label{Z} \\
&=&\int D\left[ \phi \right] D\left[ \phi ^{\dagger }\right] D\left[ A_{+}%
\right] D\left[ A_{-}\right] \delta \left( \Phi \right) \   \nonumber \\
&&\times \exp \left\{ i\int d^{2}x\left[ 4\rho _{1}\left\vert \frac{\partial 
}{\partial z}\ln \phi _{1}+a\right\vert ^{2}+4\rho _{2}\left\vert \frac{%
\partial }{\partial z}\ln \phi _{2}-a\right\vert ^{2}\right] \right\} \  
\nonumber
\end{eqnarray}%
with Jacobian $1$ for the change of variables $\left( A^{\mu },A^{\mu
\dagger }\right) \rightarrow \left( A_{+},A_{-}\right) \rightarrow \left(
a,a^{\ast }\right) $ and $\delta \left( \Phi \right) $ is the Dirac
functional expressing the Gauss constraint, denoted for simplicity $\Phi
\left( \phi ,A_{\mu }\right) =0$. The following associations are suggested%
\begin{equation}
\hspace*{-1cm}
\frac{N!}{\prod\limits_{i}N_{i}^{+}!}\rightarrow \int^{\left( 1\right) }D%
\left[ \phi _{1}\right] D\left[ \phi _{1}^{\ast }\right] D\left[ a\right] D%
\left[ a^{\ast }\right] \exp \left\{ i\int d^{2}x\ 4\rho _{1}\left\vert 
\frac{\partial }{\partial z}\ln \phi _{1}+a\right\vert ^{2}\right\}
\label{comb1}
\end{equation}%
and 
\begin{equation}
\hspace*{-1cm}
\frac{N!}{\prod\limits_{i}N_{i}^{-}!}\rightarrow \int^{\left( 2\right) }D%
\left[ \phi _{2}\right] D\left[ \phi _{2}^{\ast }\right] D\left[ a\right] D%
\left[ a^{\ast }\right] \exp \left\{ i\int d^{2}x\ 4\rho _{2}\left\vert 
\frac{\partial }{\partial z}\ln \phi _{2}-a\right\vert ^{2}\right\}
\label{comb2}
\end{equation}%
The upperscripts $\left( 1\right) $ and $\left( 2\right) $ have the meaning
that the integrations extends over function sub-space restricted by the
Gauss law, which means that the two integrals are not independent factors in
the product leading to (\ref{Z}). The same is the case in Eq.(\ref{W}) where
the two factors are connected by the the constraints $\sum%
\limits_{i}N_{i}^{+}=N^{+}=N$ and $\sum\limits_{j}N_{j}^{-}=N^{-}=N$ and by
fixed total energy $E$. The self-duality necessarly calls for the equality
of total positive and total negative vorticities (see Appendix \ref{App:AppendixA}).

The Gauss constraint is%
\begin{equation}
\delta \left( \Phi \right) \equiv \delta \left[ \left( \partial
_{+}a+\partial _{-}a^{\ast }\right) -\frac{1}{\kappa }\left( \rho _{1}-\rho
_{2}\right) \right]  \label{deltafunc}
\end{equation}%
The partition function is calculated taking the saddle point solution, which
is equivalent with Eqs.(\ref{64a}) and (\ref{64b}) leading to the \emph{sinh}%
-Poisson equation: the argument in (\ref{deltafunc}) of the $\delta $
function vanishes.

In the Eq.(\ref{comb1}) the left hand side is the number of the possible
configurations that the system of $N^{+}$ indiscernable point-like objects
can take in $i$ states, \emph{i.e.} with occupation numbers $N_{i}^{+}$. In
the right hand side we have, at SD when the exponent is zero, the volume of
the functional subspace formed by states that fulfill the first equation
that leads to SD. The same is valid for the second equation, for $N^{-}$.
The vacuum is the state with the energy of the discrete system as%
\begin{equation}
N_{i}^{+}=N_{i}^{-}  \label{vac}
\end{equation}%
which corresponds to the \emph{vacuum} in FT at $\rho _{i}=1$. This is
equivalent with \emph{pairing} of opposite vortices.

\bigskip

\section{The equations of the field theoretical model close to the self-dual
states}

\subsection{The equations for the matter field components}

The Euler - Lagrange equations resulting from the Lagrangian (\ref{Lagrange}%
) are%
\begin{equation}
iD_{0}\phi =-\frac{1}{2m}D_{+}D_{-}\phi -\frac{1}{4m\kappa }\left[ \left[
\phi ,\phi ^{\dagger }\right] ,\phi \right]   \label{eqmotion1}
\end{equation}%
and (the Gauss constraint)%
\begin{equation}
\kappa \varepsilon ^{\mu \nu \rho }F_{\nu \rho }=iJ^{\mu }  \label{gausscon}
\end{equation}%
The calculations are detailed in Appendix \ref{App:AppendixB}. These equations are valid in general,
not only at self - duality. In contrast to the latter they are difficult to
study since an explicit solution is not available. We will try to
investigate the equations in a regime that is close to the SD state. We
retain the time dependence (which necessarly is slow close to stationarity $%
\partial _{0}\rightarrow 0$) maintain $\rho _{1}$ and $\rho _{2}$ unrelated (%
$\rho _{1}\rho _{2}=1$ exists only at SD) and assume the same algebraic
structure as for SD states (see Appendices C and D).

We start by examining what can be obtained from the Gauss constraint since
it is always valid 
\begin{equation}
F_{12}=\frac{i}{2\kappa }\left[ \phi ,\phi ^{\dagger }\right]   \label{f12}
\end{equation}%
It provides a formal expression for the gauge potential components $A_{x,y}$%
. Inserting the algebraic ansatz the left hand side is%
\begin{eqnarray}
F_{12} &=&\partial _{x}A_{y}-\partial _{y}A_{x}+\left[ A_{x},A_{y}\right] 
\label{f12bar} \\
\overline{F}_{12} &=&\partial _{x}\overline{A}_{y}-\partial _{y}\overline{A}%
_{x}\ 
\end{eqnarray}%
where we denote by \emph{bar} the amplitudes along the gauge group generator 
$H$, $A_{\pm }=\overline{A}_{\pm }H$ and their combinations. The Gauss
constraint becoms an equation for the field of vectors $\overline{\mathbf{A}}%
\mathbf{\equiv }\left( \overline{A}_{x},\overline{A}_{y}\right) $ 
\begin{equation}
\mathit{curl}\overline{\mathbf{A}}=\frac{i}{2\kappa }\left( \rho _{1}-\rho
_{2}\right)   \label{Gauss7}
\end{equation}

The general solution contains the rotational of a vector field, which we
take $\frac{i}{4}g\widehat{\mathbf{e}}_{z}$ with $g$ a scalar function, plus
the gradient of another scalar function, $\frac{i}{2}h$.%
\begin{equation}
\overline{A}_{x}=\frac{i}{4}\frac{\partial g}{\partial y}+\frac{i}{2}\frac{%
\partial }{\partial x}h\ ,\ \overline{A}_{y}=-\frac{i}{4}\frac{\partial g}{%
\partial x}+\frac{i}{2}\frac{\partial }{\partial y}h\   \label{1316d}
\end{equation}

If the scalar function $g$ is found such that%
\begin{equation}
-\frac{1}{4}i\frac{\partial ^{2}g}{\partial x^{2}}-\frac{1}{4}i\frac{%
\partial ^{2}g}{\partial y^{2}}=\frac{i}{2\kappa }\left( \rho _{1}-\rho
_{2}\right)  \label{1318d}
\end{equation}%
or%
\begin{equation}
\Delta g=-\frac{2}{\kappa }\left( \rho _{1}-\rho _{2}\right)  \label{lapg}
\end{equation}%
then the Gauss law is verified and we dispose of formal expressions for $%
\overline{A}_{x,y}$ in terms of $\rho _{1}-\rho _{2}$. What we have done is
just to eliminate the gauge field components in view of reducing the problem
to only the matter field equation, Eq.(\ref{eqmotion1}).

The equation of motion (\ref{eqmotion1}) is expanded and, matching the
coefficients of each generator $E_{\pm }$ we obtain two equations for the
scalar function $\phi _{1,2}$. This is shown in detail in Appendix \ref{App:AppendixC}. The equation
resulting from $E_{+}$. 
\begin{eqnarray}
&&i\frac{\partial \phi _{1}}{\partial t}-2ib\phi _{1}  \label{579} \\
&=&-\frac{1}{2}\frac{\partial ^{2}\phi _{1}}{\partial x^{2}}-\frac{1}{2}%
\left[ \frac{\partial \left( a-a^{\ast }\right) }{\partial x}\phi
_{1}+\left( a-a^{\ast }\right) \frac{\partial \phi _{1}}{\partial x}\right]
\   \nonumber \\
&&-\frac{1}{2}\frac{\partial \phi _{1}}{\partial x}\left( a-a^{\ast }\right)
-\frac{1}{2}\left( a-a^{\ast }\right) ^{2}\phi _{1}\   \nonumber \\
&&-\frac{1}{2}\frac{\partial ^{2}\phi _{1}}{\partial y^{2}}-\frac{i}{2}\left[
\frac{\partial \left( a+a^{\ast }\right) }{\partial y}\phi _{1}+\left(
a+a^{\ast }\right) \frac{\partial \phi _{1}}{\partial y}\right] \   \nonumber \\
&&-\frac{i}{2}\frac{\partial \phi _{1}}{\partial y}\left( a+a^{\ast }\right)
+\frac{1}{2}\left( a+a^{\ast }\right) ^{2}\phi _{1}\   \nonumber \\
&&-\frac{1}{m\kappa }\left( \rho _{1}-\rho _{2}\right) \phi _{1}\   \nonumber
\end{eqnarray}

The equation resulting from $E_{-}$. 
\begin{eqnarray}
&&i\frac{\partial \phi _{2}}{\partial t}+2ib\phi _{2}  \label{580} \\
&=&-\frac{1}{2}\frac{\partial ^{2}\phi _{2}}{\partial x^{2}}+\frac{1}{2}%
\left[ \frac{\partial \left( a-a^{\ast }\right) }{\partial x}\phi
_{2}+\left( a-a^{\ast }\right) \frac{\partial \phi _{2}}{\partial x}\right]
\   \nonumber \\
&&+\frac{1}{2}\frac{\partial \phi _{2}}{\partial x}\left( a-a^{\ast }\right)
-\frac{1}{2}\left( a-a^{\ast }\right) ^{2}\phi _{2} \   \nonumber \\
&&-\frac{1}{2}\frac{\partial ^{2}\phi _{2}}{\partial y^{2}}+\frac{i}{2}\left[
\frac{\partial \left( a+a^{\ast }\right) }{\partial y}\phi _{2}+\left(
a+a^{\ast }\right) \frac{\partial \phi _{2}}{\partial y}\right] \   \nonumber \\
&&+\frac{i}{2}\frac{\partial \phi _{2}}{\partial y}\left( a+a^{\ast }\right)
+\frac{1}{2}\left( a+a^{\ast }\right) ^{2}\phi _{2} \   \nonumber \\
&&+\frac{1}{m\kappa }\left( \rho _{1}-\rho _{2}\right) \phi _{2} \   \nonumber
\end{eqnarray}%
With them we will derive equations for the two amplitudes $\rho _{1,2}$ and
also for their combinations $\rho _{1}\pm \rho _{2}$. For this we first
introduce explicit expressions for the two functions $\phi _{1}$ and $\phi
_{2}$,%
\begin{eqnarray}
\phi _{1} &=&\sqrt{\rho _{1}}\exp \left( i\chi \right) =\exp \left( \frac{%
\psi _{1}}{2}+i\chi \right)  \label{phi1phi2} \\
\phi _{2} &=&\sqrt{\rho _{2}}\exp \left( i\eta \right) =\exp \left( \frac{%
\psi _{2}}{2}+i\eta \right) \ 
\end{eqnarray}%
It is now useful to look for the SD case, such as to get an orientation of
what will be the structure of the equations amenable to the SD state. At SD
we have a unique $\psi $, $\rho _{1}=\exp \left( \psi \right) =\rho
_{2}^{-1} $ and%
\begin{eqnarray}
a &=&-\frac{\partial }{\partial z}\ln \phi _{1}=-\frac{\partial }{\partial z}%
\left( \frac{\psi }{2}+i\chi \right)  \label{aa} \\
a &=&\frac{\partial }{\partial z}\ln \phi _{2}=\frac{\partial }{\partial z}%
\left( \frac{\psi }{2}+i\eta \right) \ 
\end{eqnarray}

From Eq.(\ref{ansatz2}) the expressions of the gauge potentials at SD are%
\begin{eqnarray}
A_{x} &=&\frac{1}{2}\left( a-a^{\ast }\right) H=\frac{i}{2}\left( \frac{1}{2}%
\frac{\partial \psi }{\partial y}-\frac{\partial \chi }{\partial x}\right) H=%
\frac{i}{2}\left( -\frac{1}{2}\frac{\partial \psi }{\partial y}+\frac{%
\partial \eta }{\partial x}\right) H  \label{axysd} \\
A_{y} &=&\frac{i}{2}\left( a+a^{\ast }\right) H=-\frac{i}{2}\left( \frac{1}{2%
}\frac{\partial \psi }{\partial x}+\frac{\partial \chi }{\partial y}\right)
H=\frac{i}{2}\left( -\frac{1}{2}\frac{\partial \psi }{\partial x}+\frac{%
\partial \eta }{\partial y}\right) H\  \\
A_{0} &=&-\frac{i}{4m\kappa }\left[ \phi ,\phi ^{\dagger }\right] =-\frac{i}{%
4m\kappa }\left( \rho _{1}-\rho _{2}\right) H=\left( \frac{i}{8m}\omega
\right) H\equiv bH\ 
\end{eqnarray}%
We get the indication that at SD the $\left( x,y\right) $ gauge components
are purely imaginary and the first contribution in each of them is the \emph{%
curl} of $\psi \widehat{\mathbf{e}}_{z}$. This part is the physical
velocity, $-\mathbf{\nabla }\psi \times \widehat{\mathbf{e}}_{z}$, if $\psi $
is the streamfunction. Since all components of the gauge potential are
laying along the Cartan generator $H$ in the space of the gauge algebra the
convection $\left[ A_{\pm },\right] $ part of the covariant derivative
operator does not affect the algebraic content of the matter field, $\phi $,
assumed to be a combination of the other two generators.

Returning to Eqs.(\ref{579}) and (\ref{580}) we introduce the definitions 
\begin{equation}
v_{x}^{\left( 1\right) }=\frac{2\overline{A}_{x}}{i}+\frac{\partial \chi }{%
\partial x}\ ,\ v_{y}^{\left( 1\right) }=\frac{2\overline{A}_{y}}{i}+\frac{%
\partial \chi }{\partial y}  \label{v1def}
\end{equation}%
\begin{equation}
v_{x}^{\left( 2\right) }=-\frac{2\overline{A}_{x}}{i}+\frac{\partial \eta }{%
\partial x}\ ,\ v_{y}^{\left( 2\right) }=-\frac{2\overline{A}_{y}}{i}+\frac{%
\partial \eta }{\partial y}  \label{v2def}
\end{equation}%
and taking into account that $b+b^{\ast }=0$ we derive the equations for the
difference and for the sum $\rho _{1}\pm \rho _{2}$.%
\begin{equation}
\frac{\partial }{\partial t}\left( \rho _{1}-\rho _{2}\right) +\frac{%
\partial }{\partial x}\left[ v_{x}^{\left( 1\right) }\rho _{1}-v_{x}^{\left(
2\right) }\rho _{2}\right] +\frac{\partial }{\partial y}\left[ v_{y}^{\left(
1\right) }\rho _{1}-v_{y}^{\left( 2\right) }\rho _{2}\right] =0
\label{rhodif1}
\end{equation}%
and similarly%
\begin{equation}
\frac{\partial }{\partial t}\left( \rho _{1}+\rho _{2}\right) +\frac{%
\partial }{\partial x}\left[ v_{x}^{\left( 1\right) }\rho _{1}+v_{x}^{\left(
2\right) }\rho _{2}\right] +\frac{\partial }{\partial y}\left[ v_{y}^{\left(
1\right) }\rho _{1}+v_{y}^{\left( 2\right) }\rho _{2}\right] =0
\label{rhosum1}
\end{equation}%
(The calculations are presented in detail in Appendix \ref{App:AppendixD}). These equations generalize those of
the Abelian model of Ref.\cite{JackiwPi}.

\bigskip

We also derive equations for the two functions $\rho _{1,2}$.%
\begin{equation}
\frac{\partial }{\partial t}\rho _{1}+\mathit{div}\left( \mathbf{v}^{\left(
1\right) }\rho _{1}\right) =0  \label{1327}
\end{equation}%
\begin{equation}
\frac{\partial }{\partial t}\rho _{2}+\mathit{div}\left( \mathbf{v}^{\left(
2\right) }\rho _{2}\right) =0  \label{1328}
\end{equation}

\subsection{The velocity fields}

The first velocity field 
\begin{equation}
\mathbf{v}^{\left( 1\right) }=\frac{1}{2}\mathbf{\nabla }g\times \widehat{%
\mathbf{e}}_{z}+\mathbf{\nabla }\left( h+\chi \right)  \label{1329}
\end{equation}%
and the second velocity field%
\begin{equation}
\mathbf{v}^{\left( 2\right) }=-\frac{1}{2}\mathbf{\nabla }g\times \widehat{%
\mathbf{e}}_{z}+\mathbf{\nabla }\left( -h+\eta \right)  \label{1330}
\end{equation}%
differ by the phases of the functions $\phi _{1}$ and $\phi _{2}$, \emph{i.e.%
} by $\chi $ and $\eta $. We try to learn more about the velocity fields $%
\mathbf{v}^{\left( 1,2\right) }$ by taking the limit to SD.

The formal solutions of the equation%
\begin{equation*}
\mathit{curl}\overline{\mathbf{A}}=\frac{i}{2\kappa }\left( \rho _{1}-\rho
_{2}\right) 
\end{equation*}%
is expressed as 
\begin{eqnarray}
\overline{A}_{x} &=&-\frac{\partial }{\partial y}\int d\mathbf{r}^{\prime
}G\left( \mathbf{r-r}^{\prime }\right) \left\{ \frac{i}{2\kappa }\left[ \rho
_{1}\left( \mathbf{r}^{\prime },t\right) -\rho _{2}\left( \mathbf{r}^{\prime
},t\right) \right] \right\}  \label{Abarx3} \\
&&+gauge \ term \   \nonumber
\end{eqnarray}%
\begin{eqnarray}
\overline{A}_{y} &=&\frac{\partial }{\partial x}\int d\mathbf{r}^{\prime
}G\left( \mathbf{r-r}^{\prime }\right) \left\{ \frac{i}{2\kappa }\left[ \rho
_{1}\left( \mathbf{r}^{\prime },t\right) -\rho _{2}\left( \mathbf{r}^{\prime
},t\right) \right] \right\}  \label{Abary3} \\
&&+ gauge \ term \   \nonumber
\end{eqnarray}%
and at SD, where we have a unique $\psi $, 
\begin{equation*}
\rho _{1}\left( \mathbf{r}^{\prime },t\right) -\rho _{2}\left( \mathbf{r}%
^{\prime },t\right) \rightarrow \frac{-\kappa }{2}\omega \left( x,y\right) =-%
\frac{\kappa }{2}\Delta \psi \left( x,y\right) 
\end{equation*}%
We can choose the gauge terms such as to cancel the gradients in Eqs.(\ref%
{1316d}). Alternatively we can use Eq.(\ref{axysd})%
\begin{equation}
v_{x}^{\left( 1\right) }=\frac{2}{i}\overline{A}_{x}+\frac{\partial \chi }{%
\partial x}\rightarrow \frac{1}{2}\frac{\partial \psi }{\partial y}\ ,\
v_{y}^{\left( 1\right) }=\frac{2}{i}\overline{A}_{y}+\frac{\partial \chi }{%
\partial x}\rightarrow -\frac{1}{2}\frac{\partial \psi }{\partial x}
\label{v1}
\end{equation}%
Similarly for the second velocity field%
\begin{equation}
v_{x}^{\left( 2\right) }=-\frac{2}{i}\overline{A}_{x}+\frac{\partial \eta }{%
\partial x}\rightarrow -\frac{1}{2}\frac{\partial \psi }{\partial y}\ ,\
v_{y}^{\left( 2\right) }=-\frac{2}{i}\overline{A}_{y}+\frac{\partial \eta }{%
\partial y}\rightarrow \frac{1}{2}\frac{\partial \psi }{\partial x}
\label{v2}
\end{equation}

At SD both velocity fields become divergenceless $\mathbf{\nabla \cdot v}%
^{\left( 1\right) }=0$, $\mathbf{\nabla \cdot v}^{\left( 2\right) }=0$ and
they are opposite%
\begin{equation}
\mathbf{v}^{\left( 2\right) }=-\mathbf{v}^{\left( 1\right) }  \label{v1v2neg}
\end{equation}

If we assume that these properties are approximately fulfilled in the states
close (but not at) SD, we get%
\begin{equation}
\frac{\partial }{\partial t}\left( \rho _{1}-\rho _{2}\right) +\frac{%
\partial }{\partial x}\left[ v_{x}^{\left( 1\right) }\left( \rho _{1}+\rho
_{2}\right) \right] +\frac{\partial }{\partial y}\left[ v_{y}^{\left(
1\right) }\left( \rho _{1}+\rho _{2}\right) \right] \approx 0  \label{spin}
\end{equation}%
and respectively%
\begin{equation}
\frac{\partial }{\partial t}\left( \rho _{1}+\rho _{2}\right) +\frac{%
\partial }{\partial x}\left[ v_{x}^{\left( 1\right) }\left( \rho _{1}-\rho
_{2}\right) \right] +\frac{\partial }{\partial y}\left[ v_{y}^{\left(
1\right) }\left( \rho _{1}-\rho _{2}\right) \right] \approx 0  \label{chir}
\end{equation}%
After replacing the SD expression of $\mathbf{v}^{\left( 1\right) }$ and
taking into account that at SD $\rho _{1}=\exp \left( \psi \right) $, $\rho
_{2}=\exp \left( -\psi \right) $, we see that both equations become a simple
statement of the stationarity $\partial \left( \rho \pm 1/\rho \right)
/\partial t=0$.

\subsection{The current of the matter field}

The expressions of the matter current will help us to prove that the FT
reproduces in the continuum limit the equations of the point-like vortices
Eqs.(\ref{statiseqs}). In field theory $J^{\mu }$ is calculated according to
standard procedures%
\begin{eqnarray}
J^{0} &=&\left[ \phi ,\phi ^{\dagger }\right]   \label{curr} \\
J^{i} &=&-\frac{i}{2m}\left( \left[ \phi ^{\dagger },D_{i}\phi \right] -%
\left[ \left( D_{i}\phi \right) ^{\dagger },\phi \right] \right) \ 
\end{eqnarray}

Using the \emph{algebraic ansatz for }$\phi $ and $A_{\mu }$ we obtain the
following expressions%
\begin{eqnarray}
m\overline{J}^{x} &=&-\rho _{1}\frac{\partial \chi }{\partial x}+\rho _{2}%
\frac{\partial \eta }{\partial x}+i(a-a^{\ast })\left( \rho _{1}+\rho
_{2}\right)   \label{c1} \\
&=&-\rho _{1}\frac{\partial \chi }{\partial x}+\rho _{2}\frac{\partial \eta 
}{\partial x}-\frac{2\overline{A}_{x}}{i}\left( \rho _{1}+\rho _{2}\right) \ 
\nonumber
\end{eqnarray}%
\begin{eqnarray}
m\overline{J}^{y} &=&-\rho _{1}\frac{\partial \chi }{\partial y}+\rho _{2}%
\frac{\partial \eta }{\partial y}-\left( a+a^{\ast }\right) \left( \rho
_{1}+\rho _{2}\right)   \label{c2} \\
&=&-\rho _{1}\frac{\partial \chi }{\partial y}+\rho _{2}\frac{\partial \eta 
}{\partial y}-\frac{2\overline{A}_{y}}{i}\left( \rho _{1}+\rho _{2}\right) \ 
\nonumber
\end{eqnarray}%
\begin{equation}
\overline{J}^{0}=\rho _{1}-\rho _{2}  \label{c0}
\end{equation}%
in which the gauge potentials $\overline{A}_{x,y}$ appear. The detailed
calculations are in the Appendices E and F.

We now examine these expressions close to SD. From the first equation of
self-duality, $D_{-}\phi =0$ we obtain the combinations of $a$ and $a^{\ast }
$ as%
\begin{equation}
a+a^{\ast }=-\frac{1}{2}\frac{\partial \psi }{\partial x}-\frac{\partial
\chi }{\partial y}  \label{apa}
\end{equation}%
\begin{equation}
a-a^{\ast }=i\left( \frac{1}{2}\frac{\partial \psi }{\partial y}-\frac{%
\partial \chi }{\partial x}\right)   \label{ama}
\end{equation}%
Further, we take $\rho _{1}\rightarrow \exp \left( \psi \right) $ and $\rho
_{2}\rightarrow \exp \left( -\psi \right) $. At SD the phases of $\phi _{1}$
and $\phi _{2}$ are opposite $\chi =-\eta $. \ Then it is obtained, close to
SD%
\begin{equation}
m\overline{J}^{x}\approx -\left( \rho _{1}+\rho _{2}\right) v_{x}^{\left(
1\right) }=-\frac{\partial }{\partial y}\frac{1}{2}\left( \rho _{1}-\rho
_{2}\right) =\frac{\kappa }{4}\frac{\partial }{\partial y}\omega 
\label{jxom}
\end{equation}%
and%
\begin{equation}
m\overline{J}^{y}\approx -\left( \rho _{1}+\rho _{2}\right) v_{y}^{\left(
1\right) }=\frac{\partial }{\partial x}\frac{1}{2}\left( \rho _{1}-\rho
_{2}\right) =-\frac{\kappa }{4}\frac{\partial }{\partial x}\omega 
\label{jyom}
\end{equation}

To this we have to add%
\begin{equation}
\overline{J}^{0}\approx \rho _{1}-\rho _{2}=-\frac{\kappa }{2}\omega
\label{j0m}
\end{equation}

\bigskip

The formulas can be written in the form%
\begin{eqnarray}
\overline{J}^{x} &\rightarrow &-\frac{1}{2m}\left( \rho _{1}+\rho
_{2}\right) \frac{\partial \psi }{\partial y}  \label{eqmot} \\
\overline{J}^{y} &\rightarrow &\frac{1}{2m}\left( \rho _{1}+\rho _{2}\right) 
\frac{\partial \psi }{\partial x}\ 
\end{eqnarray}%
We note that these expression for $\overline{J}^{x,y}/\left( \rho _{1}+\rho
_{2}\right) $ coincide at SD with Eqs.(\ref{statiseqs}).

\section{Discussion}

Detailed calculations regarding the properties of the velocity fields and the
currents can be found in Appendixes A to F. One may find that the field-theoretical formulation of the $2D$ Euler
fluid has a consistent background that justifies applications and/or extension.

\subsection{Few comments}

The FT is based on a dual representation of the same physical object: the
vorticity. It is the density of matter $J^{0}=\left[ \phi ,\phi ^{\dagger }%
\right] $ and is the magnetic field $F_{12}=B\sim \left[ \phi ,\phi
^{\dagger }\right] $ ; the Gauss law constrains them to be equal. This
representation unfolds the nonlinearity of Eq.(\ref{eq1}) but expresses it
in a different way: the gauge-field-induced repulsion between elements of
vorticity (part of the kinetic energy) is balanced by the two-body $\delta $%
-function attraction represented by the last term in the Lagrangian (it is
true for vortices of each sign; in addition, we must have made the option
Eq.(\ref{match})). This permits that at self-duality the differential degree
in the equations of motion to be decreased: the first SD equation (\ref%
{eqmotsd}) is \emph{first-order} differential in contrast with Eq.(\ref%
{eqmotion}) which is second order.

The FT reveals that the essential nature of self-organization is
topological. Less visible in the case of the (present) Euler  model, it is
explicit in the FT models for fluids of single-sign vorticity (leading to
the Liouville equation), etc. where the asymptotic states are mappings
between compact manifolds and the energy is bounded from below by an integer
multiple of the magnetic flux of a single vortex. Since $B\sim \omega $ the
suggestion is clear: only the vorticity can self-organize, the combinations
like the potential vorticity do not have this property. Essentially $B$ and $%
\omega $ are flux-like quantities, we must think to them as $Bdx\wedge dy$
and $\omega dx\wedge dy$, \emph{i.e.} they are differential two-forms. The integral over the plane is the degree of the topological mappings mentioned above. We
note however that for the fluids with short range interaction like $2D$
plasma and the $2D$ atmosphere the self-organization (inherited from $\omega 
$) is approximative and the potential vorticity dominates the dynamics via
Ertel's theorem.

\subsection{The approach to SD through states where the parameters do not
obey the constraint Eq.(\protect\ref{match})}

The CS term and the matter self-interaction term combine to give a
contribution to the energy, the second term in Eq.(\ref{energia}). When the
parameters (coefficients of the CS respectively matter self-interaction
terms) are not chosen as in Eq.(\ref{match}) the energy of the system is
non-zero even if we take the SD condition $D_{-}\phi =0$. Approaching the SD
state means that these two parameters must progressively become equal.
Compared with the preceding part of this work, this gives another meaning to
\textquotedblleft being close to self-duality\textquotedblright\ but a FT
description still remains to be elaborated. Few qualitative aspects of such
a FT description are however available and we draw a parallel with the
evolution of the physical fluid in the late phases of approaching stationary
and coherent flow solutions of Eq.(\ref{eq2}).

As is well known (and reviewed in the Introduction) in the late phase of
fluid relaxation (equivalently, vorticity self-organization) the process of
separation of opposite-sign elements of vorticity and coalescence of
like-sign has led to formation of mesoscopic vortices of both signs. Their
motion in plane is much slower than the rate of rotation of the fluid on the
closed streamlines. The FT equivalent is that the energy term 
\begin{equation}
\delta E\equiv \left( -\frac{g}{2}+\frac{1}{4m\kappa }\right) \mathrm{tr}%
\left( \left[ \phi ^{\dagger },\phi \right] ^{2}\right)   \label{enerplus}
\end{equation}%
is very small. The merging of mesoscopic vortices is possible due to
dissipation-mediated reconnections of streamlines. In the physical fluid, in
such an event part of the energy is lost by dissipation and part of the
energy related to the motion of the centres of the mesoscopic vortices that
merge, is transferred to motion on streamlines. In FT we must see the term (%
\ref{enerplus}) approaching zero.

\bigskip

When the two parameters are not equal there is interaction between vortices.
This has been studied for similar FT systems (\cite{JacobsRebbi}, \cite%
{DunneBook}, \cite{Arthur}, \cite{9812103FuertesGuilarte}). When the system
is very close to SD one assumes that the mesoscopic vortices are not too
different of the exact SD vortices. Then one inserts exact solutions of Eq.(%
\ref{eq2}) into the expression of the energy (\ref{energia}), without
assuming SD (Eq.(\ref{eqmotsd}) and (\ref{match})). Taking as parameters the
positions of the centers of these exact SD vortices, it is possible to
determine the force of interaction from variation of the energy to these
parameters. The result depends decisively on the sign of the term (\ref%
{enerplus}). It is also possible to derive the relative motion of the
vortices from their geodesic flow on the manifold generated by the positions
in plane \cite{Manton}. This argument works for several FT systems but the
application to the present case is not straightforward: we have both
positive and negative vortices and the energy is bounded from below by $E=0$%
. We anticipate a more careful analysis and just mention the argument for
the present case. At SD (\emph{i.e.} $g-1/\left( 2m\kappa \right) =0$) the
total energy is zero and the solution consists of a dipole. This exact
solution approximates the one of the phase just before reaching SD, when the
field consisted of two mesoscopic vortices of opposite signs, in slow
relative motion.  We note that when $\delta E<0$ (in Eq.(\ref{enerplus})) this supplementary energy
being negative means that there is attraction between vortices. We say that
there is a predominance of the CS term ($\kappa $ is large) from which it
arises the second term in the paranthesis. Qualitatively, we say that the
evolution toward SD must involve a decay of this attraction energy, \emph{%
i.e.} at every reconnection a certain amount of the absolute magnitude of
the CS term ($\sim $ helicity) must be removed. Since we know that at SD the
CS part in the Lagrangian is%
\begin{equation*}
\mathcal{L}_{CS}^{SD}=-\kappa \mathrm{tr}\left( A_{0}F_{12}\right) =-\kappa 
\frac{1}{16m}\omega ^{2}
\end{equation*}%
we can reformulate, saying that at every reconnection event a certain amount
of enstrophy is removed. This seems to be compatible with the numerical
simulations, where the evolution toward order is associated with decrease of
the enstrophy.

We understand that the approach to SD and suppression of (\ref{enerplus})
implies the decrease of the topological content that is due to the
Chern-Simons term. This is mediated by dissipative mechanisms which are
missing from the basic formulation (\ref{Lagrange}). We can get a hint on
the necessary extension of the model from the \emph{baryogenesis}, which
involves the change of Chern-Simons topological number by transitions
between states with different topological content \cite{Nauta}, \cite%
{ArnoldSonYaffe}. A simple application is prevented by the absence of the
Higgs vacua and implicitely of the \emph{sphaleron} solutions. This study is
underway.

\subsection{The conformal transformation as mappings between solutions of
the FT equations of motion}

The FT model inherits the conformal invariance of the the $2D$ Euler fluid (%
\ref{eq1}): there is no intrinsic length in the physical system and the
length of the side of the box $L$ is just an arbitrary parameter. The
Lagrangian (\ref{Lagrange}) is invariant to conformal transformations \cite%
{DunneJackiwTrugenberg}, \cite{JackiwPiSymm}, \cite{DunneBook} and their
generators verify the following relation ($t$ is the time)%
\begin{equation*}
\mathcal{E}t^{2}-2\mathcal{D}t+\mathcal{K}>0
\end{equation*}%
where $\mathcal{E}$ is energy \emph{i.e.} the integral of Eq.(\ref{enersd}), 
$\mathcal{D}$ and $\mathcal{K}>0$ are generators of the dilation and special
conformal transformations, $\mathbf{x}\rightarrow \mathbf{x}/(1+at)$, where $a=$ const., explained in Ref.\cite{DunneJackiwTrugenberg}.
The conformal transformations allow to find new, time-dependent solutions of
the equations of motion (\ref{eqmotion}), starting from the static solutions
of the SD equation (\ref{eq2}).%
 These new solutions have energy $E>0$ which means that
they cannot spontaneously evolve from the static SD solutions without an
external input of energy. Each conformal transformation is a map in the
function space connecting solutions of (\ref{eqmotion}). It is not a
necessary dynamic change of the behavior of the system but, since each
function obtained by the conformal transformation is an extremum of the
action, the path in the function space connecting such solutions is the most
economic way for the system to access a particular type of behavior.

As noted in \cite{collapse} when $\mathcal{E}>0$ and $\mathcal{D}>\sqrt{%
\mathcal{EK}}$ there is a finite time $t^{\ast }$ such that for $%
t\rightarrow t^{\ast }$ the amplitude of the solution $\phi $ becomes zero
anywhere on the plane with the exception of $r=0$ where diverges. In
particular, when the system is initialized in this region of parameters $%
\left( \mathcal{E}=0,\mathcal{D}>0,\mathcal{K}>0\right) $ the two
opposite-sign vortices evolve to cuasi-singular concentrated spikes. When
there is no spontaneous evolution toward singularity, we note that, for a
one-dimensional solution of (\ref{eq2}), the profile of $\psi \left(
x\right) $ can be mapped to another solution $\psi ^{\prime }\left(
x,t\right) $ which, for fixed $t$ and $a>0$ is more narrow, closer to the
symmetry axis $x=0$. The velocity $v_{y}^{\prime }\left( x,t\right) =-d\psi
^{\prime }/dx$ is higher so there is need of energy for the system to evolve
from the static solution to the time-dependent one. The shear increases and, with just small external drive, the sheared layer can evolve to onset of the Kelvin-Helmholtz instability. 

\subsection{The dynamics of the $2D$ physical fluid and its FT model}

The ideal incompressible fluid in two dimensions evolves from a turbulent
initial state to a stationary, highly ordered flow pattern via mergings of
vortices and concentration of the vorticities of both signs into separate
large scale vortices. The evolution has two components:

(1) isotopological motion with preservation of all streamlines and exact
conservation of the energy

(2) fast events consisting of breaking up and reconnection of streamlines
leading to change in topology of the flow. In particular merging of vortices 
\emph{i.e.} generation of larger scale flow from two smaller vortices at
their encounter is only possible by reconnection. A dissipative mechanism is
necessary like molecular viscosity or collisions. However the amount of
energy that is lost (by heat) in this way is very small and the total energy
is approximately conserved. The events of reconnections (equivalently: the
dissipative events) take place in a set with very small measure \cite%
{Levich1987}. The main importance of reconnections is obviously the
topological re-arrangement they make possible. In this way the system get
closer to the state of SD which has a simple topological structure \cite%
{Montgomery1992}.

If we exclude any dissipative process and initialise the state such that its
energy is not minimal (zero at SD) the fluid will continue to move, never
reaching stationarity. This happens because the processes that would allow
the system to access states of lower energy, and finally the lowest of all,
the SD state, are forbidden since reconnections are not allowed.

For very small positive energy the system has only few mesoscopic vortices
moving slowly as this state only precedes the full organization into the
stationary vortex dipole solution of Eq.(\ref{eq2}). Then the motion can be
seen as consisting of the fast rotation in the vortices and the slow
displacement of their centres. In the energy-plateau states of
isotopological motion (between two reconnection events) the system creates
accumulation of streamlines in few narrow regions and these generate
conditions favorable for reconnection. The narrow regions are characterised
by high values of the gradients of vorticity and any dissipation, if exists,
will be easier exploited to start a reconnection event. The asymptotic SD
state has all motion in the vortical rotation with no displacement of the
centres of vortices.

\bigskip

The action functional reduces at stationarity to the square of an expression
of $\left( \phi ,A_{\mu }\right) $ and the states extremizing the action are
identified by taking to zero this expression. They are characterised by
equality of the total amount of positive and negative vorticity, although
the Lagrangian does not include this explicitely. By comparison, the
statistical approach based on the variational treatment of the entropy must
impose these properties and include them via Lagrange multipliers
supplementing the entropy functional extremization.

Regarding the \emph{negative temperature} determined in Taylor \cite%
{Taylor1972}, it has been shown by Joyce and Montgomery \cite{JoyceMontgomery}
and by Edwards and Taylor \cite{EdwardsTaylor} that the threshold energy is $%
E=0$ and for any \emph{positive} energy the temperature is negative. The
FT model finds indeed that the SD state has $E=0$ which must be interpreted as follows: the
energy corresponds to the situation where there is no motion of the centres
of the remaining vortices (the dipole) and the only motion is rotation along
the streamlines of the two vortices. Since the \ system of point-like
vortices is purely kinematic, the energy of the displacement along the
streamlines is zero. It means that the only change in the matter function $%
\phi $ is given by the phase modification which is due to the potential $%
A_{x,y}$. This corresponds to the rotation of the fluid on the streamlines
of the dipole. Is just an indefinite increase of the angular phase and this
is expressed by $D_{-}\phi =0$ .

\section{Conclusions}

The field theoretical formalism for the Euler fluid finds that the
asymptotic, highly organized, states are due to the property of
self-duality. It derives in a very transparent way the \emph{sinh}-Poisson
equation. It implies that all other states, either with $E\neq 0$ or
non-doubly periodic or with $\int d^{2}r\omega \neq 0$ cannot be stationary.

The fact that the asymptotic states exist due to the self-duality (as shown
by the field theoretical formulation) may help to better understand the
universal character of the vorticity concentration \cite{HopfingerVanHeijst}%
, \cite{CorcosSherman}. In fluids with similar properties ($2D$ atmosphere,
plasma in magnetic field) highly organized flows are observed \cite%
{Mcwilliams}. We must remember that the evolution of the $2D$ Euler fluid to
the coherent flow pattern [solution of Eq.(\ref{eq2})] takes place in the
absence of gradients of pressure, of gradients of temperature, of buoyancy,
of centrifugal forces, etc. Nothing was needed for the vorticity separation
and concentration, except for the nature of the nonlinearity which supports
inverse cascade, \emph{i.e. }the intrinsic tendency to self-organization of
the flow toward large scales. This process is similar to the Widom -
Rowlinson phase transition by its universality and by the fact that besides
the equation itself the input is quasi-inexistent. When formation of
structures is described, as for example the tropical cyclones and tornadoes
in atmosphere or the convection cells in plasma, etc. the necessary use of
the conservation laws as dynamical equations should not make us to forget
that inside the final pattern of flow there is also a \emph{universal}
structure. This tendency to self-organization is revealed or made more
visible at relaxation but it does not depend on any particular circumstance.
Also, the drive and dissipation in real systems can alter substantially the
structure and actually can dominate the system's behavior but there is no
way to simply suppress the tendency to self-organization, which will always
be present. We may neglect the self-organization, on quantitative basis, but
we should not ignore it \cite{FlorinMadi2005}, \cite{FlorinMadiGAFD}, \cite%
{FlorinmadiEqualRadiixxx}.

Although the field theoretical formulation of the $2D$ Euler fluid proposes
an interesting perspective on the fluid dynamics, it also has limitations:
it cannot (simply) accomodate dissipation therefore the evolution of the FT
variables actually reproduces isotopological motions of the fluid. If the
energy in the initial state is not zero the FT system does not reach
self-duality and the \emph{sinh}-Poisson solutions.

The interest for the FT formulation also comes from the developments that it
suggests: the connection with the Constant Mean Curvature (CMC) surfaces (a
flow in the SD state has an associated CMC surface); the representation of
the fluid "contour dynamics" as sections in a Riemann surface which is the
solution of a supersymmetric extension of the model; the role of the Anti-de
Sitter metrics in associating to the ideal fluid the geometric-algebraic
structure that underlies the self-duality; etc. All these are certainly
attractive fields of investigation.

\bigskip

\textbf{Acknowledgement} This work has been partly supported by the grant
ERC - Like 4/2012 of UEFISCDI Romania. 

\bigskip

%

\begin{appendices}
\section{Appendix A. The condition of zero total vorticity} \label{App:AppendixA}

\renewcommand{\theequation}{A.\arabic{equation}} \setcounter{equation}{0}

In the statistical approach (SA) it is adopted from the start the condition
that the total number of positive vortices equals the total number of
negative vortices 
\begin{equation}
N^{+}\equiv \sum\limits_{i}N_{i}^{+}=\text{const\ \ ,\ \ }N^{-}\equiv
\sum\limits_{i}N_{i}^{-}=\text{const}  \label{npmc}
\end{equation}%
and the \emph{balance} 
\begin{equation}
N_{+}=N_{-}  \label{bala}
\end{equation}%
This is equivalent to the assumption that in the surface of interest the
total amount of vorticity is zero. In FT there is no such assumption from
the beginning and we can inquire if the system identifies as extremum (the
SD state) the same situation \emph{i.e.} zero total vorticity%
\begin{equation}
\int d^{2}r\ \omega =0  \label{omzer}
\end{equation}%
This would mean 
\begin{equation}
\int d^{2}r\ \rho _{1}=\int d^{2}r\ \rho _{2}  \label{rhoi}
\end{equation}%
at SD, where $\rho _{1}=\rho =\exp \left( \psi \right) $ and $\rho _{2}=\rho
^{-1}=\exp \left( -\psi \right) $. We consider that the sign of $\kappa $ is
fixed and from the equation at SD%
\begin{equation}
\omega +\frac{2}{\kappa }\left( \rho -\frac{1}{\rho }\right) =0  \label{471}
\end{equation}%
we obtain in the regions where $\kappa \omega =+\left\vert \kappa \omega
\right\vert $ 
\begin{equation}
\rho ^{+}=\frac{1}{4}\left( -\left\vert \kappa \omega \right\vert +\sqrt{%
\left\vert \kappa \omega \right\vert ^{2}+16}\right)  \label{475}
\end{equation}%
with only the \emph{positive} root $\rho \equiv \exp \left( \psi \right) $
being retained. The upperscript means that the result is valid in the
regions with positive vorticity. In the same regions $1/\rho ^{+}=\exp
\left( -\psi \right) $ is%
\begin{equation}
\frac{1}{\rho ^{+}}=\frac{1}{4}\left( \left\vert \kappa \omega \right\vert +%
\sqrt{\left\vert \kappa \omega \right\vert ^{2}+16}\right)  \label{476}
\end{equation}

In the regions where the vorticity is negative $\kappa \omega =-\left\vert
\kappa \omega \right\vert $ we have, taking the positive root%
\begin{equation}
\rho ^{-}=\frac{1}{4}\left( \left\vert \kappa \omega \right\vert +\sqrt{%
\left\vert \kappa \omega \right\vert ^{2}+16}\right)  \label{477}
\end{equation}%
and the inverse%
\begin{equation}
\frac{1}{\rho ^{-}}=\frac{1}{4}\left( -\left\vert \kappa \omega \right\vert +%
\sqrt{\left\vert \kappa \omega \right\vert ^{2}+16}\right)  \label{478}
\end{equation}

We have to prove Eq.(\ref{rhoi}), \emph{i.e.}%
\begin{equation}
\int d^{2}r\ \rho -\int d^{2}r\ \left( 1/\rho \right) =0  \label{479}
\end{equation}%
Writting such as to exhibit the domains $\kappa \omega \gtrless 0$, 
\begin{equation}
\int^{+}d^{2}r\ \rho ^{+}+\int^{-}d^{2}r\ \rho ^{-}=\int^{+}d^{2}r\ \frac{1}{%
\rho ^{+}}+\int^{-}d^{2}r\ \frac{1}{\rho ^{-}}  \label{1400}
\end{equation}%
we have%
\begin{eqnarray*}
&&\int^{+}d^{2}r\frac{1}{4}\left( -\left\vert \kappa \omega \right\vert +%
\sqrt{\left\vert \kappa \omega \right\vert ^{2}+16}\right) +\int^{-}d^{2}r%
\frac{1}{4}\left( \left\vert \kappa \omega \right\vert +\sqrt{\left\vert
\kappa \omega \right\vert ^{2}+16}\right) \\
&=&\int^{+}d^{2}r\frac{1}{4}\left( \left\vert \kappa \omega \right\vert +%
\sqrt{\left\vert \kappa \omega \right\vert ^{2}+16}\right) +\int^{-}d^{2}r%
\frac{1}{4}\left( -\left\vert \kappa \omega \right\vert +\sqrt{\omega ^{2}+16%
}\right)
\end{eqnarray*}%
where the upper sign at the integrals labels the regions where $\kappa
\omega $ is positive respectively negative. After cancellations%
\begin{equation}
\int^{+}d^{2}r\ \left( -\frac{1}{2}\left\vert \omega \right\vert \right)
+\int^{-}d^{2}r\ \left( \frac{1}{2}\left\vert \omega \right\vert \right) =0
\label{1368}
\end{equation}%
and this indeed means that the integrals of the vorticity over the region
where it is positive equals the integral of the vorticity over the region
where it is negative%
\begin{equation}
\int^{+}d^{2}r\ \left( \frac{1}{2}\left\vert \omega \right\vert \right)
=\int^{-}d^{2}r\ \left( \frac{1}{2}\left\vert \omega \right\vert \right)
\label{1401}
\end{equation}%
equivalent with%
\begin{equation}
N^{+}=\sum\limits_{i}N_{i}^{+}=N^{-}=\sum\limits_{i}N_{i}^{-}  \label{1402}
\end{equation}%
In other words the SD gives that the total vorticity in the field is \emph{%
zero}. We note that in FT this is not an assumption but a result.

\section{Appendix B. Derivation of the equations of motion} \label{App:AppendixB}

\renewcommand{\theequation}{B.\arabic{equation}} \setcounter{equation}{0}

The Lagrangian of the model is 
\begin{eqnarray}
L &=&-\kappa \varepsilon ^{\mu \nu \rho }\mathrm{tr}\left( \left( \partial
_{\mu }A_{\nu }\right) A_{\rho }+\frac{2}{3}A_{\mu }A_{\nu }A_{\rho }\right)
\label{a1} \\
&&+i\mathrm{tr}\left( \phi ^{\dagger }\left( D_{0}\phi \right) \right) -%
\frac{1}{2m}\mathrm{tr}\left( \left( D_{k}\phi \right) ^{\dagger }\left(
D^{k}\phi \right) \right)  \notag \\
&&+\frac{1}{4m\kappa }\mathrm{tr}\left( \left[ \phi ,\phi ^{\dagger }\right]
^{2}\right)  \notag
\end{eqnarray}%
where 
\begin{equation}
D_{\mu }=\partial _{\mu }+\left[ A_{\mu },\right]  \label{a2}
\end{equation}%
and the metric is%
\begin{equation}
g_{\mu \nu }=g^{\mu \nu }=\left( 
\begin{array}{ccc}
-1 & 0 & 0 \\ 
0 & 1 & 0 \\ 
0 & 0 & 1%
\end{array}%
\right)  \label{27}
\end{equation}

\bigskip

\subsection{Preparation for the derivation of the equation of motion
equivalent to the Gauss constraint}

\subsubsection{The Chern-Simons term}

This part is presented in detail in \cite{FlorinMadiXXX} and here we only
mention the principal steps. The Chern - Simons part of the gauge Lagrangean
is%
\begin{equation}
\mathcal{L}_{CS}=-\frac{1}{2}\kappa \varepsilon ^{\mu \nu \rho }\mathrm{tr}%
\left( A_{\mu }\left( \partial _{\nu }A_{\rho }-\partial _{\rho }A_{\nu
}\right) +\frac{2}{3}A_{\mu }\left[ A_{\nu },A_{\rho }\right] \right)
\label{fb18}
\end{equation}

and expanded 
\begin{eqnarray}
\mathcal{L}_{CS} &=&-\kappa \mathrm{tr}\left\{ A_{0}\left( \partial
_{1}A_{2}\right) -A_{0}\left( \partial _{2}A_{1}\right) -A_{1}\left(
\partial _{0}A_{2}\right) \right.  \label{fb26} \\
&&+A_{1}\left( \partial _{2}A_{0}\right) -A_{2}\left( \partial
_{1}A_{0}\right) +A_{2}\left( \partial _{0}A_{1}\right)  \notag \\
&&+\frac{2}{3}A_{0}A_{1}A_{2}-\frac{2}{3}A_{0}A_{2}A_{1}-\frac{2}{3}%
A_{1}A_{0}A_{2}  \notag \\
&&\left. +\frac{2}{3}A_{1}A_{2}A_{0}-\frac{2}{3}A_{2}A_{1}A_{0}+\frac{2}{3}%
A_{2}A_{0}A_{1}\right\}  \notag
\end{eqnarray}

Using the properties of the \textrm{Trace} operator we obtain 
\begin{eqnarray}
\mathcal{L}_{CS} &=&-\kappa \mathrm{tr}\left\{ A_{0}\left( \partial
_{1}A_{2}\right) -A_{0}\left( \partial _{2}A_{1}\right) -A_{1}\left(
\partial _{0}A_{2}\right) \right.  \label{fb3310} \\
&&+A_{1}\left( \partial _{2}A_{0}\right) -A_{2}\left( \partial
_{1}A_{0}\right) +A_{2}\left( \partial _{0}A_{1}\right)  \notag \\
&&\left. +2A_{0}A_{1}A_{2}-2A_{0}A_{2}A_{1}\right\}  \notag
\end{eqnarray}

or 
\begin{eqnarray}
\mathcal{L}_{CS} &=&-\kappa \mathrm{tr}\left\{ -A_{1}\partial
_{0}A_{2}+A_{2}\partial _{0}A_{1}+2A_{0}\partial _{1}A_{2}-2A_{0}\partial
_{2}A_{1}\right.  \label{fb46} \\
&&\left. +2A_{0}A_{1}A_{2}-2A_{0}A_{2}A_{1}\right\}  \notag
\end{eqnarray}

This will be used for functional derivatives of the Lagrangian density.

\subsubsection{The matter Lagrangean}

This part is 
\begin{eqnarray}
\mathcal{L}_{m} &=&i\mathrm{tr}\left( \phi ^{\dagger }\left( D_{0}\phi
\right) \right) -\frac{1}{2m}\mathrm{tr}\left( D_{\kappa }\phi \right)
^{\dagger }\left( D^{k}\phi \right)  \label{fb50} \\
&\equiv &\mathcal{L}_{m}^{\left( 1\right) }+\mathcal{L}_{m}^{\left( 2\right)
}
\end{eqnarray}

The first term is%
\begin{eqnarray*}
\mathcal{L}_{m}^{\left( 1\right) } &=&i\mathrm{tr}\left( \phi ^{\dagger
}\left( D_{0}\phi \right) \right) \\
&=&i\mathrm{tr}\left\{ \phi ^{\dagger }\left( \frac{\partial \phi }{\partial
t}+\left[ A_{0},\phi \right] \right) \right\} =i\mathrm{tr}\left( \phi
^{\dagger }\frac{\partial \phi }{\partial t}+\phi ^{\dagger }A_{0}\phi -\phi
^{\dagger }\phi A_{0}\right)
\end{eqnarray*}%
and this is the form that we will use for functional variation to $A_{0}$.

Now the other term 
\begin{eqnarray}
&&\mathcal{L}_{m}^{\left( 2\right) }\equiv -\frac{1}{2m}\mathrm{tr}\left[
\left( D^{k}\phi \right) ^{\dagger }\left( D_{k}\phi \right) \right]
\label{fb53} \\
&=&-\frac{1}{2m}\mathrm{tr}\left[ \left( \frac{\partial \phi ^{\dagger }}{%
\partial x}+\phi ^{\dagger }A^{1\dagger }-A^{1\dagger }\phi ^{\dagger
}\right) \left( \frac{\partial \phi }{\partial x}+A_{1}\phi -\phi
A_{1}\right) \right.  \notag \\
&&\left. +\left( \frac{\partial \phi ^{\dagger }}{\partial y}+\phi ^{\dagger
}A^{2\dagger }-A^{2\dagger }\phi ^{\dagger }\right) \left( \frac{\partial
\phi }{\partial y}+A_{2}\phi -\phi A_{2}\right) \right]  \notag
\end{eqnarray}%
We expand the products 
\begin{eqnarray}
\mathcal{L}_{m}^{\left( 2\right) } &=&-\frac{1}{2m}\mathrm{tr}\left\{ \frac{%
\partial \phi ^{\dagger }}{\partial x}\frac{\partial \phi }{\partial x}+%
\frac{\partial \phi ^{\dagger }}{\partial x}A_{1}\phi -\frac{\partial \phi
^{\dagger }}{\partial x}\phi A_{1}\right.  \label{fb54} \\
&&+\phi ^{\dagger }A^{1\dagger }\frac{\partial \phi }{\partial x}+\phi
^{\dagger }A^{1\dagger }A_{1}\phi -\phi ^{\dagger }A^{1\dagger }\phi A_{1} 
\notag \\
&&-A^{1\dagger }\phi ^{\dagger }\frac{\partial \phi }{\partial x}%
-A^{1\dagger }\phi ^{\dagger }A_{1}\phi +A^{1\dagger }\phi ^{\dagger }\phi
A_{1}  \notag \\
&&+\frac{\partial \phi ^{\dagger }}{\partial y}\frac{\partial \phi }{%
\partial y}+\frac{\partial \phi ^{\dagger }}{\partial y}A_{2}\phi -\frac{%
\partial \phi ^{\dagger }}{\partial y}\phi A_{2}  \notag \\
&&+\phi ^{\dagger }A^{2\dagger }\frac{\partial \phi }{\partial y}+\phi
^{\dagger }A^{2\dagger }A_{2}\phi -\phi ^{\dagger }A^{2\dagger }\phi A_{2} 
\notag \\
&&\left. -A^{2\dagger }\phi ^{\dagger }\frac{\partial \phi }{\partial y}%
-A^{2\dagger }\phi ^{\dagger }A_{2}\phi +A^{2\dagger }\phi ^{\dagger }\phi
A_{2}\right\}  \notag
\end{eqnarray}%
and this form will be used in the functional derivations.

\subsection{The Euler-Lagrange equations for the gauge field}

The Euler-Lagrange equations 
\begin{equation}
\frac{\partial }{\partial x^{\mu }}\frac{\delta \mathcal{L}}{\delta \left( 
\frac{\partial A_{\alpha }}{\partial x^{\mu }}\right) }-\frac{\delta 
\mathcal{L}}{\delta A_{\alpha }}=0  \label{fb55}
\end{equation}%
We use distinct notations for the three components of the Lagrangean
density, $\mathcal{L}=\mathcal{L}_{CS}+\mathcal{L}_{m}+V$ where $\mathcal{L}%
_{CS}$ is the gauge field (Chern - Simons) part, $\mathcal{L}_{m}$ is the
\textquotedblleft matter\textquotedblright\ part and $V$ is the nonlinear
self-interaction potential for the \textquotedblleft
matter\textquotedblright\ field. We use the detailed expressions for $%
\mathcal{L}_{CS}$ from Eq.(\ref{fb3310}) and $\mathcal{L}_{m}$ is given by
the Eq.(\ref{fb54}). The functional derivations are done separately on these
two parts.

\subsubsection{The variation to $A_{0}$}

The equation of motion resulting from the variation to $A_{0}$ is 
\begin{equation}
\frac{\partial }{\partial x^{\mu }}\frac{\delta \mathcal{L}}{\delta \left( 
\frac{\partial A_{0}}{\partial x^{\mu }}\right) }-\frac{\delta \mathcal{L}}{%
\delta A_{0}}=0  \label{fb58}
\end{equation}
or 
\begin{equation}
\frac{\partial }{\partial x^{0}}\frac{\delta \mathcal{L}}{\delta \left(
\partial _{0}A_{0}\right) }+\frac{\partial }{\partial x^{1}}\frac{\delta 
\mathcal{L}}{\delta \left( \partial _{1}A_{0}\right) }+\frac{\partial }{%
\partial x^{2}}\frac{\delta \mathcal{L}}{\delta \left( \partial
_{2}A_{0}\right) }-\frac{\delta \mathcal{L}}{\delta A_{0}}=0  \label{fb5810}
\end{equation}

\paragraph{Functional derivations to $A_{0}$ of the gauge field
(Chern-Simons) Lagrangean}

The gauge field Lagrangean is Eq.(\ref{fb3310}) 
\begin{eqnarray}
\mathcal{L}_{CS} &=&\left( -\kappa \right) \mathrm{tr}\left\{ A_{0}\left(
\partial _{1}A_{2}\right) -A_{0}\left( \partial _{2}A_{1}\right)
-A_{1}\left( \partial _{0}A_{2}\right) \right.  \label{fb6010} \\
&&+A_{1}\left( \partial _{2}A_{0}\right) -A_{2}\left( \partial
_{1}A_{0}\right) +A_{2}\left( \partial _{0}A_{1}\right)  \notag \\
&&\left. +2A_{0}A_{1}A_{2}-2A_{0}A_{2}A_{1}\right\}  \notag
\end{eqnarray}%
and we have to calculate 
\begin{equation*}
\frac{\partial }{\partial x^{0}}\frac{\delta \mathcal{L}_{CS}}{\delta \left(
\partial _{0}A_{0}\right) }+\frac{\partial }{\partial x^{1}}\frac{\delta 
\mathcal{L}_{CS}}{\delta \left( \partial _{1}A_{0}\right) }+\frac{\partial }{%
\partial x^{2}}\frac{\delta \mathcal{L}_{CS}}{\delta \left( \partial
_{2}A_{0}\right) }-\frac{\delta \mathcal{L}_{CS}}{\delta A_{0}}
\end{equation*}

The calculations have been presented in detail in Ref.\cite{FlorinMadiXXX}.
The result is 
\begin{equation}
\kappa \varepsilon ^{0\nu \rho }F_{\nu \rho }=iJ^{0}  \label{fb7822}
\end{equation}

and the general form%
\begin{equation}
\kappa \varepsilon ^{\mu \nu \rho }F_{\nu \rho }=iJ^{\mu }  \label{fb7822a}
\end{equation}

\bigskip

\subsection{Euler-Lagrange equations for the \emph{matter} field}

We start from the Euler-Lagrange equation resulting from variation of the
functional variable $\phi ^{\dagger }$.%
\begin{equation}
\frac{\partial }{\partial x^{0}}\frac{\delta \mathcal{L}}{\delta \left(
\partial _{0}\phi ^{\dagger }\right) }+\frac{\partial }{\partial x^{1}}\frac{%
\delta \mathcal{L}}{\delta \left( \partial _{1}\phi ^{\dagger }\right) }+%
\frac{\partial }{\partial x^{2}}\frac{\delta \mathcal{L}}{\delta \left(
\partial _{2}\phi ^{\dagger }\right) }-\frac{\delta \mathcal{L}}{\delta \phi
^{\dagger }}=0  \label{c10}
\end{equation}%
where $\mathcal{L=L}_{CS}\mathcal{+L}_{m}+\mathcal{V}$. The Chern-Simons
term is in Eq.(\ref{fb46}) and the other two are%
\begin{eqnarray}
\mathcal{L}_{m} &=&i\mathrm{tr}\left( \phi ^{\dagger }\frac{\partial \phi }{%
\partial t}+\phi ^{\dagger }A_{0}\phi -\phi ^{\dagger }\phi A_{0}\right)
\label{c11} \\
&&-\frac{1}{2m}\mathrm{tr}\left\{ \frac{\partial \phi ^{\dagger }}{\partial x%
}\frac{\partial \phi }{\partial x}+\frac{\partial \phi ^{\dagger }}{\partial
x}A_{1}\phi -\frac{\partial \phi ^{\dagger }}{\partial x}\phi A_{1}\right. 
\notag \\
&&+\phi ^{\dagger }A^{1\dagger }\frac{\partial \phi }{\partial x}+\phi
^{\dagger }A^{1\dagger }A_{1}\phi -\phi ^{\dagger }A^{1\dagger }\phi A_{1} 
\notag \\
&&-A^{1\dagger }\phi ^{\dagger }\frac{\partial \phi }{\partial x}%
-A^{1\dagger }\phi ^{\dagger }A_{1}\phi +A^{1\dagger }\phi ^{\dagger }\phi
A_{1}  \notag \\
&&+\frac{\partial \phi ^{\dagger }}{\partial y}\frac{\partial \phi }{%
\partial y}+\frac{\partial \phi ^{\dagger }}{\partial y}A_{2}\phi -\frac{%
\partial \phi ^{\dagger }}{\partial y}\phi A_{2}  \notag \\
&&+\phi ^{\dagger }A^{2\dagger }\frac{\partial \phi }{\partial y}+\phi
^{\dagger }A^{2\dagger }A_{2}\phi -\phi ^{\dagger }A^{2\dagger }\phi A_{2} 
\notag \\
&&\left. -A^{2\dagger }\phi ^{\dagger }\frac{\partial \phi }{\partial y}%
-A^{2\dagger }\phi ^{\dagger }A_{2}\phi +A^{2\dagger }\phi ^{\dagger }\phi
A_{2}\right\}  \notag
\end{eqnarray}%
\begin{equation}
\mathcal{V}=\frac{1}{4m\kappa }\mathrm{tr}\left( \left[ \phi ^{\dagger
},\phi \right] ^{2}\right)  \label{c12}
\end{equation}

\paragraph{The contribution of $\mathcal{L}_{CS}$ (Chern-Simons) to the
Euler Lagrange equation for the functional variable $\protect\phi ^{\dagger
} $}

This means%
\begin{equation}
\frac{\partial }{\partial x^{0}}\frac{\delta \mathcal{L}_{CS}}{\delta \left(
\partial _{0}\phi ^{\dagger }\right) }+\frac{\partial }{\partial x^{1}}\frac{%
\delta \mathcal{L}_{CS}}{\delta \left( \partial _{1}\phi ^{\dagger }\right) }%
+\frac{\partial }{\partial x^{2}}\frac{\delta \mathcal{L}_{CS}}{\delta
\left( \partial _{2}\phi ^{\dagger }\right) }-\frac{\delta \mathcal{L}_{CS}}{%
\delta \phi ^{\dagger }}=0  \label{c13}
\end{equation}%
The Lagrangian $\mathcal{L}_{CS}$ is the Chern-Simons Lagrangian and does
not contain matter fields, $\phi $ and/or $\phi ^{\dagger }$. It results
that there is no contribution from it.

\paragraph{The contribution of $\mathcal{L}_{m}$ to the Euler Lagrange
equation for the functional variable $\protect\phi ^{\dagger }$}

The contribution from the "matter" Lagrangian is

\begin{equation}
\frac{\partial }{\partial x^{0}}\frac{\delta \mathcal{L}_{m}}{\delta \left(
\partial _{0}\phi ^{\dagger }\right) }+\frac{\partial }{\partial x^{1}}\frac{%
\delta \mathcal{L}_{m}}{\delta \left( \partial _{1}\phi ^{\dagger }\right) }+%
\frac{\partial }{\partial x^{2}}\frac{\delta \mathcal{L}_{m}}{\delta \left(
\partial _{2}\phi ^{\dagger }\right) }-\frac{\delta \mathcal{L}_{m}}{\delta
\phi ^{\dagger }}  \label{c14}
\end{equation}%
The first term%
\begin{equation}
\frac{\partial }{\partial x^{0}}\frac{\delta \mathcal{L}_{m}}{\delta \left(
\partial _{0}\phi ^{\dagger }\right) }  \label{c15}
\end{equation}%
Before calculating it we have to symetrise the roles of $\phi $ and $\phi
^{\dagger }$ by integrating by parts the first term%
\begin{equation}
i\mathrm{tr}\left( \phi ^{\dagger }\frac{\partial \phi }{\partial t}\right)
\label{c16}
\end{equation}%
using%
\begin{equation}
\frac{\partial }{\partial t}\left( \phi ^{\dagger }\phi \right) =\frac{%
\partial \phi ^{\dagger }}{\partial t}\phi +\phi ^{\dagger }\frac{\partial
\phi }{\partial t}  \label{c17}
\end{equation}%
Then%
\begin{eqnarray}
\phi ^{\dagger }\frac{\partial \phi }{\partial t} &=&\frac{\partial }{%
\partial t}\left( \phi ^{\dagger }\phi \right) -\frac{\partial \phi
^{\dagger }}{\partial t}\phi  \label{c18} \\
&\rightarrow &-\frac{\partial \phi ^{\dagger }}{\partial t}\phi  \notag
\end{eqnarray}%
and the first part of the matter Lagrangian now looks%
\begin{equation}
i\mathrm{tr}\left( -\frac{\partial \phi ^{\dagger }}{\partial t}\phi +\phi
^{\dagger }A_{0}\phi -\phi ^{\dagger }\phi A_{0}\right)  \label{c19}
\end{equation}%
and%
\begin{eqnarray}
\frac{\partial }{\partial x^{0}}\frac{\delta \mathcal{L}_{m}}{\delta \left(
\partial _{0}\phi ^{\dagger }\right) } &=&\frac{\partial }{\partial x^{0}}i%
\mathrm{tr}\frac{\delta }{\delta \left( \partial _{0}\phi ^{\dagger }\right) 
}\left[ -\left( \partial _{0}\phi ^{\dagger }\right) \phi \right]
\label{c20} \\
&=&-i\frac{\partial }{\partial x^{0}}\left( \phi \right) ^{T}  \notag
\end{eqnarray}%
There is no other contribution from $\mathcal{L}_{m}$ to this functional
variation to $\left( \partial _{0}\phi ^{\dagger }\right) $.

\bigskip

The next term is calculating after retaining from the full expression of $%
\mathcal{L}_{m}$ the part that has a nonvanishing contribution 
\begin{eqnarray}
&&\frac{\partial }{\partial x^{1}}\frac{\delta \mathcal{L}_{m}}{\delta
\left( \partial _{1}\phi ^{\dagger }\right) }  \label{c21} \\
&=&\frac{\partial }{\partial x^{1}}\frac{\delta }{\delta \left( \partial
_{1}\phi ^{\dagger }\right) }\left( -\frac{1}{2m}\right) \mathrm{tr}\left\{ 
\frac{\partial \phi ^{\dagger }}{\partial x^{1}}\frac{\partial \phi }{%
\partial x^{1}}+\frac{\partial \phi ^{\dagger }}{\partial x^{1}}A_{1}\phi -%
\frac{\partial \phi ^{\dagger }}{\partial x^{1}}\phi A_{1}\right\}  \notag
\end{eqnarray}%
We have%
\begin{equation}
\left( -\frac{1}{2m}\right) \frac{\partial }{\partial x^{1}}\frac{\delta }{%
\delta \left( \partial _{1}\phi ^{\dagger }\right) }\mathrm{tr}\left\{ \frac{%
\partial \phi ^{\dagger }}{\partial x^{1}}\frac{\partial \phi }{\partial
x^{1}}\right\} =\left( -\frac{1}{2m}\right) \frac{\partial }{\partial x^{1}}%
\left( \frac{\partial \phi }{\partial x^{1}}\right) ^{T}  \label{c22}
\end{equation}%
\begin{eqnarray}
\frac{\partial }{\partial x^{1}}\frac{\delta }{\delta \left( \partial
_{1}\phi ^{\dagger }\right) }\left( -\frac{1}{2m}\right) \mathrm{tr}\left\{ 
\frac{\partial \phi ^{\dagger }}{\partial x}A_{1}\phi \right\} &=&\left( -%
\frac{1}{2m}\right) \frac{\partial }{\partial x^{1}}\left( A_{1}\phi \right)
^{T}  \label{c23} \\
&=&\left( -\frac{1}{2m}\right) \left( \frac{\partial \phi ^{T}}{\partial
x^{1}}A_{1}^{T}+\phi ^{T}\frac{\partial A_{1}^{T}}{\partial x^{1}}\right) 
\notag
\end{eqnarray}%
\begin{eqnarray}
\frac{\partial }{\partial x^{1}}\frac{\delta }{\delta \left( \partial
_{1}\phi ^{\dagger }\right) }\left( -\frac{1}{2m}\right) \mathrm{tr}\left\{ -%
\frac{\partial \phi ^{\dagger }}{\partial x^{1}}\phi A_{1}\right\} &=&\left( 
\frac{1}{2m}\right) \frac{\partial }{\partial x^{1}}\left( \phi A_{1}\right)
^{T}  \label{c24} \\
&=&\left( \frac{1}{2m}\right) \left( \frac{\partial A_{1}^{T}}{\partial x^{1}%
}\phi ^{T}+A_{1}^{T}\frac{\partial \phi ^{T}}{\partial x^{1}}\right)  \notag
\end{eqnarray}%
The result from this term is%
\begin{eqnarray}
&&\frac{\partial }{\partial x^{1}}\frac{\delta \mathcal{L}_{m}}{\delta
\left( \partial _{1}\phi ^{\dagger }\right) }  \label{c25} \\
&=&\left( -\frac{1}{2m}\right) \left( \frac{\partial ^{2}\phi }{\partial
\left( x^{1}\right) ^{2}}\right) ^{T}  \notag \\
&&+\left( -\frac{1}{2m}\right) \left( \frac{\partial \phi ^{T}}{\partial
x^{1}}A_{1}^{T}+\phi ^{T}\frac{\partial A_{1}^{T}}{\partial x^{1}}\right) 
\notag \\
&&+\left( \frac{1}{2m}\right) \left( \frac{\partial A_{1}^{T}}{\partial x^{1}%
}\phi ^{T}+A_{1}^{T}\frac{\partial \phi ^{T}}{\partial x^{1}}\right)  \notag
\end{eqnarray}%
We still can transform this expression%
\begin{eqnarray}
&&\frac{\partial }{\partial x^{1}}\frac{\delta \mathcal{L}_{m}}{\delta
\left( \partial _{1}\phi ^{\dagger }\right) }  \label{c26} \\
&=&\left( -\frac{1}{2m}\right) \left( \left( \frac{\partial ^{2}\phi }{%
\partial \left( x^{1}\right) ^{2}}\right) ^{T}+\frac{\partial \phi ^{T}}{%
\partial x^{1}}A_{1}^{T}-\frac{\partial A_{1}^{T}}{\partial x^{1}}\phi
^{T}+\phi ^{T}\frac{\partial A_{1}^{T}}{\partial x^{1}}-A_{1}^{T}\frac{%
\partial \phi ^{T}}{\partial x^{1}}\right)  \notag \\
&=&\left( -\frac{1}{2m}\right) \left\{ \left( \frac{\partial ^{2}\phi }{%
\partial \left( x^{1}\right) ^{2}}\right) ^{T}+\left[ A_{1},\frac{\partial
\phi }{\partial x^{1}}\right] ^{T}-\left[ \phi ,\frac{\partial A_{1}}{%
\partial x^{1}}\right] ^{T}\right\}  \notag
\end{eqnarray}

\bigskip

We repeat the calculation for $x^{2}\left( \equiv y\right) $.%
\begin{eqnarray}
&&\frac{\partial }{\partial x^{2}}\frac{\delta \mathcal{L}_{m}}{\delta
\left( \partial _{2}\phi ^{\dagger }\right) }  \label{c27} \\
&=&\frac{\partial }{\partial x^{2}}\frac{\delta }{\delta \left( \partial
_{2}\phi ^{\dagger }\right) }\left( -\frac{1}{2m}\right) \mathrm{tr}\left\{ 
\frac{\partial \phi ^{\dagger }}{\partial x^{2}}\frac{\partial \phi }{%
\partial x^{2}}+\frac{\partial \phi ^{\dagger }}{\partial x^{2}}A_{2}\phi -%
\frac{\partial \phi ^{\dagger }}{\partial x^{2}}\phi A_{2}\right\}  \notag
\end{eqnarray}%
We take separately the terms%
\begin{eqnarray}
\frac{\partial }{\partial x^{2}}\frac{\delta }{\delta \left( \partial
_{2}\phi ^{\dagger }\right) }\left( -\frac{1}{2m}\right) \mathrm{tr}\left\{ 
\frac{\partial \phi ^{\dagger }}{\partial x^{2}}\frac{\partial \phi }{%
\partial x^{2}}\right\} &=&\left( -\frac{1}{2m}\right) \frac{\partial }{%
\partial x^{2}}\left( \frac{\partial \phi }{\partial x^{2}}\right) ^{T}
\label{c28} \\
&=&\left( -\frac{1}{2m}\right) \left( \frac{\partial ^{2}\phi }{\partial
\left( x^{2}\right) ^{2}}\right) ^{T}  \notag
\end{eqnarray}%
\begin{eqnarray}
\frac{\partial }{\partial x^{2}}\frac{\delta }{\delta \left( \partial
_{2}\phi ^{\dagger }\right) }\left( -\frac{1}{2m}\right) \mathrm{tr}\left\{ 
\frac{\partial \phi ^{\dagger }}{\partial x^{2}}A_{2}\phi \right\} &=&\left(
-\frac{1}{2m}\right) \frac{\partial }{\partial x^{2}}\left( A_{2}\phi
\right) ^{T}  \label{c29} \\
&=&\left( -\frac{1}{2m}\right) \left( \frac{\partial \phi ^{T}}{\partial
x^{2}}A_{2}^{T}+\phi ^{T}\frac{\partial A_{2}^{T}}{\partial x^{2}}\right) 
\notag
\end{eqnarray}%
\begin{eqnarray}
\frac{\partial }{\partial x^{2}}\frac{\delta }{\delta \left( \partial
_{2}\phi ^{\dagger }\right) }\left( -\frac{1}{2m}\right) \mathrm{tr}\left\{ -%
\frac{\partial \phi ^{\dagger }}{\partial x^{2}}\phi A_{2}\right\} &=&\left( 
\frac{1}{2m}\right) \frac{\partial }{\partial x^{2}}\left( \phi A_{2}\right)
^{T}  \label{c30} \\
&=&\left( \frac{1}{2m}\right) \left( \frac{\partial A_{2}^{T}}{\partial x^{2}%
}\phi ^{T}+A_{2}^{T}\frac{\partial \phi ^{T}}{\partial x^{2}}\right)  \notag
\end{eqnarray}%
Adding the three parts 
\begin{eqnarray}
&&\frac{\partial }{\partial x^{2}}\frac{\delta }{\delta \left( \partial
_{2}\phi ^{\dagger }\right) }\mathcal{L}_{m}  \label{c31} \\
&=&\left( -\frac{1}{2m}\right) \left( \frac{\partial ^{2}\phi }{\partial
\left( x^{2}\right) ^{2}}\right) ^{T}  \notag \\
&&+\left( -\frac{1}{2m}\right) \left( \frac{\partial \phi ^{T}}{\partial
x^{2}}A_{2}^{T}+\phi ^{T}\frac{\partial A_{2}^{T}}{\partial x^{2}}\right) 
\notag \\
&&+\left( \frac{1}{2m}\right) \left( \frac{\partial A_{2}^{T}}{\partial x^{2}%
}\phi ^{T}+A_{2}^{T}\frac{\partial \phi ^{T}}{\partial x^{2}}\right)  \notag
\end{eqnarray}

This expression can be transformed as%
\begin{eqnarray}
&&\frac{\partial }{\partial x^{2}}\frac{\delta }{\delta \left( \partial
_{2}\phi ^{\dagger }\right) }\mathcal{L}_{m}  \label{c32} \\
&=&\left( -\frac{1}{2m}\right) \left( \frac{\partial ^{2}\phi }{\partial
\left( x^{2}\right) ^{2}}\right) ^{T}  \notag \\
&&+\left( -\frac{1}{2m}\right) \left\{ \left( A_{2}\frac{\partial \phi }{%
\partial x^{2}}\right) ^{T}+\left( \frac{\partial A_{2}}{\partial x^{2}}\phi
\right) ^{T}-\left( \phi \frac{\partial A_{2}}{\partial x^{2}}\right)
^{T}-\left( \frac{\partial \phi }{\partial x^{2}}A_{2}\right) ^{T}\right\} 
\notag \\
&=&\left( -\frac{1}{2m}\right) \left\{ \left( \frac{\partial ^{2}\phi }{%
\partial \left( x^{2}\right) ^{2}}\right) ^{T}+\left[ A_{2},\frac{\partial
\phi }{\partial x^{2}}\right] ^{T}-\left[ \phi ,\frac{\partial A_{2}}{%
\partial x^{2}}\right] ^{T}\right\}  \notag
\end{eqnarray}

\bigskip

Now the last term, retaining in the lagrangian $\mathcal{L}_{m}$ only the
terms that can contribute to the functional derivative%
\begin{eqnarray}
&&-\frac{\delta \mathcal{L}_{m}}{\delta \phi ^{\dagger }}  \label{c37} \\
&=&-\frac{\delta }{\delta \phi ^{\dagger }}i\mathrm{tr}\left\{ \phi
^{\dagger }A_{0}\phi -\phi ^{\dagger }\phi A_{0}\right\}  \notag \\
&&-\frac{\delta }{\delta \phi ^{\dagger }}\left( -\frac{1}{2m}\right) 
\mathrm{tr}\left\{ \phi ^{\dagger }A^{1\dagger }\frac{\partial \phi }{%
\partial x^{1}}+\phi ^{\dagger }A^{1\dagger }A_{1}\phi -\phi ^{\dagger
}A^{1\dagger }\phi A_{1}\right.  \notag \\
&&\ \ \ \ \ \ \ \ \ \ \ \ \ \ \ \ \ \ \ \ \ \ \ \ -A^{1\dagger }\phi
^{\dagger }\frac{\partial \phi }{\partial x^{1}}-A^{1\dagger }\phi ^{\dagger
}A_{1}\phi +A^{1\dagger }\phi ^{\dagger }\phi A_{1}  \notag \\
&&\ \ \ \ \ \ \ \ \ \ \ \ \ \ \ \ \ \ \ \ \ \ \ \ \ +\phi ^{\dagger
}A^{2\dagger }\frac{\partial \phi }{\partial x^{2}}+\phi ^{\dagger
}A^{2\dagger }A_{2}\phi -\phi ^{\dagger }A^{2\dagger }\phi A_{2}  \notag \\
&&\ \ \ \ \ \ \ \ \ \ \ \ \ \ \ \ \ \ \ \ \ \ \ \ \ \left. -A^{2\dagger
}\phi ^{\dagger }\frac{\partial \phi }{\partial x^{2}}-A^{2\dagger }\phi
^{\dagger }A_{2}\phi +A^{2\dagger }\phi ^{\dagger }\phi A_{2}\right\}  \notag
\end{eqnarray}%
The first two terms are%
\begin{equation}
-\frac{\delta }{\delta \phi ^{\dagger }}i\mathrm{tr}\left\{ \phi ^{\dagger
}A_{0}\phi -\phi ^{\dagger }\phi A_{0}\right\} =-\frac{\delta }{\delta \phi
^{\dagger }}i\mathrm{tr}\left\{ \phi ^{\dagger }\left[ A_{0},\phi \right]
\right\} =-i\left( \left[ A_{0},\phi \right] \right) ^{T}  \label{c38}
\end{equation}

Derivation of the first line of the part $\left( D^{k}\phi \right) ^{\dagger
}\left( D_{k}\phi \right) $. 
\begin{eqnarray}
&&-\frac{\delta }{\delta \phi ^{\dagger }}\left( -\frac{1}{2m}\right) 
\mathrm{tr}\left\{ \phi ^{\dagger }A^{1\dagger }\frac{\partial \phi }{%
\partial x^{1}}+\phi ^{\dagger }A^{1\dagger }A_{1}\phi -\phi ^{\dagger
}A^{1\dagger }\phi A_{1}\right\}  \label{c39} \\
&=&\left( \frac{1}{2m}\right) \frac{\delta }{\delta \phi ^{\dagger }}\mathrm{%
tr}\left\{ \phi ^{\dagger }A^{1\dagger }\frac{\partial \phi }{\partial x^{1}}%
+\phi ^{\dagger }A^{1\dagger }A_{1}\phi -\phi ^{\dagger }A^{1\dagger }\phi
A_{1}\right\}  \notag \\
&=&\left( \frac{1}{2m}\right) \frac{\delta }{\delta \phi ^{\dagger }}\mathrm{%
tr}\left\{ \phi ^{\dagger }A^{1\dagger }\left( \frac{\partial \phi }{%
\partial x^{1}}+\left[ A_{1},\phi \right] \right) \right\} =\left( \frac{1}{%
2m}\right) \frac{\delta }{\delta \phi ^{\dagger }}\mathrm{tr}\left\{ \phi
^{\dagger }A^{1\dagger }\left( D_{1}\phi \right) \right\}  \notag \\
&=&\left( \frac{1}{2m}\right) \left[ A^{1\dagger }\left( D_{1}\phi \right) %
\right] ^{T}=\left( \frac{1}{2m}\right) \left( D_{1}\phi \right) ^{T}\left(
A^{1\dagger }\right) ^{T}  \notag
\end{eqnarray}

Derivation of the second line of the part $\left( D^{k}\phi \right)
^{\dagger }\left( D_{k}\phi \right) $.%
\begin{eqnarray}
&&-\frac{\delta }{\delta \phi ^{\dagger }}\left( -\frac{1}{2m}\right) 
\mathrm{tr}\left\{ -A^{1\dagger }\phi ^{\dagger }\frac{\partial \phi }{%
\partial x^{1}}-A^{1\dagger }\phi ^{\dagger }A_{1}\phi +A^{1\dagger }\phi
^{\dagger }\phi A_{1}\right\}  \label{c40} \\
&=&\left( \frac{1}{2m}\right) \left\{ -\left( A^{1\dagger }\right)
^{T}\left( \frac{\partial \phi }{\partial x^{1}}\right) ^{T}-\left(
A^{1\dagger }\right) ^{T}\left( A_{1}\phi \right) ^{T}+\left( A^{1\dagger
}\right) ^{T}\left( \phi A_{1}\right) ^{T}\right\}  \notag \\
&=&\left( \frac{1}{2m}\right) \left( -\right) \left( A^{1\dagger }\right)
^{T}\left\{ \left( \frac{\partial \phi }{\partial x^{1}}\right) ^{T}+\left(
A_{1}\phi \right) ^{T}-\left( \phi A_{1}\right) ^{T}\right\}  \notag \\
&=&\left( -\frac{1}{2m}\right) \left( A^{1\dagger }\right) ^{T}\left(
D_{1}\phi \right) ^{T}  \notag
\end{eqnarray}

Derivation of the third line of the part $\left( D^{k}\phi \right) ^{\dagger
}\left( D_{k}\phi \right) $.%
\begin{eqnarray}
&&-\frac{\delta }{\delta \phi ^{\dagger }}\left( -\frac{1}{2m}\right) 
\mathrm{tr}\left\{ \phi ^{\dagger }A^{2\dagger }\frac{\partial \phi }{%
\partial x^{2}}+\phi ^{\dagger }A^{2\dagger }A_{2}\phi -\phi ^{\dagger
}A^{2\dagger }\phi A_{2}\right\}  \label{c41} \\
&=&\left( \frac{1}{2m}\right) \left\{ \left( A^{2\dagger }\frac{\partial
\phi }{\partial x^{2}}\right) ^{T}+\left( A^{2\dagger }A_{2}\phi \right)
^{T}-\left( A^{2\dagger }\phi A_{2}\right) ^{T}\right\}  \notag \\
&=&\left( \frac{1}{2m}\right) \left\{ \left( \frac{\partial \phi }{\partial
x^{2}}\right) ^{T}+\left( A_{2}\phi \right) ^{T}-\left( \phi A_{2}\right)
^{T}\right\} \left( A^{2\dagger }\right) ^{T}  \notag \\
&=&\left( \frac{1}{2m}\right) \left( D_{2}\phi \right) ^{T}\left(
A^{2\dagger }\right) ^{T}  \notag
\end{eqnarray}

Derivation of the fourth (last) line of $\left( D^{k}\phi \right) ^{\dagger
}\left( D_{k}\phi \right) $. 
\begin{eqnarray}
&&-\frac{\delta }{\delta \phi ^{\dagger }}\left( -\frac{1}{2m}\right) 
\mathrm{tr}\left\{ -A^{2\dagger }\phi ^{\dagger }\frac{\partial \phi }{%
\partial y}-A^{2\dagger }\phi ^{\dagger }A_{2}\phi +A^{2\dagger }\phi
^{\dagger }\phi A_{2}\right\}  \label{c42} \\
&=&\left( \frac{1}{2m}\right) \left\{ -\left( A^{2\dagger }\right)
^{T}\left( \frac{\partial \phi }{\partial y}\right) ^{T}-\left( A^{2\dagger
}\right) ^{T}\left( A_{2}\phi \right) ^{T}+\left( A^{2\dagger }\right)
^{T}\left( \phi A_{2}\right) ^{T}\right\}  \notag \\
&=&\left( \frac{1}{2m}\right) \left( -\right) \left( A^{2\dagger }\right)
^{T}\left\{ \left( \frac{\partial \phi }{\partial y}\right) ^{T}+\left(
A_{2}\phi \right) ^{T}-\left( \phi A_{2}\right) ^{T}\right\}  \notag \\
&=&\left( \frac{1}{2m}\right) \left( -\right) \left( A^{2\dagger }\right)
^{T}\left( D_{2}\phi \right) ^{T}  \notag
\end{eqnarray}

Putting together the four results on the five lines above:%
\begin{eqnarray}
&&-\frac{\delta \mathcal{L}_{m}}{\delta \phi ^{\dagger }}  \label{c45} \\
&=&-i\left( \left[ A_{0},\phi \right] \right) ^{T}  \notag \\
&&+\left( \frac{1}{2m}\right) \left( D_{1}\phi \right) ^{T}\left(
A^{1\dagger }\right) ^{T}  \notag \\
&&-\left( \frac{1}{2m}\right) \left( A^{1\dagger }\right) ^{T}\left(
D_{1}\phi \right) ^{T}  \notag \\
&&+\left( \frac{1}{2m}\right) \left( D_{2}\phi \right) ^{T}\left(
A^{2\dagger }\right) ^{T}  \notag \\
&&-\left( \frac{1}{2m}\right) \left( A^{2\dagger }\right) ^{T}\left(
D_{2}\phi \right) ^{T}  \notag
\end{eqnarray}%
or%
\begin{equation}
-\frac{\delta \mathcal{L}_{m}}{\delta \phi ^{\dagger }}=-i\left( \left[
A_{0},\phi \right] \right) ^{T}+\left( \frac{1}{2m}\right) \left\{ \left[
D_{1}\phi ,A^{1\dagger }\right] ^{T}+\left[ D_{2}\phi ,A^{2\dagger }\right]
^{T}\right\}  \label{c46}
\end{equation}

\bigskip

Now we can write all the terms of the equation Euler-Lagrange resulting from
the variation to the function $\phi ^{\dagger }$. 
\begin{eqnarray}
&&\text{contribution from the "matter" Lagrangian }\mathcal{L}_{m}
\label{c47} \\
&=&\frac{\partial }{\partial x^{0}}\frac{\delta \mathcal{L}_{m}}{\delta
\left( \partial _{0}\phi ^{\dagger }\right) }+\frac{\partial }{\partial x^{1}%
}\frac{\delta \mathcal{L}_{m}}{\delta \left( \partial _{1}\phi ^{\dagger
}\right) }+\frac{\partial }{\partial x^{2}}\frac{\delta \mathcal{L}_{m}}{%
\delta \left( \partial _{2}\phi ^{\dagger }\right) }-\frac{\delta \mathcal{L}%
_{m}}{\delta \phi ^{\dagger }}\text{ }  \notag
\end{eqnarray}%
is%
\begin{eqnarray}
&&-i\frac{\partial }{\partial x^{0}}\left( \phi \right) ^{T}  \label{c48} \\
&&\left( -\frac{1}{2m}\right) \left\{ \left( \frac{\partial ^{2}\phi }{%
\partial \left( x^{1}\right) ^{2}}\right) ^{T}+\left[ A_{1},\frac{\partial
\phi }{\partial x^{1}}\right] ^{T}-\left[ \phi ,\frac{\partial A_{1}}{%
\partial x^{1}}\right] ^{T}\right\}  \notag \\
&&+\left( -\frac{1}{2m}\right) \left\{ \left( \frac{\partial ^{2}\phi }{%
\partial \left( x^{2}\right) ^{2}}\right) ^{T}+\left[ A_{2},\frac{\partial
\phi }{\partial x^{2}}\right] ^{T}-\left[ \phi ,\frac{\partial A_{2}}{%
\partial x^{2}}\right] ^{T}\right\}  \notag \\
&&-i\left( \left[ A_{0},\phi \right] \right) ^{T}+\left( \frac{1}{2m}\right)
\left\{ \left[ D_{1}\phi ,A^{1\dagger }\right] ^{T}+\left[ D_{2}\phi
,A^{2\dagger }\right] ^{T}\right\}  \notag
\end{eqnarray}%
We take off the transpose operator $T$ and try to recollect the expressions
in a simpler form%
\begin{eqnarray}
&&-i\frac{\partial }{\partial x^{0}}\phi -i\left[ A_{0},\phi \right]
\label{c49} \\
&&+\left( -\frac{1}{2m}\right) \left\{ \frac{\partial ^{2}\phi }{\partial
x^{2}}+\left[ A_{1},\frac{\partial \phi }{\partial x^{1}}\right] -\left[
\phi ,\frac{\partial A_{1}}{\partial x^{1}}\right] -\left[ D_{1}\phi
,A^{1\dagger }\right] \right.  \notag \\
&&\ \ \ \ \ \ \ \ \ \ \ \ \ \ \ \ \left. \frac{\partial ^{2}\phi }{\partial
y^{2}}+\left[ A_{2},\frac{\partial \phi }{\partial x^{2}}\right] -\left[
\phi ,\frac{\partial A_{2}}{\partial x^{2}}\right] -\left[ D_{2}\phi
,A^{2\dagger }\right] \right\}  \notag
\end{eqnarray}%
The terms that contain the \emph{time} are%
\begin{equation}
-iD_{0}\phi  \label{c50}
\end{equation}%
The first group of terms (those that refers to the variable $x^{1}$). 
\begin{eqnarray}
&&\frac{\partial ^{2}\phi }{\partial \left( x^{1}\right) ^{2}}+\left[ A_{1},%
\frac{\partial \phi }{\partial x^{1}}\right] -\left[ \phi ,\frac{\partial
A_{1}}{\partial x^{1}}\right] -\left[ D_{1}\phi ,A^{1\dagger }\right]
\label{c51} \\
&=&\frac{\partial ^{2}\phi }{\partial \left( x^{1}\right) ^{2}}+\frac{%
\partial }{\partial x^{1}}\left[ A_{1},\phi \right] -\left[ D_{1}\phi
,A^{1\dagger }\right]  \notag \\
&=&\frac{\partial }{\partial x^{1}}\left( \frac{\partial \phi }{\partial
x^{1}}+\left[ A_{1},\phi \right] \right) -\left[ D_{1}\phi ,A^{1\dagger }%
\right]  \notag \\
&=&\frac{\partial }{\partial x^{1}}D_{1}\phi +\left[ A^{1\dagger },D_{1}\phi %
\right]  \notag
\end{eqnarray}%
(to be multiplied by $\left( -\frac{1}{2m}\right) $. The second group of
terms (those that refers to the variable $x^{2}$).%
\begin{eqnarray}
&&\frac{\partial ^{2}\phi }{\partial \left( x^{2}\right) ^{2}}+\left[ A_{2},%
\frac{\partial \phi }{\partial x^{2}}\right] -\left[ \phi ,\frac{\partial
A_{2}}{\partial x^{2}}\right] -\left[ D_{2}\phi ,A^{2\dagger }\right]
\label{c53} \\
&=&\frac{\partial ^{2}\phi }{\partial \left( x^{2}\right) ^{2}}+\frac{%
\partial }{\partial x^{2}}\left[ A_{2},\phi \right] -\left[ D_{2}\phi
,A^{2\dagger }\right]  \notag \\
&=&\frac{\partial }{\partial x^{2}}\left( \frac{\partial \phi }{\partial
x^{2}}+\left[ A_{2},\phi \right] \right) -\left[ D_{2}\phi ,A^{2\dagger }%
\right]  \notag \\
&=&\frac{\partial }{\partial x^{2}}D_{2}\phi +\left[ A^{2\dagger },D_{2}\phi %
\right]  \notag
\end{eqnarray}%
(to be multiplied by $\left( -\frac{1}{2m}\right) $.

\bigskip

\paragraph{The contribution of $\mathcal{V}$ to the Euler Lagrange equation
for the functional variable $\protect\phi ^{\dagger }$}

We recall that the full Lagrangian was 
\begin{equation}
\mathcal{L}=\mathcal{L}_{CS}+\mathcal{L}_{m}+\mathcal{V}  \label{c60}
\end{equation}%
where the potential is%
\begin{equation}
\mathcal{V}=\frac{1}{4m\kappa }\mathrm{tr}\left( \left[ \phi ^{\dagger
},\phi \right] ^{2}\right)  \label{c61}
\end{equation}%
we have to calculate%
\begin{eqnarray}
&&\text{contribution from the potential }\mathcal{V}  \label{c62} \\
&=&\frac{\partial }{\partial x^{0}}\frac{\delta \mathcal{V}}{\delta \left(
\partial _{0}\phi ^{\dagger }\right) }+\frac{\partial }{\partial x^{1}}\frac{%
\delta \mathcal{V}}{\delta \left( \partial _{1}\phi ^{\dagger }\right) }+%
\frac{\partial }{\partial x^{2}}\frac{\delta \mathcal{V}}{\delta \left(
\partial _{2}\phi ^{\dagger }\right) }-\frac{\delta \mathcal{V}}{\delta \phi
^{\dagger }}  \notag
\end{eqnarray}%
We find%
\begin{equation}
\frac{\partial }{\partial x^{0}}\frac{\delta \mathcal{V}}{\delta \left(
\partial _{0}\phi ^{\dagger }\right) }=0  \label{c63}
\end{equation}%
\begin{equation}
\frac{\partial }{\partial x^{1}}\frac{\delta \mathcal{V}}{\delta \left(
\partial _{1}\phi ^{\dagger }\right) }=0  \label{c64}
\end{equation}%
\begin{equation}
\frac{\partial }{\partial x^{2}}\frac{\delta \mathcal{V}}{\delta \left(
\partial _{2}\phi ^{\dagger }\right) }=0  \label{c65}
\end{equation}%
\begin{eqnarray}
&&-\frac{\delta \mathcal{V}}{\delta \phi ^{\dagger }}  \label{c66} \\
&=&\frac{1}{4m\kappa }\left( -\frac{\delta }{\delta \phi ^{\dagger }}\right) 
\mathrm{tr}\left( \left[ \phi ^{\dagger },\phi \right] ^{2}\right)  \notag \\
&=&\frac{1}{4m\kappa }\left( -\frac{\delta }{\delta \phi ^{\dagger }}\right) 
\mathrm{tr}\left[ \left( \phi ^{\dagger }\phi -\phi \phi ^{\dagger }\right)
^{2}\right]  \notag \\
&=&\frac{1}{4m\kappa }\left( -\frac{\delta }{\delta \phi ^{\dagger }}\right) 
\mathrm{tr}\left( \phi ^{\dagger }\phi \phi ^{\dagger }\phi -\phi ^{\dagger
}\phi \phi \phi ^{\dagger }-\phi \phi ^{\dagger }\phi ^{\dagger }\phi +\phi
\phi ^{\dagger }\phi \phi ^{\dagger }\right)  \notag
\end{eqnarray}%
The derivations use$\ \ $%
\begin{equation}
\frac{d}{d\mathbf{X}}\left( \mathbf{AXBX}\right) =\mathbf{A}^{T}\mathbf{X}%
^{T}\mathbf{B}^{T}+\mathbf{B}^{T}\mathbf{X}^{T}\mathbf{A}^{T}  \label{c67}
\end{equation}

The first term%
\begin{eqnarray}
&&\left( -\frac{1}{4m\kappa }\right) \left( \frac{\delta }{\delta \phi
^{\dagger }}\right) \mathrm{tr}\left( \phi ^{\dagger }\phi \phi ^{\dagger
}\phi \right)  \label{c68} \\
&=&\left( -\frac{1}{4m\kappa }\right) \left( \frac{\delta }{\delta \phi
^{\dagger }}\right) \mathrm{tr}\left( \phi \phi ^{\dagger }\phi \phi
^{\dagger }\right) \ \ \left( \text{applying cyclic permutation under }%
\mathrm{tr}\right)  \notag \\
&=&\left( -\frac{1}{4m\kappa }\right) \left( \phi ^{T}\phi ^{\dagger T}\phi
^{T}+\phi ^{T}\phi ^{\dagger T}\phi ^{T}\right)  \notag \\
&=&\left( -\frac{1}{2m\kappa }\right) \left( \phi ^{T}\phi ^{\dagger T}\phi
^{T}\right)  \notag
\end{eqnarray}

The second term%
\begin{eqnarray}
&&\frac{1}{4m\kappa }\left( -\frac{\delta }{\delta \phi ^{\dagger }}\right) 
\mathrm{tr}\left( -\phi ^{\dagger }\phi \phi \phi ^{\dagger }\right)
\label{c69} \\
&=&\left( \frac{1}{4m\kappa }\right) \left( \frac{\delta }{\delta \phi
^{\dagger }}\right) \mathrm{tr}\left( \phi ^{\dagger }\phi \phi \phi
^{\dagger }\right)  \notag
\end{eqnarray}%
The type of this term is%
\begin{equation}
\frac{\delta }{\delta \mathbf{X}}\left( \mathbf{XAX}\right) =\left( \mathbf{%
AX}\right) ^{T}+\left( \mathbf{XA}\right) ^{T}  \label{c70}
\end{equation}%
where%
\begin{eqnarray}
\mathbf{X} &\mathbf{\equiv }&\phi ^{\dagger }  \label{c71} \\
\mathbf{A} &\equiv &\phi \phi  \notag
\end{eqnarray}%
then it results%
\begin{equation}
\left( \frac{1}{4m\kappa }\right) \left( \frac{\delta }{\delta \phi
^{\dagger }}\right) \mathrm{tr}\left( \phi ^{\dagger }\phi \phi \phi
^{\dagger }\right) =\left( \frac{1}{4m\kappa }\right) \left[ \left( \phi
\phi \phi ^{\dagger }\right) ^{T}+\left( \phi ^{\dagger }\phi \phi \right)
^{T}\right]  \label{c72}
\end{equation}

The third term%
\begin{eqnarray}
&&\frac{1}{4m\kappa }\left( -\frac{\delta }{\delta \phi ^{\dagger }}\right) 
\mathrm{tr}\left( -\phi \phi ^{\dagger }\phi ^{\dagger }\phi \right)
\label{c73} \\
&=&\left( \frac{1}{4m\kappa }\right) \left( \frac{\delta }{\delta \phi
^{\dagger }}\right) \mathrm{tr}\left( \phi \phi ^{\dagger }\phi ^{\dagger
}\phi \right) =\left( \frac{1}{4m\kappa }\right) \left( \frac{\delta }{%
\delta \phi ^{\dagger }}\right) \mathrm{tr}\left( \phi ^{\dagger }\phi
^{\dagger }\phi \phi \right)  \notag
\end{eqnarray}%
This derivation has the type%
\begin{eqnarray}
\frac{\delta }{\delta \mathbf{X}}\left( \mathbf{XXA}\right) &=&T_{1}+T_{2}
\label{c74} \\
T_{1} &=&\frac{\delta }{\delta \mathbf{X}}\left[ \mathbf{X}\left( \mathbf{XA}%
\right) \right] =\left( \mathbf{XA}\right) ^{T}  \notag \\
T_{2} &=&\frac{\delta }{\delta \mathbf{X}}\left[ \mathbf{XAX}\right] =\frac{%
\delta }{\delta \mathbf{X}}\left[ \mathbf{X}\left( \mathbf{AX}\right) \right]
=\left( \mathbf{AX}\right) ^{T}  \notag
\end{eqnarray}%
where%
\begin{eqnarray}
\mathbf{X} &\equiv &\phi ^{\dagger }  \label{c75} \\
\mathbf{A} &\equiv &\phi \phi  \notag
\end{eqnarray}%
and we write%
\begin{equation}
\left( \frac{1}{4m\kappa }\right) \left( \frac{\delta }{\delta \phi
^{\dagger }}\right) \mathrm{tr}\left( \phi ^{\dagger }\phi ^{\dagger }\phi
\phi \right) =\left( \frac{1}{4m\kappa }\right) \left[ \left( \phi ^{\dagger
}\phi \phi \right) ^{T}+\left( \phi \phi \phi ^{\dagger }\right) ^{T}\right]
\label{c76}
\end{equation}%
The fourth term%
\begin{eqnarray}
&&\frac{1}{4m\kappa }\left( -\frac{\delta }{\delta \phi ^{\dagger }}\right) 
\mathrm{tr}\left( \phi \phi ^{\dagger }\phi \phi ^{\dagger }\right)
\label{c77} \\
&=&\left( -\frac{1}{4m\kappa }\right) \left( \frac{\delta }{\delta \phi
^{\dagger }}\right) \mathrm{tr}\left( \phi ^{\dagger }\phi \phi ^{\dagger
}\phi \right)  \notag \\
&=&\left( -\frac{1}{4m\kappa }\right) \left( \phi ^{T}\phi ^{\dagger T}\phi
^{T}+\phi ^{T}\phi ^{\dagger T}\phi ^{T}\right)  \notag \\
&=&\left( -\frac{1}{2m\kappa }\right) \left( \phi ^{T}\phi ^{\dagger T}\phi
^{T}\right)  \notag
\end{eqnarray}

Now let us collect all terms%
\begin{eqnarray}
-\frac{\delta \mathcal{V}}{\delta \phi ^{\dagger }} &=&\frac{1}{4m\kappa }%
\left( -\frac{\delta }{\delta \phi ^{\dagger }}\right) \mathrm{tr}\left(
\phi ^{\dagger }\phi \phi ^{\dagger }\phi -\phi ^{\dagger }\phi \phi \phi
^{\dagger }-\phi \phi ^{\dagger }\phi ^{\dagger }\phi +\phi \phi ^{\dagger
}\phi \phi ^{\dagger }\right)  \notag \\
&=&\left( -\frac{1}{2m\kappa }\right) \left( \phi ^{T}\phi ^{\dagger T}\phi
^{T}\right)  \label{c78} \\
&&+\left( \frac{1}{4m\kappa }\right) \left[ \left( \phi \phi \phi ^{\dagger
}\right) ^{T}+\left( \phi ^{\dagger }\phi \phi \right) ^{T}\right]  \notag \\
&&+\left( \frac{1}{4m\kappa }\right) \left[ \left( \phi ^{\dagger }\phi \phi
\right) ^{T}+\left( \phi \phi \phi ^{\dagger }\right) ^{T}\right]  \notag \\
&&\left( -\frac{1}{2m\kappa }\right) \left( \phi ^{T}\phi ^{\dagger T}\phi
^{T}\right)  \notag
\end{eqnarray}%
This can be written%
\begin{eqnarray}
-\frac{\delta \mathcal{V}}{\delta \phi ^{\dagger }} &=&\left( -\frac{1}{%
2m\kappa }\right) \left\{ \phi ^{T}\phi ^{\dagger T}\phi ^{T}-\left( \phi
\phi \phi ^{\dagger }\right) ^{T}-\left( \phi ^{\dagger }\phi \phi \right)
^{T}+\phi ^{T}\phi ^{\dagger T}\phi ^{T}\right\}  \notag \\
&=&\left( -\frac{1}{2m\kappa }\right) \left\{ \phi \phi ^{\dagger }\phi
-\phi \phi \phi ^{\dagger }-\phi ^{\dagger }\phi \phi +\phi \phi ^{\dagger
}\phi \right\} ^{T}  \label{c80} \\
&=&\left( -\frac{1}{2m\kappa }\right) \left\{ \phi \left( \phi ^{\dagger
}\phi -\phi \phi ^{\dagger }\right) -\left( \phi ^{\dagger }\phi -\phi \phi
^{\dagger }\right) \phi \right\} ^{T}  \notag \\
&=&\left( -\frac{1}{2m\kappa }\right) \left\{ \phi \left[ \phi ^{\dagger
},\phi \right] -\left[ \phi ^{\dagger },\phi \right] \phi \right\} ^{T} 
\notag \\
&=&\left( -\frac{1}{2m\kappa }\right) \left[ \phi ,\left[ \phi ^{\dagger
},\phi \right] \right] ^{T}  \notag \\
&=&\left( \frac{1}{2m\kappa }\right) \left[ \left[ \phi ^{\dagger },\phi %
\right] ,\phi \right] ^{T}  \notag
\end{eqnarray}%
And finally%
\begin{eqnarray}
&&\text{contribution from the potential }\mathcal{V}  \label{c81} \\
&=&\frac{\partial }{\partial x^{0}}\frac{\delta \mathcal{V}}{\delta \left(
\partial _{0}\phi ^{\dagger }\right) }+\frac{\partial }{\partial x^{1}}\frac{%
\delta \mathcal{V}}{\delta \left( \partial _{1}\phi ^{\dagger }\right) }+%
\frac{\partial }{\partial x^{2}}\frac{\delta \mathcal{V}}{\delta \left(
\partial _{2}\phi ^{\dagger }\right) }-\frac{\delta \mathcal{V}}{\delta \phi
^{\dagger }}  \notag \\
&=&\left( \frac{1}{2m\kappa }\right) \left[ \left[ \phi ^{\dagger },\phi %
\right] ,\phi \right] ^{T}  \notag
\end{eqnarray}

\bigskip

\paragraph{All contributions}

We collect all results%
\begin{eqnarray}
&&0  \label{c82} \\
&&\left[ -iD_{0}\phi +\left( -\frac{1}{2m}\right) \left( D^{k\dagger
}D_{k}\phi \right) \right] ^{T}  \notag \\
&&+\left( \frac{1}{2m\kappa }\right) \left[ \left[ \phi ^{\dagger },\phi %
\right] ,\phi \right] ^{T}  \notag \\
&=&0  \notag
\end{eqnarray}%
or%
\begin{equation}
iD_{0}\phi =-\frac{1}{2m}\left( D^{k\dagger }D_{k}\phi \right) +\frac{1}{%
2m\kappa }\left[ \left[ \phi ^{\dagger },\phi \right] ,\phi \right]
\label{c83}
\end{equation}

\paragraph{Final form of the equations of motion as derived from
Euler-Lagrange eqs.}

The equations of motion that represent the Euler-Lagrange equations for the
Lagrangian are 
\begin{eqnarray}
iD_{0}\phi &=&-\frac{1}{2m}\left( D^{k}D_{\kappa }\right) \phi
\label{euler1} \\
&&-\frac{1}{2m\kappa }\left[ \left[ \phi ,\phi ^{\dagger }\right] ,\phi %
\right]  \notag
\end{eqnarray}%
\begin{equation}
\kappa \varepsilon ^{\mu \nu \rho }F_{\nu \rho }=iJ^{\mu }  \label{euler2}
\end{equation}

\bigskip

\section{Appendix C. Detailed form of the equation of motion for the \emph{%
matter} field} \label{App:AppendixC}

\renewcommand{\theequation}{C.\arabic{equation}} \setcounter{equation}{0}

The first equation of motion is%
\begin{equation}
iD_{0}\phi =-\frac{1}{2m}\left( D^{k}D_{\kappa }\right) \phi -\frac{1}{%
2m\kappa }\left[ \left[ \phi ,\phi ^{\dagger }\right] ,\phi \right]
\label{d12}
\end{equation}

We have to calculate%
\begin{equation}
D_{0}\phi =\frac{\partial \phi }{\partial t}+A_{0}\phi -\phi A_{0}
\label{540}
\end{equation}%
\begin{equation}
\mathbf{D}^{2}\phi =D_{k}D^{k}\phi  \label{541}
\end{equation}

\bigskip

We write explicitely the covariant derivative operators 
\begin{eqnarray}
&&i\left( \frac{\partial \phi }{\partial t}+A_{0}\phi -\phi A_{0}\right)
\label{542} \\
&=&-\frac{1}{2m}\left\{ \left( \frac{\partial }{\partial x}+\left[ A_{1},%
\right] \right) \left( \frac{\partial }{\partial x}+\left[ A^{1},\right]
\right) +\left( \frac{\partial }{\partial y}+\left[ A_{2},\right] \right)
\left( \frac{\partial }{\partial y}+\left[ A^{2},\right] \right) \right\}
\phi  \notag \\
&&-\frac{1}{2m\kappa }\left[ \left[ \phi ,\phi ^{\dagger }\right] ,\phi %
\right]  \notag
\end{eqnarray}

\bigskip

\subsection{Calculation of the term $D_{k}D^{k}\protect\phi $}

We calculate separately the terms in the RHS.

First the $x$ term, $D_{x}D^{x}\phi $ is expanded%
\begin{eqnarray}
&&\left( \frac{\partial }{\partial x}+\left[ A_{1},\right] \right) \left( 
\frac{\partial }{\partial x}+\left[ A^{1},\right] \right) \phi  \label{543}
\\
&=&\left( \frac{\partial }{\partial x}+\left[ A_{1},\right] \right) \left( 
\frac{\partial \phi }{\partial x}+A^{1}\phi -\phi A^{1}\right)  \notag \\
&=&\frac{\partial ^{2}\phi }{\partial x^{2}}+\frac{\partial }{\partial x}%
\left( A^{1}\phi \right) -\frac{\partial }{\partial x}\left( \phi
A^{1}\right)  \notag \\
&&+A_{1}\left( \frac{\partial \phi }{\partial x}+A^{1}\phi -\phi
A^{1}\right) -\left( \frac{\partial \phi }{\partial x}+A^{1}\phi -\phi
A^{1}\right) A_{1}  \notag \\
&=&\frac{\partial ^{2}\phi }{\partial x^{2}}+\frac{\partial A^{1}}{\partial x%
}\phi +A^{1}\frac{\partial \phi }{\partial x}-\frac{\partial \phi }{\partial
x}A^{1}-\phi \frac{\partial A^{1}}{\partial x}  \notag \\
&&+A_{1}\frac{\partial \phi }{\partial x}+\left( A_{1}\right) ^{2}\phi
-A_{1}\phi A^{1}  \notag \\
&&-\frac{\partial \phi }{\partial x}A_{1}-A^{1}\phi A_{1}+\phi A^{1}A_{1} 
\notag
\end{eqnarray}%
Since we have 
\begin{equation}
A^{1}=A_{1}\equiv A_{x}  \label{544}
\end{equation}%
we can simplify the expression 
\begin{eqnarray}
&&\left( \frac{\partial }{\partial x}+\left[ A_{1},\right] \right) \left( 
\frac{\partial }{\partial x}+\left[ A^{1},\right] \right) \phi  \label{545}
\\
&=&\frac{\partial ^{2}\phi }{\partial x^{2}}+\frac{\partial A_{x}}{\partial x%
}\phi -\phi \frac{\partial A_{x}}{\partial x}  \notag \\
&&+2A_{x}\frac{\partial \phi }{\partial x}-2\frac{\partial \phi }{\partial x}%
A_{x}  \notag \\
&&+A_{x}^{2}\phi +\phi A_{x}^{2}-2A_{x}\phi A_{x}  \notag
\end{eqnarray}

\bigskip

The same calculation is made for the $y$ term%
\begin{eqnarray}
&&\left( \frac{\partial }{\partial y}+\left[ A_{2},\right] \right) \left( 
\frac{\partial }{\partial y}+\left[ A^{2},\right] \right) \phi  \label{546}
\\
&=&\left( \frac{\partial }{\partial y}+\left[ A_{2},\right] \right) \left( 
\frac{\partial \phi }{\partial y}+A^{2}\phi -\phi A^{2}\right)  \notag \\
&=&\frac{\partial ^{2}\phi }{\partial y^{2}}+\frac{\partial }{\partial y}%
\left( A^{2}\phi \right) -\frac{\partial }{\partial y}\left( \phi
A^{2}\right)  \notag \\
&&+A_{2}\left( \frac{\partial \phi }{\partial y}+A^{2}\phi -\phi
A^{2}\right) -\left( \frac{\partial \phi }{\partial y}+A^{2}\phi -\phi
A^{2}\right) A_{2}  \notag \\
&=&\frac{\partial ^{2}\phi }{\partial y^{2}}+\frac{\partial A^{2}}{\partial y%
}\phi +A^{2}\frac{\partial \phi }{\partial y}-\frac{\partial \phi }{\partial
y}A^{2}-\phi \frac{\partial A^{2}}{\partial y}  \notag \\
&&+A_{2}\frac{\partial \phi }{\partial y}+\left( A_{2}\right) ^{2}\phi
-A_{2}\phi A^{2}  \notag \\
&&-\frac{\partial \phi }{\partial y}A_{2}-A^{2}\phi A_{2}+\phi A^{2}A_{2} 
\notag
\end{eqnarray}

Since we have 
\begin{equation}
A^{2}=A_{2}\equiv A_{y}  \label{547}
\end{equation}%
we can simplify the expression 
\begin{eqnarray}
&&\left( \frac{\partial }{\partial y}+\left[ A_{2},\right] \right) \left( 
\frac{\partial }{\partial y}+\left[ A^{2},\right] \right) \phi  \label{548}
\\
&=&\frac{\partial ^{2}\phi }{\partial y^{2}}+\frac{\partial A_{y}}{\partial y%
}\phi -\phi \frac{\partial A_{y}}{\partial y}  \notag \\
&&+2A_{y}\frac{\partial \phi }{\partial y}-2\frac{\partial \phi }{\partial y}%
A_{y}  \notag \\
&&+A_{y}^{2}\phi +\phi A_{y}^{2}-2A_{y}\phi A_{y}  \notag
\end{eqnarray}

\bigskip

Now we can sum the two terms%
\begin{equation*}
D_{k}D^{k}\phi =\left( D_{x}D^{x}+D_{y}D^{y}\right) \phi
\end{equation*}%
\begin{eqnarray}
&&\left[ \left( \frac{\partial }{\partial x}+\left[ A_{1},\right] \right)
\left( \frac{\partial }{\partial x}+\left[ A^{1},\right] \right) +\left( 
\frac{\partial }{\partial y}+\left[ A_{2},\right] \right) \left( \frac{%
\partial }{\partial y}+\left[ A^{2},\right] \right) \right] \phi  \notag \\
&=&\frac{\partial ^{2}\phi }{\partial x^{2}}+\frac{\partial A_{x}}{\partial x%
}\phi -\phi \frac{\partial A_{x}}{\partial x}+\frac{\partial ^{2}\phi }{%
\partial y^{2}}+\frac{\partial A_{y}}{\partial y}\phi -\phi \frac{\partial
A_{y}}{\partial y}  \label{549} \\
&&+2A_{x}\frac{\partial \phi }{\partial x}-2\frac{\partial \phi }{\partial x}%
A_{x}+2A_{y}\frac{\partial \phi }{\partial y}-2\frac{\partial \phi }{%
\partial y}A_{y}  \notag \\
&&+A_{x}^{2}\phi +\phi A_{x}^{2}-2A_{x}\phi A_{x}+A_{y}^{2}\phi +\phi
A_{y}^{2}-2A_{y}\phi A_{y}  \notag
\end{eqnarray}%
This expression must be multiplied by the numerical factor $-\frac{1}{2m}$.

\bigskip

This expresion of $D_{k}D^{k}\phi $ will be compared later with $%
D_{+}D_{-}\phi $.

\subsubsection{Calculation of $D_{+}D_{-}\protect\phi $ in terms of $A_{x,y}$%
}

We will find the detailed expression of the term%
\begin{eqnarray}
&&D_{+}D_{-}\phi  \label{d20} \\
&=&\left( \partial _{+}+\left[ A_{+},\right] \right) \left( \partial _{-}+%
\left[ A_{-},\right] \right) \phi  \notag
\end{eqnarray}%
where%
\begin{eqnarray}
A_{+} &=&A_{x}+iA_{y}  \label{d22} \\
A_{-} &=&A_{x}-iA_{y}  \notag
\end{eqnarray}%
\begin{eqnarray}
\partial _{+} &=&\frac{\partial }{\partial x}+i\frac{\partial }{\partial y}
\label{d23} \\
\partial _{-} &=&\frac{\partial }{\partial x}-i\frac{\partial }{\partial y} 
\notag
\end{eqnarray}%
\begin{equation}
\left( \frac{\partial }{\partial x}+i\frac{\partial }{\partial y}+\left[
A_{x}+iA_{y},\right] \right) \left( \frac{\partial }{\partial x}-i\frac{%
\partial }{\partial y}+\left[ A_{x}-iA_{y},\right] \right) \phi  \label{d24}
\end{equation}%
Separately, the second operator in the product (second paranthesis) is%
\begin{eqnarray}
&&\left( \frac{\partial }{\partial x}-i\frac{\partial }{\partial y}+\left[
A_{x}-iA_{y},\right] \right) \phi  \label{d25} \\
&=&\frac{\partial \phi }{\partial x}-i\frac{\partial \phi }{\partial y}%
+A_{x}\phi -\phi A_{x}-iA_{y}\phi +i\phi A_{y}  \notag
\end{eqnarray}%
Now we apply the first operator (first paranthesis, $\left( \frac{\partial }{%
\partial x}+i\frac{\partial }{\partial y}+\left[ A_{x}+iA_{y},\right]
\right) $) on this expression%
\begin{eqnarray}
&&\left( \frac{\partial }{\partial x}+i\frac{\partial }{\partial y}+\left[
A_{x}+iA_{y},\right] \right) \left( \frac{\partial \phi }{\partial x}-i\frac{%
\partial \phi }{\partial y}+A_{x}\phi -\phi A_{x}-iA_{y}\phi +i\phi
A_{y}\right)  \label{d26} \\
&=&\frac{\partial ^{2}\phi }{\partial x^{2}}-\underbrace{i\frac{\partial
^{2}\phi }{\partial x\partial y}}+\frac{\partial A_{x}}{\partial x}\phi
+A_{x}\frac{\partial \phi }{\partial x}-\frac{\partial \phi }{\partial x}%
A_{x}-\phi \frac{\partial A_{x}}{\partial x}-i\frac{\partial A_{y}}{\partial
x}\phi -\underline{iA_{y}\frac{\partial \phi }{\partial x}}+\underleftarrow{i%
\frac{\partial \phi }{\partial x}A_{y}}+i\phi \frac{\partial A_{y}}{\partial
x}  \notag \\
&&+\underbrace{i\frac{\partial ^{2}\phi }{\partial y\partial x}}+\frac{%
\partial ^{2}\phi }{\partial y^{2}}+i\frac{\partial A_{x}}{\partial y}\phi +%
\underline{\underline{iA_{x}\frac{\partial \phi }{\partial y}}}-%
\underleftrightarrow{i\frac{\partial \phi }{\partial y}A_{x}}-i\phi \frac{%
\partial A_{x}}{\partial y}+\frac{\partial A_{y}}{\partial y}\phi +A_{y}%
\frac{\partial \phi }{\partial y}-\frac{\partial \phi }{\partial y}%
A_{y}-\phi \frac{\partial A_{y}}{\partial y}  \notag \\
&&+A_{x}\frac{\partial \phi }{\partial x}-\underline{\underline{iA_{x}\frac{%
\partial \phi }{\partial y}}}+A_{x}^{2}\phi -A_{x}\phi A_{x}-iA_{x}A_{y}\phi
+\underbrace{\underbrace{iA_{x}\phi A_{y}}}  \notag \\
&&-\frac{\partial \phi }{\partial x}A_{x}+\underleftrightarrow{i\frac{%
\partial \phi }{\partial y}A_{x}}-A_{x}\phi A_{x}+\phi A_{x}^{2}%
\underleftrightarrow{\underleftrightarrow{+iA_{y}\phi A_{x}}}-i\phi
A_{y}A_{x}  \notag \\
&&+\underline{iA_{y}\frac{\partial \phi }{\partial x}}+A_{y}\frac{\partial
\phi }{\partial y}+iA_{y}A_{x}\phi \underleftrightarrow{\underleftrightarrow{%
-iA_{y}\phi A_{x}}}+A_{y}^{2}\phi -A_{y}\phi A_{y}  \notag \\
&&-\underleftarrow{i\frac{\partial \phi }{\partial x}A_{y}}-\frac{\partial
\phi }{\partial y}A_{y}-\underbrace{\underbrace{iA_{x}\phi A_{y}}}+i\phi
A_{x}A_{y}-A_{y}\phi A_{y}+\phi A_{y}^{2}  \notag
\end{eqnarray}%
There are $44$ terms. Few terms, $14$, cancel and others $30$ are grouped.

\bigskip

The result is 
\begin{eqnarray}
&&D_{+}D_{-}\phi =  \label{dpdm2} \\
&=&\frac{\partial ^{2}\phi }{\partial x^{2}}+\frac{\partial ^{2}\phi }{%
\partial y^{2}}  \notag \\
&&+2A_{x}\frac{\partial \phi }{\partial x}-2\frac{\partial \phi }{\partial x}%
A_{x}+2A_{y}\frac{\partial \phi }{\partial y}-2\frac{\partial \phi }{%
\partial y}A_{y}  \notag \\
&&+\frac{\partial A_{x}}{\partial x}\phi -\phi \frac{\partial A_{x}}{%
\partial x}-i\frac{\partial A_{y}}{\partial x}\phi +i\phi \frac{\partial
A_{y}}{\partial x}+i\frac{\partial A_{x}}{\partial y}\phi -i\phi \frac{%
\partial A_{x}}{\partial y}-\phi \frac{\partial A_{y}}{\partial y}+\frac{%
\partial A_{y}}{\partial y}\phi  \notag \\
&&+A_{x}^{2}\phi -2A_{x}\phi A_{x}-iA_{x}A_{y}\phi +\phi A_{x}^{2}-i\phi
A_{y}A_{x}+iA_{y}A_{x}\phi +A_{y}^{2}\phi -2A_{y}\phi A_{y}  \notag \\
&&+i\phi A_{x}A_{y}+\phi A_{y}^{2}  \notag
\end{eqnarray}

This is $D_{+}D_{-}\phi $.

Now we compare this with $D_{k}D^{k}\phi $ from Eq.(\ref{549}) (we do not
multiply yet by $-\frac{1}{2m}$) and substract 
\begin{eqnarray}
&&D_{k}D^{k}\phi -D_{+}D_{-}\phi  \label{d31} \\
&=&\left[ -i\frac{\partial A_{y}}{\partial x}\phi +i\phi \frac{\partial A_{y}%
}{\partial x}+i\frac{\partial A_{x}}{\partial y}\phi -i\phi \frac{\partial
A_{x}}{\partial y}\right.  \notag \\
&&\left. -iA_{x}A_{y}\phi -i\phi A_{y}A_{x}+iA_{y}A_{x}\phi +i\phi
A_{x}A_{y} \right]  \notag
\end{eqnarray}%
We have%
\begin{eqnarray}
&&D_{+}D_{-}-D_{k}D^{k}  \label{d32} \\
&=&-i\frac{\partial A_{y}}{\partial x}\phi +i\phi \frac{\partial A_{y}}{%
\partial x}+i\frac{\partial A_{x}}{\partial y}\phi -i\phi \frac{\partial
A_{x}}{\partial y}  \notag \\
&&-iA_{x}A_{y}\phi -i\phi A_{y}A_{x}+iA_{y}A_{x}\phi +i\phi A_{x}A_{y} 
\notag \\
&=&\left( -i\frac{\partial A_{y}}{\partial x}+i\frac{\partial A_{x}}{%
\partial y}-iA_{x}A_{y}+iA_{y}A_{x}\right) \phi  \notag \\
&&+\phi \left( i\frac{\partial A_{y}}{\partial x}-i\frac{\partial A_{x}}{%
\partial y}-iA_{y}A_{x}+iA_{x}A_{y}\right)  \notag \\
&=&-i\left( \frac{\partial A_{y}}{\partial x}-\frac{\partial A_{x}}{\partial
y}+A_{x}A_{y}-A_{y}A_{x}\right) \phi  \notag \\
&&+i\phi \left( \frac{\partial A_{y}}{\partial x}-\frac{\partial A_{x}}{%
\partial y}+A_{x}A_{y}-A_{y}A_{x}\right)  \notag
\end{eqnarray}%
This can be written%
\begin{eqnarray}
&&D_{+}D_{-}-D_{k}D^{k}  \label{d33} \\
&=&-iF_{xy}\phi +i\phi F_{xy}  \notag \\
&=&-i\left[ F_{xy},\phi \right]  \notag
\end{eqnarray}%
or%
\begin{equation}
D_{k}D^{k}=D_{+}D_{-}+i\left[ F_{xy},\phi \right]  \label{d34}
\end{equation}

\bigskip

Now we replace with the formula derived by us for $F_{12}$,%
\begin{equation}
F_{xy}=F_{12}=\frac{i}{2\kappa }\left[ \phi ,\phi ^{\dagger }\right]
\label{d35}
\end{equation}%
and obtain%
\begin{eqnarray}
D_{k}D^{k}\phi &=&D_{+}D_{-}\phi +i\left[ \frac{i}{2\kappa }\left[ \phi
,\phi ^{\dagger }\right] ,\phi \right]  \label{d36} \\
&=&D_{+}D_{-}\phi -\frac{1}{2\kappa }\left[ \left[ \phi ,\phi ^{\dagger }%
\right] ,\phi \right]  \notag
\end{eqnarray}

At this moment the first equation of motion can be written%
\begin{eqnarray}
iD_{0}\phi &=&-\frac{1}{2m}\mathbf{D}^{2}\phi -\frac{1}{2m\kappa }\left[ %
\left[ \phi ,\phi ^{\dagger }\right] ,\phi \right]  \label{d37} \\
iD_{0}\phi &=&-\frac{1}{2m}\left\{ D_{+}D_{-}\phi -\frac{1}{2\kappa }\left[ %
\left[ \phi ,\phi ^{\dagger }\right] ,\phi \right] \right\} -\frac{1}{%
2m\kappa }\left[ \left[ \phi ,\phi ^{\dagger }\right] ,\phi \right]  \notag
\\
&=&-\frac{1}{2m}D_{+}D_{-}\phi +\frac{1}{4m\kappa }\left[ \left[ \phi ,\phi
^{\dagger }\right] ,\phi \right] -\frac{1}{2m\kappa }\left[ \left[ \phi
,\phi ^{\dagger }\right] ,\phi \right]  \notag \\
&=&-\frac{1}{2m}D_{+}D_{-}\phi -\frac{1}{4m\kappa }\left[ \left[ \phi ,\phi
^{\dagger }\right] ,\phi \right]  \notag
\end{eqnarray}

\bigskip

The last term in the right hand side of the expression of $D_{k}D^{k}$ is%
\begin{equation}
-\frac{1}{2\kappa }\left[ \left[ \phi ,\phi ^{\dagger }\right] ,\phi \right]
\label{550}
\end{equation}

The full expression of the \textbf{first equation of motion} in detailed
form is obtained from the Eqs.(\ref{540}), (\ref{549}) and (\ref{550}).

We have 
\begin{eqnarray}
iD_{0}\phi &=&-\frac{1}{2m}\mathbf{D}^{2}\phi -\frac{1}{2m\kappa }\left[ %
\left[ \phi ,\phi ^{\dagger }\right] ,\phi \right]  \label{d41} \\
&=&-\frac{1}{2m}\left\{ D_{+}D_{-}\phi -\frac{1}{2\kappa }\left[ \left[ \phi
,\phi ^{\dagger }\right] ,\phi \right] \right\} -\frac{1}{2m\kappa }\left[ %
\left[ \phi ,\phi ^{\dagger }\right] ,\phi \right]  \notag \\
&=&-\frac{1}{2m}D_{+}D_{-}\phi -\frac{1}{4m\kappa }\left[ \left[ \phi ,\phi
^{\dagger }\right] ,\phi \right]  \notag
\end{eqnarray}%
\begin{equation}
iD_{0}\phi =-\frac{1}{2m}D_{+}D_{-}\phi -\frac{1}{4m\kappa }\left[ \left[
\phi ,\phi ^{\dagger }\right] ,\phi \right]  \label{d42}
\end{equation}

\bigskip

We \textbf{NOTE} that this equation is valid in general not only at
self-duality.

\subsection{An expression for the time-component of the gauge potential $%
A_{0}$ at SD}

We note that in the derivation of the Bogomolnyi form of the energy it was
not necessary to impose the static states. Then at this moment the states
may still have a time evolution, although they verify the lowest energy
condition 
\begin{equation}
D_{-}\phi =0  \label{d51}
\end{equation}%
In this case we can combine the spatial components of the current density 
\begin{eqnarray}
J^{+} &=&J^{x}+iJ^{y}  \label{d52} \\
&=&-\frac{i}{2m}\left( \left[ \phi ^{\dagger },\left( D^{x}\phi \right) %
\right] -\left[ \left( D^{x}\phi \right) ^{\dagger },\phi \right] \right)
+i\left\{ -\frac{i}{2m}\left( \left[ \phi ^{\dagger },\left( D^{y}\phi
\right) \right] -\left[ \left( D^{y}\phi \right) ^{\dagger },\phi \right]
\right) \right\}  \notag \\
&=&-\frac{i}{2m}\left\{ \left[ \phi ^{\dagger },\left( D^{x}\phi \right) %
\right] +i\left[ \phi ^{\dagger },\left( D^{y}\phi \right) \right] \right. 
\notag \\
&&\ \ \ \ \ \ \ \ \left. -\left[ \left( D^{x}\phi \right) ^{\dagger },\phi %
\right] -i\left[ \left( D^{y}\phi \right) ^{\dagger },\phi \right] \right\} 
\notag \\
&=&-\frac{i}{2m}\left( \left[ \phi ^{\dagger },\left( D^{+}\phi \right) %
\right] -\left[ \left( D^{-}\phi \right) ^{\dagger },\phi \right] \right) 
\notag
\end{eqnarray}%
and inserting in the equation written above the equation at Self-Duality $%
D_{-}\phi =0$ we get 
\begin{equation}
J^{+}=-\frac{i}{2m}\left( \left[ \phi ^{\dagger },\left( D^{+}\phi \right) %
\right] \right) \ \ \text{at Self-Duality}  \label{d53}
\end{equation}

We return to the expression of the current in the second (gauge-field)
equation of motion, which is the Gauss law 
\begin{equation}
\kappa \varepsilon ^{\mu \nu \rho }F_{\nu \rho }=iJ^{\mu }  \label{d54}
\end{equation}%
and take the $x$ and $y$ components 
\begin{eqnarray}
\kappa \varepsilon ^{x\mu \nu }F_{\mu \nu } &=&iJ^{x}  \label{d55} \\
\kappa \varepsilon ^{y\mu \nu }F_{\mu \nu } &=&iJ^{y}  \notag
\end{eqnarray}%
\begin{eqnarray}
\kappa \left( \varepsilon ^{xy0}F_{y0}+\varepsilon ^{x0y}F_{0y}\right)
&=&iJ^{x}  \label{d57} \\
2\kappa F_{y0} &=&iJ^{x}  \notag \\
2\kappa \left( \partial _{y}A_{0}-\partial _{0}A_{y}+\left[ A_{y},A_{0}%
\right] \right) &=&iJ^{x}  \notag
\end{eqnarray}%
and analogous 
\begin{eqnarray}
\kappa \left( \varepsilon ^{yx0}F_{x0}+\varepsilon ^{y0x}F_{0x}\right)
&=&iJ^{y}  \label{d58} \\
-2\kappa F_{x0} &=&iJ^{y}  \notag \\
-2\kappa \left( \partial _{x}A_{0}-\partial _{0}A_{x}+\left[ A_{x},A_{0}%
\right] \right) &=&iJ^{y}  \notag
\end{eqnarray}%
Now we combine them 
\begin{eqnarray}
i\left( J^{x}+iJ^{y}\right) &=&2\kappa \left( \partial _{y}A_{0}-\partial
_{0}A_{y}+\left[ A_{y},A_{0}\right] \right.  \label{d59} \\
&&\left. -i\left( \partial _{x}A_{0}-\partial _{0}A_{x}+\left[ A_{x},A_{0}%
\right] \right) \right)  \notag \\
&=&2\kappa \left( \left( \partial _{y}-i\partial _{x}\right) A_{0}\right. 
\notag \\
&&-\partial _{0}\left( A_{y}-iA_{x}\right)  \notag \\
&&\left. +\left[ A_{y}-iA_{x},A_{0}\right] \right)  \notag \\
&=&\frac{2\kappa }{i}\left( \left( \partial _{x}+i\partial _{y}\right)
A_{0}\right.  \notag \\
&&-\partial _{0}\left( A_{x}+iA_{y}\right)  \notag \\
&&\left. +\left[ A_{x}+iA_{y},A_{0}\right] \right)  \notag
\end{eqnarray}%
\begin{eqnarray}
-J^{+} &=&2\kappa \left( \partial _{+}A_{0}-\partial _{0}A_{+}+\left[
A_{+},A_{0}\right] \right)  \label{d60} \\
&=&2\kappa \left( D_{+}A_{0}-\partial _{0}A_{+}\right)  \notag
\end{eqnarray}%
where we have introduced the notation 
\begin{equation}
D_{+}\equiv \partial _{+}+\left[ A_{+},\right]  \label{d63}
\end{equation}%
Now we have two expressions for the current density $J^{+}$ at Self-Duality%
\begin{eqnarray}
J^{+} &=&-\frac{i}{2m}\left( \left[ \phi ^{\dagger },\left( D^{+}\phi
\right) \right] \right)  \label{d65} \\
J^{+} &=&-2\kappa \left( D^{+}A_{0}-\partial _{0}A_{+}\right)  \notag
\end{eqnarray}%
At stationarity 
\begin{equation}
\partial _{0}A_{+}=0  \label{d66}
\end{equation}%
and from the two expressions of the current we have%
\begin{equation}
J^{+}=-\frac{i}{2m}\left( \left[ \phi ^{\dagger },\left( D^{+}\phi \right) %
\right] \right) =-2\kappa \left( D^{+}A_{0}\right) \ \ \text{at SD}
\label{d67}
\end{equation}%
This allows us to identify the expression of the time-component of the
potential, $A_{0}$%
\begin{equation}
A_{0}=\frac{i}{4m\kappa }\left[ \phi ,\phi ^{\dagger }\right] \ \ \text{at SD%
}  \label{d68}
\end{equation}

We can replace%
\begin{equation}
\left[ \phi ,\phi ^{\dagger }\right] =\left( \rho _{1}-\rho _{2}\right) H
\label{d69}
\end{equation}%
and further, since at SD we have introduced $\omega =\Delta \ln \rho
_{1}=\Delta \ln \left( 1/\rho _{2}\right) =\Delta \psi $, 
\begin{equation}
\rho _{1}-\rho _{2}=-\frac{\kappa }{2}\omega \ \ \ \text{at SD}  \label{d70}
\end{equation}%
This shows that the zero component of the potential of interaction has
algebraic content reduced to the Cartan generator 
\begin{equation}
A_{0}\sim \frac{i}{4m\kappa }\left( \rho _{1}-\rho _{2}\right) H  \label{d71}
\end{equation}%
and that it is purely imaginary. The magnitude of $A_{0}$ at SD is given by
the \emph{vorticity}.

Using this suggestion and following the text \cite{FlorinMadiXXX} we take
the temporal component of the potential in the form 
\begin{eqnarray}
A^{0} &\equiv &bH  \label{55130} \\
A^{0\dagger } &=&A_{0}^{\ast T}\equiv -b^{\ast }H  \notag
\end{eqnarray}

Taking into account the metric we have%
\begin{equation}
A_{0}=-A^{0}=-bH  \label{55132}
\end{equation}%
and we can identify, at \emph{self-duality}:%
\begin{eqnarray}
A_{0} &=&\frac{i}{4m\kappa }\left[ \phi ,\phi ^{\dagger }\right] =\frac{i}{%
4m\kappa }\left( \rho _{1}-\rho _{2}\right) H  \label{d79} \\
&=&-bH  \notag
\end{eqnarray}%
or%
\begin{eqnarray}
b &=&-\frac{i}{4m\kappa }\left( \rho _{1}-\rho _{2}\right)  \label{d82} \\
&=&\text{imaginary\ \ }\left( b^{\ast }+b=0\right)  \notag
\end{eqnarray}%
and, after identifications at \emph{SD},%
\begin{equation}
b=\frac{i}{8m\kappa }\omega \ \ \text{at SD}  \label{b84}
\end{equation}

\bigskip

In detail, the operator of covariant derivative to time%
\begin{eqnarray}
&&i\left( \frac{\partial \phi }{\partial t}+A_{0}\phi -\phi A_{0}\right)
\label{55135} \\
&=&i\left\{ \frac{\partial }{\partial t}\left( \phi _{1}E_{+}+\phi
_{2}E_{-}\right) +\left[ A_{0},\phi _{1}E_{+}+\phi _{2}E_{-}\right] \right\}
\notag \\
&=&i\frac{\partial \phi _{1}}{\partial t}E_{+}+i\frac{\partial \phi _{2}}{%
\partial t}E_{-}+i\left( -b\right) \phi _{1}\left[ H,E_{+}\right] +i\left(
-b\right) \phi _{2}\left[ H,E_{-}\right]  \notag \\
&=&i\frac{\partial \phi _{1}}{\partial t}E_{+}+i\frac{\partial \phi _{2}}{%
\partial t}E_{-}-2ib\phi _{1}E_{+}+2ib\phi _{2}E_{-}  \notag
\end{eqnarray}%
Collecting the factors of the ladder generators%
\begin{eqnarray}
&&i\left( \frac{\partial \phi }{\partial t}+A_{0}\phi -\phi A_{0}\right)
\label{55139} \\
&=&\left( i\frac{\partial \phi _{1}}{\partial t}-2ib\phi _{1}\right)
E_{+}+\left( i\frac{\partial \phi _{2}}{\partial t}+2ib\phi _{2}\right) E_{-}
\notag
\end{eqnarray}

\subsubsection{The first part of the first term in the RHS of the FIRST
equation of motion, adopting the algebraic \emph{ansatz}}

We can try to replace the algebraic ansatz in the first term (for the $x$
component) of the Eq.(\ref{542}) 
\begin{equation}
\left( \frac{\partial }{\partial x}+\left[ A_{1},\right] \right) \left( 
\frac{\partial \phi }{\partial x}+A^{1}\phi -\phi A^{1}\right)  \label{551}
\end{equation}%
taking 
\begin{equation}
\phi =\phi _{1}E_{+}+\phi _{2}E_{-}  \label{552}
\end{equation}%
and 
\begin{eqnarray}
A_{x} &=&\frac{1}{2}\left( a-a^{\ast }\right) H  \label{axyeuler} \\
A_{y} &=&\frac{i}{2}\left( a+a^{\ast }\right) H  \notag
\end{eqnarray}

Then the second paranthesis is 
\begin{eqnarray}
&&\frac{\partial \phi }{\partial x}+A^{1}\phi -\phi A^{1}  \label{553} \\
&=&\frac{\partial }{\partial x}\left( \phi _{1}E_{+}+\phi _{2}E_{-}\right) +%
\left[ \frac{1}{2}\left( a-a^{\ast }\right) H,\phi _{1}E_{+}+\phi _{2}E_{-}%
\right]  \notag \\
&=&\frac{\partial \phi _{1}}{\partial x}E_{+}+\frac{\partial \phi _{2}}{%
\partial x}E_{-}+\frac{1}{2}\left( a-a^{\ast }\right) \phi _{1}\left[ H,E_{+}%
\right] +\frac{1}{2}\left( a-a^{\ast }\right) \phi _{2}\left[ H,E_{-}\right]
\notag
\end{eqnarray}

Here we must use the commutators of the generators and obtain%
\begin{eqnarray}
&&\frac{\partial \phi }{\partial x}+A^{1}\phi -\phi A^{1}  \label{554} \\
&=&\frac{\partial \phi _{1}}{\partial x}E_{+}+\frac{\partial \phi _{2}}{%
\partial x}E_{-}+\frac{1}{2}\left( a-a^{\ast }\right) \phi _{1}2E_{+}-\frac{1%
}{2}\left( a-a^{\ast }\right) \phi _{2}2E_{-}  \notag \\
&=&\left[ \frac{\partial \phi _{1}}{\partial x}+\left( a-a^{\ast }\right)
\phi _{1}\right] E_{+}+\left[ \frac{\partial \phi _{2}}{\partial x}-\left(
a-a^{\ast }\right) \phi _{2}\right] E_{-}  \notag
\end{eqnarray}%
The first paranthesis%
\begin{equation}
\frac{\partial }{\partial x}+\left[ A_{1},\right]  \label{555}
\end{equation}%
is an operator which is applied on the second paranthesis%
\begin{eqnarray}
&&\left( \frac{\partial }{\partial x}+\left[ A_{1},\right] \right) \left\{ %
\left[ \frac{\partial \phi _{1}}{\partial x}+\left( a-a^{\ast }\right) \phi
_{1}\right] E_{+}+\left[ \frac{\partial \phi _{2}}{\partial x}-\left(
a-a^{\ast }\right) \phi _{2}\right] E_{-}\right\}  \notag \\
&=&\left( \frac{\partial }{\partial x}+\left[ A_{1},\right] \right) \left[ 
\frac{\partial \phi _{1}}{\partial x}+\left( a-a^{\ast }\right) \phi _{1}%
\right] E_{+}+\left( \frac{\partial }{\partial x}+\left[ A_{1},\right]
\right) \left[ \frac{\partial \phi _{2}}{\partial x}-\left( a-a^{\ast
}\right) \phi _{2}\right] E_{-}  \notag \\
&\equiv &I^{1}+II^{1}  \label{556}
\end{eqnarray}%
The first part is%
\begin{equation*}
I^{1}\equiv \left( \frac{\partial }{\partial x}+\left[ A_{1},\right] \right)
\left\{ \left[ \frac{\partial \phi _{1}}{\partial x}+\left( a-a^{\ast
}\right) \phi _{1}\right] E_{+}\right\}
\end{equation*}%
and is written in detail%
\begin{eqnarray}
I^{1} &=&\left( \frac{\partial }{\partial x}+\left[ A_{1},\right] \right) %
\left[ \frac{\partial \phi _{1}}{\partial x}+\left( a-a^{\ast }\right) \phi
_{1}\right] E_{+}  \label{557} \\
&=&\frac{\partial ^{2}\phi _{1}}{\partial x^{2}}E_{+}+\left[ \frac{\partial
\left( a-a^{\ast }\right) }{\partial x}\phi _{1}+\left( a-a^{\ast }\right) 
\frac{\partial \phi _{1}}{\partial x}\right] E_{+}  \notag \\
&&+\frac{\partial \phi _{1}}{\partial x}\left[ A_{1},E_{+}\right]  \notag \\
&&+\left( a-a^{\ast }\right) \phi _{1}\left[ A_{1},E_{+}\right]  \notag
\end{eqnarray}%
and we have 
\begin{eqnarray}
\left[ A_{1},E_{+}\right] &=&\frac{1}{2}\left( a-a^{\ast }\right) \left[
H,E_{+}\right]  \label{558} \\
&=&\frac{1}{2}\left( a-a^{\ast }\right) 2E_{+}  \notag \\
&=&\left( a-a^{\ast }\right) E_{+}  \notag
\end{eqnarray}

Then the first part $I$ becomes%
\begin{eqnarray}
I^{1} &=&\left( \frac{\partial }{\partial x}+\left[ A_{1},\right] \right) %
\left[ \frac{\partial \phi _{1}}{\partial x}+\left( a-a^{\ast }\right) \phi
_{1}\right] E_{+}  \label{559} \\
&=&\frac{\partial ^{2}\phi _{1}}{\partial x^{2}}E_{+}+\left[ \frac{\partial
\left( a-a^{\ast }\right) }{\partial x}\phi _{1}+\left( a-a^{\ast }\right) 
\frac{\partial \phi _{1}}{\partial x}\right] E_{+}  \notag \\
&&+\frac{\partial \phi _{1}}{\partial x}\left( a-a^{\ast }\right)
E_{+}+\left( a-a^{\ast }\right) ^{2}\phi _{1}E_{+}  \notag
\end{eqnarray}

\bigskip

\bigskip

The second part is%
\begin{equation}
II^{1}\equiv \left( \frac{\partial }{\partial x}+\left[ A_{1},\right]
\right) \left\{ \left[ \frac{\partial \phi _{2}}{\partial x}-\left(
a-a^{\ast }\right) \phi _{2}\right] E_{-}\right\}  \label{d110}
\end{equation}%
Now we expand the second part $II^{1}$%
\begin{eqnarray}
II^{1} &=&\left( \frac{\partial }{\partial x}+\left[ A_{1},\right] \right) %
\left[ \frac{\partial \phi _{2}}{\partial x}-\left( a-a^{\ast }\right) \phi
_{2}\right] E_{-}  \label{560} \\
&=&\frac{\partial ^{2}\phi _{2}}{\partial x^{2}}E_{-}-\left[ \frac{\partial
\left( a-a^{\ast }\right) }{\partial x}\phi _{2}+\left( a-a^{\ast }\right) 
\frac{\partial \phi _{2}}{\partial x}\right] E_{-}  \notag \\
&&+\frac{\partial \phi _{2}}{\partial x}\left[ A_{1},E_{-}\right] -\left(
a-a^{\ast }\right) \phi _{2}\left[ A_{1},E_{-}\right]  \notag
\end{eqnarray}%
The commutator is%
\begin{eqnarray}
\left[ A_{1},E_{-}\right] &=&\frac{1}{2}\left( a-a^{\ast }\right) \left[
H,E_{-}\right]  \label{561} \\
&=&\frac{1}{2}\left( a-a^{\ast }\right) \left( -2E_{-}\right)  \notag \\
&=&-\left( a-a^{\ast }\right) E_{-}  \notag
\end{eqnarray}%
and the second term becomes%
\begin{eqnarray}
II^{1} &=&\left( \frac{\partial }{\partial x}+\left[ A_{1},\right] \right) %
\left[ \frac{\partial \phi _{2}}{\partial x}-\left( a-a^{\ast }\right) \phi
_{2}\right] E_{-}  \label{562} \\
&=&\frac{\partial ^{2}\phi _{2}}{\partial x^{2}}E_{-}-\left[ \frac{\partial
\left( a-a^{\ast }\right) }{\partial x}\phi _{2}+\left( a-a^{\ast }\right) 
\frac{\partial \phi _{2}}{\partial x}\right] E_{-}  \notag \\
&&-\frac{\partial \phi _{2}}{\partial x}\left( a-a^{\ast }\right)
E_{-}+\left( a-a^{\ast }\right) ^{2}\phi _{2}E_{-}  \notag
\end{eqnarray}

\bigskip

Now we collect the two terms $I^{1}$ from Eq.(\ref{559}) and $II^{1}$ from
Eq.(\ref{562})%
\begin{eqnarray}
&&D_{x}D^{x}\phi  \label{d210} \\
&=&\left( \frac{\partial }{\partial x}+\left[ A_{1},\right] \right) \left( 
\frac{\partial \phi }{\partial x}+A^{1}\phi -\phi A^{1}\right)  \notag \\
&=&I^{1}+II^{1}  \notag
\end{eqnarray}%
\begin{eqnarray}
&&\left( \frac{\partial }{\partial x}+\left[ A_{1},\right] \right) \left( 
\frac{\partial \phi }{\partial x}+A^{1}\phi -\phi A^{1}\right)  \label{563}
\\
&=&\left( \frac{\partial }{\partial x}+\left[ A_{1},\right] \right) \left\{ %
\left[ \frac{\partial \phi _{1}}{\partial x}+\left( a-a^{\ast }\right) \phi
_{1}\right] E_{+}+\left[ \frac{\partial \phi _{2}}{\partial x}-\left(
a-a^{\ast }\right) \phi _{2}\right] E_{-}\right\}  \notag \\
&=&\frac{\partial ^{2}\phi _{1}}{\partial x^{2}}E_{+}+\left[ \frac{\partial
\left( a-a^{\ast }\right) }{\partial x}\phi _{1}+\left( a-a^{\ast }\right) 
\frac{\partial \phi _{1}}{\partial x}\right] E_{+}  \notag \\
&&+\frac{\partial \phi _{1}}{\partial x}\left( a-a^{\ast }\right)
E_{+}+\left( a-a^{\ast }\right) ^{2}\phi _{1}E_{+}  \notag \\
&&+\frac{\partial ^{2}\phi _{2}}{\partial x^{2}}E_{-}-\left[ \frac{\partial
\left( a-a^{\ast }\right) }{\partial x}\phi _{2}+\left( a-a^{\ast }\right) 
\frac{\partial \phi _{2}}{\partial x}\right] E_{-}  \notag \\
&&-\frac{\partial \phi _{2}}{\partial x}\left( a-a^{\ast }\right)
E_{-}+\left( a-a^{\ast }\right) ^{2}\phi _{2}E_{-}  \notag
\end{eqnarray}%
This is the first part of the first term in the RHS of the FIRST equation of
motion.

\subsubsection{The second part of the first term in the RHS of the FIRST
equation of motion, with the algebraic \emph{ansatz}}

This part is very similar to the previous one, with $x$ replaced by $y$ and $%
A_{1}$ replaced by $A_{2}$.%
\begin{eqnarray}
&&D_{y}D^{y}\phi  \label{564} \\
&=&\left( \frac{\partial }{\partial y}+\left[ A_{2},\right] \right) \left( 
\frac{\partial }{\partial y}+\left[ A^{2},\right] \right) \phi \\
&=&\left( \frac{\partial }{\partial y}+\left[ A_{2},\right] \right) \left( 
\frac{\partial \phi }{\partial y}+\left[ A^{2},\phi \right] \right)  \notag
\end{eqnarray}%
The second paranthesis can be written in more detail, using the \emph{%
algebraic ansatz}:%
\begin{eqnarray}
A_{-} &=&A_{x}-iA_{y}=aH  \label{d215} \\
A_{+} &=&A_{x}+iA_{y}=-a^{\ast }H  \notag
\end{eqnarray}%
\begin{eqnarray}
\phi &=&\phi _{1}E_{+}+\phi _{2}E_{-}  \label{565} \\
A_{y} &\equiv &A_{2}=\frac{i}{2}\left( a+a^{\ast }\right) H  \notag
\end{eqnarray}%
It is%
\begin{eqnarray}
&&\frac{\partial \phi }{\partial y}+\left[ A^{2},\phi \right]  \label{566} \\
&=&\frac{\partial }{\partial y}\left( \phi _{1}E_{+}+\phi _{2}E_{-}\right) +%
\left[ \frac{i}{2}\left( a+a^{\ast }\right) H,\phi _{1}E_{+}+\phi _{2}E_{-}%
\right]  \notag \\
&=&\frac{\partial \phi _{1}}{\partial y}E_{+}+\frac{\partial \phi _{2}}{%
\partial y}E_{-}+\frac{i}{2}\left( a+a^{\ast }\right) \phi _{1}\left[ H,E_{+}%
\right] +\frac{i}{2}\left( a+a^{\ast }\right) \phi _{2}\left[ H,E_{-}\right]
\notag \\
&=&\frac{\partial \phi _{1}}{\partial y}E_{+}+\frac{\partial \phi _{2}}{%
\partial y}E_{-}+i\left( a+a^{\ast }\right) \phi _{1}E_{+}-i\left( a+a^{\ast
}\right) \phi _{2}E_{-}  \notag \\
&=&\left[ \frac{\partial \phi _{1}}{\partial y}+i\left( a+a^{\ast }\right)
\phi _{1}\right] E_{+}+\left[ \frac{\partial \phi _{2}}{\partial y}-i\left(
a+a^{\ast }\right) \phi _{2}\right] E_{-}  \notag
\end{eqnarray}%
On this expression we have to apply the operator from the first paranthesis%
\begin{eqnarray}
&&\left( \frac{\partial }{\partial y}+\left[ A_{2},\right] \right) \left\{ %
\left[ \frac{\partial \phi _{1}}{\partial y}+i\left( a+a^{\ast }\right) \phi
_{1}\right] E_{+}+\left[ \frac{\partial \phi _{2}}{\partial y}-i\left(
a+a^{\ast }\right) \phi _{2}\right] E_{-}\right\}  \notag \\
&\equiv &I^{2}+II^{2}  \label{d219}
\end{eqnarray}%
The first part%
\begin{eqnarray}
I^{2} &=&\left( \frac{\partial }{\partial y}+\left[ A_{2},\right] \right)
\left\{ \left[ \frac{\partial \phi _{1}}{\partial y}+i\left( a+a^{\ast
}\right) \phi _{1}\right] E_{+}\right\}  \label{d221} \\
&=&\frac{\partial ^{2}\phi _{1}}{\partial y^{2}}E_{+}+i\left[ \frac{\partial
\left( a+a^{\ast }\right) }{\partial y}\phi _{1}+\left( a+a^{\ast }\right) 
\frac{\partial \phi _{1}}{\partial y}\right] E_{+}  \notag \\
&&+\frac{\partial \phi _{1}}{\partial y}\left[ A_{2},E_{+}\right] +i\left(
a+a^{\ast }\right) \phi _{1}\left[ A_{2},E_{+}\right]  \notag
\end{eqnarray}%
Here we replace%
\begin{equation}
A_{2}=\frac{i}{2}\left( a+a^{\ast }\right) H  \label{568}
\end{equation}%
and we have the commutator%
\begin{eqnarray}
\left[ A_{2},E_{+}\right] &=&\frac{i}{2}\left( a+a^{\ast }\right) \left[
H,E_{+}\right]  \label{569} \\
&=&i\left( a+a^{\ast }\right) E_{+}  \notag
\end{eqnarray}%
and obtain%
\begin{eqnarray}
I^{2} &=&\frac{\partial ^{2}\phi _{1}}{\partial y^{2}}E_{+}+i\left[ \frac{%
\partial \left( a+a^{\ast }\right) }{\partial y}\phi _{1}+\left( a+a^{\ast
}\right) \frac{\partial \phi _{1}}{\partial y}\right] E_{+}  \label{d223} \\
&&+\frac{\partial \phi _{1}}{\partial y}i\left( a+a^{\ast }\right)
E_{+}-\left( a+a^{\ast }\right) ^{2}\phi _{1}E_{+}  \notag
\end{eqnarray}

\bigskip

\bigskip

Now we expand the expression of the second part%
\begin{eqnarray}
II^{2} &=&\left( \frac{\partial }{\partial y}+\left[ A_{2},\right] \right)
\left\{ \left[ \frac{\partial \phi _{2}}{\partial y}-i\left( a+a^{\ast
}\right) \phi _{2}\right] E_{-}\right\}  \label{d224} \\
&=&\frac{\partial ^{2}\phi _{2}}{\partial y^{2}}E_{-}-i\left[ \frac{\partial
\left( a+a^{\ast }\right) }{\partial y}\phi _{2}+\left( a+a^{\ast }\right) 
\frac{\partial \phi _{2}}{\partial y}\right] E_{-}  \notag \\
&&+\frac{\partial \phi _{2}}{\partial y}\left[ A_{2},E_{-}\right] -i\left(
a+a^{\ast }\right) \phi _{2}\left[ A_{2},E_{-}\right]  \notag
\end{eqnarray}%
As before we use 
\begin{eqnarray}
\left[ A_{2},E_{-}\right] &=&\frac{i}{2}\left( a+a^{\ast }\right) \left[
H,E_{-}\right]  \label{570} \\
&=&-i\left( a+a^{\ast }\right) E_{-}  \notag
\end{eqnarray}%
to replace the commutators%
\begin{eqnarray}
II^{2} &=&\frac{\partial ^{2}\phi _{2}}{\partial y^{2}}E_{-}-i\left[ \frac{%
\partial \left( a+a^{\ast }\right) }{\partial y}\phi _{2}+\left( a+a^{\ast
}\right) \frac{\partial \phi _{2}}{\partial y}\right] E_{-}  \label{d226} \\
&&+\frac{\partial \phi _{2}}{\partial y}\left( -\right) i\left( a+a^{\ast
}\right) E_{-}-\left( a+a^{\ast }\right) ^{2}\phi _{2}E_{-}  \notag
\end{eqnarray}%
The final formula for this first part of the Right Hand Side is%
\begin{equation}
I^{2}+II^{2}  \label{d234}
\end{equation}%
\begin{eqnarray}
&&\left( \frac{\partial }{\partial y}+\left[ A_{2},\right] \right) \left\{ %
\left[ \frac{\partial \phi _{1}}{\partial y}+i\left( a+a^{\ast }\right) \phi
_{1}\right] E_{+}+\left[ \frac{\partial \phi _{2}}{\partial y}-i\left(
a+a^{\ast }\right) \phi _{2}\right] E_{-}\right\}  \notag \\
&=&\frac{\partial ^{2}\phi _{1}}{\partial y^{2}}E_{+}+i\left[ \frac{\partial
\left( a+a^{\ast }\right) }{\partial y}\phi _{1}+\left( a+a^{\ast }\right) 
\frac{\partial \phi _{1}}{\partial y}\right] E_{+}  \notag \\
&&+\frac{\partial ^{2}\phi _{2}}{\partial y^{2}}E_{-}-i\left[ \frac{\partial
\left( a+a^{\ast }\right) }{\partial y}\phi _{2}+\left( a+a^{\ast }\right) 
\frac{\partial \phi _{2}}{\partial y}\right] E_{-}  \notag \\
&&+\frac{\partial \phi _{1}}{\partial y}i\left( a+a^{\ast }\right)
E_{+}-\left( a+a^{\ast }\right) ^{2}\phi _{1}E_{+}  \notag \\
&&+\frac{\partial \phi _{2}}{\partial y}\left( -\right) i\left( a+a^{\ast
}\right) E_{-}-\left( a+a^{\ast }\right) ^{2}\phi _{2}E_{-}  \label{571}
\end{eqnarray}

\subsubsection{The full first term in the RHS of the FIRST equation of
motion with the algebraic \emph{ansatz}}

This term is%
\begin{eqnarray}
&&-\frac{1}{2}\left\{ \left( \frac{\partial }{\partial x}+\left[ A_{1},%
\right] \right) \left( \frac{\partial }{\partial x}+\left[ A^{1},\right]
\right) +\left( \frac{\partial }{\partial y}+\left[ A_{2},\right] \right)
\left( \frac{\partial }{\partial y}+\left[ A^{2},\right] \right) \right\}
\phi  \notag \\
&=&-\frac{1}{2}\left( I^{1}+II^{1}+I^{2}+II^{2}\right)  \label{572}
\end{eqnarray}%
and it is constructed on the basis of the Eqs.(\ref{563}) and (\ref{571}).
We write separately the coefficients of $E_{+}$ and of $E_{-}$.

The coefficient of $E_{+}$ (not yet multiplied by $-1/2$) is%
\begin{eqnarray}
&&\frac{\partial ^{2}\phi _{1}}{\partial x^{2}}+\left[ \frac{\partial \left(
a-a^{\ast }\right) }{\partial x}\phi _{1}+\left( a-a^{\ast }\right) \frac{%
\partial \phi _{1}}{\partial x}\right]  \label{573} \\
&&+\frac{\partial \phi _{1}}{\partial x}\left( a-a^{\ast }\right) +\left(
a-a^{\ast }\right) ^{2}\phi _{1}  \notag \\
&&+\frac{\partial ^{2}\phi _{1}}{\partial y^{2}}+i\left[ \frac{\partial
\left( a+a^{\ast }\right) }{\partial y}\phi _{1}+\left( a+a^{\ast }\right) 
\frac{\partial \phi _{1}}{\partial y}\right]  \notag \\
&&+\frac{\partial \phi _{1}}{\partial y}i\left( a+a^{\ast }\right) -\left(
a+a^{\ast }\right) ^{2}\phi _{1}  \notag
\end{eqnarray}

The coefficient of $E_{-}$ (not yet multiplied by $-1/2$) is%
\begin{eqnarray}
&&\frac{\partial ^{2}\phi _{2}}{\partial x^{2}}-\left[ \frac{\partial \left(
a-a^{\ast }\right) }{\partial x}\phi _{2}+\left( a-a^{\ast }\right) \frac{%
\partial \phi _{2}}{\partial x}\right]  \label{574} \\
&&-\frac{\partial \phi _{2}}{\partial x}\left( a-a^{\ast }\right) +\left(
a-a^{\ast }\right) ^{2}\phi _{2}  \notag \\
&&+\frac{\partial ^{2}\phi _{2}}{\partial y^{2}}-i\left[ \frac{\partial
\left( a+a^{\ast }\right) }{\partial y}\phi _{2}+\left( a+a^{\ast }\right) 
\frac{\partial \phi _{2}}{\partial y}\right]  \notag \\
&&+\frac{\partial \phi _{2}}{\partial y}\left( -\right) i\left( a+a^{\ast
}\right) -\left( a+a^{\ast }\right) ^{2}\phi _{2}  \notag
\end{eqnarray}

\subsubsection{The last term in the RHS of the first equation of motion,
with the algebraic \emph{ansatz}}

This term is%
\begin{equation}
-\frac{1}{2m\kappa }\left[ \left[ \phi ,\phi ^{\dagger }\right] ,\phi \right]
\label{575}
\end{equation}%
This is calculated in \textbf{xxx\_clean.tex}. The steps and the result are:

\begin{eqnarray}
\left[ \phi ,\phi ^{\dagger }\right] &=&\left( \phi _{1}^{\ast }\phi
_{1}-\phi _{2}^{\ast }\phi _{2}\right) H  \label{576} \\
&=&\left( \rho _{1}-\rho _{2}\right) H  \notag
\end{eqnarray}%
where we have introduced the notations 
\begin{eqnarray}
\rho _{1} &\equiv &\left\vert \phi _{1}\right\vert ^{2}  \label{577} \\
\rho _{2} &\equiv &\left\vert \phi _{2}\right\vert ^{2}  \notag
\end{eqnarray}%
The next step is to calculate 
\begin{equation}
\left[ \left[ \phi ,\phi ^{\dagger }\right] ,\phi \right] =\left[ \left(
\rho _{1}-\rho _{2}\right) H,\phi _{1}E_{+}+\phi _{2}E_{-}\right]  \label{50}
\end{equation}%
This is 
\begin{eqnarray}
\left[ \left[ \phi ,\phi ^{\dagger }\right] ,\phi \right] &=&\left( \rho
_{1}-\rho _{2}\right) \phi _{1}\left[ H,E_{+}\right] +\left( \rho _{1}-\rho
_{2}\right) \phi _{2}\left[ H,E_{-}\right]  \label{51} \\
&=&2\left( \rho _{1}-\rho _{2}\right) \left( \phi _{1}E_{+}-\phi
_{2}E_{-}\right)  \notag
\end{eqnarray}

Finally%
\begin{eqnarray}
-\frac{1}{2m\kappa }\left[ \left[ \phi ,\phi ^{\dagger }\right] ,\phi \right]
&=&-\frac{1}{2m\kappa }2\left( \rho _{1}-\rho _{2}\right) \left( \phi
_{1}E_{+}-\phi _{2}E_{-}\right)  \label{578} \\
&=&-\frac{1}{m\kappa }\left( \rho _{1}-\rho _{2}\right) \left( \phi
_{1}E_{+}-\phi _{2}E_{-}\right)  \notag
\end{eqnarray}

\subsubsection{The full equations obtained from the FIRST (matter) equation
of motion after adopting the algebraic \emph{ansatz}}

Here are the terms that results by equating the coefficients of the two
ladder generators.

The equation resulting from $E_{+}$. We use Eqs.(\ref{55139}), (\ref{573})
and (\ref{578})%
\begin{eqnarray}
&&i\frac{\partial \phi _{1}}{\partial t}-2ib\phi _{1}  \label{ap579} \\
&=&-\frac{1}{2}\frac{\partial ^{2}\phi _{1}}{\partial x^{2}}-\frac{1}{2}%
\left[ \frac{\partial \left( a-a^{\ast }\right) }{\partial x}\phi
_{1}+\left( a-a^{\ast }\right) \frac{\partial \phi _{1}}{\partial x}\right] 
\notag \\
&&-\frac{1}{2}\frac{\partial \phi _{1}}{\partial x}\left( a-a^{\ast }\right)
-\frac{1}{2}\left( a-a^{\ast }\right) ^{2}\phi _{1}  \notag \\
&&-\frac{1}{2}\frac{\partial ^{2}\phi _{1}}{\partial y^{2}}-\frac{i}{2}\left[
\frac{\partial \left( a+a^{\ast }\right) }{\partial y}\phi _{1}+\left(
a+a^{\ast }\right) \frac{\partial \phi _{1}}{\partial y}\right]  \notag \\
&&-\frac{i}{2}\frac{\partial \phi _{1}}{\partial y}\left( a+a^{\ast }\right)
+\frac{1}{2}\left( a+a^{\ast }\right) ^{2}\phi _{1}  \notag \\
&&-\frac{1}{m\kappa }\left( \rho _{1}-\rho _{2}\right) \phi _{1}  \notag
\end{eqnarray}

The equation resulting from $E_{-}$. We use Eqs.(\ref{55139}), (\ref{574})
and (\ref{578})%
\begin{eqnarray}
&&i\frac{\partial \phi _{2}}{\partial t}+2ib\phi _{2}  \label{ap580} \\
&=&-\frac{1}{2}\frac{\partial ^{2}\phi _{2}}{\partial x^{2}}+\frac{1}{2}%
\left[ \frac{\partial \left( a-a^{\ast }\right) }{\partial x}\phi
_{2}+\left( a-a^{\ast }\right) \frac{\partial \phi _{2}}{\partial x}\right] 
\notag \\
&&+\frac{1}{2}\frac{\partial \phi _{2}}{\partial x}\left( a-a^{\ast }\right)
-\frac{1}{2}\left( a-a^{\ast }\right) ^{2}\phi _{2}  \notag \\
&&-\frac{1}{2}\frac{\partial ^{2}\phi _{2}}{\partial y^{2}}+\frac{i}{2}\left[
\frac{\partial \left( a+a^{\ast }\right) }{\partial y}\phi _{2}+\left(
a+a^{\ast }\right) \frac{\partial \phi _{2}}{\partial y}\right]  \notag \\
&&+\frac{i}{2}\frac{\partial \phi _{2}}{\partial y}\left( a+a^{\ast }\right)
+\frac{1}{2}\left( a+a^{\ast }\right) ^{2}\phi _{2}  \notag \\
&&+\frac{1}{m\kappa }\left( \rho _{1}-\rho _{2}\right) \phi _{2}  \notag
\end{eqnarray}

\section{Appendix D. Applications of the equations of motion} \label{App:AppendixD}

\renewcommand{\theequation}{D.\arabic{equation}} \setcounter{equation}{0}

We examine how the equations of motion can be transformed into a form that
gives the time evolution of the vorticity, defined as%
\begin{equation}
\rho _{1}-\rho _{2}  \label{e10}
\end{equation}

\subsection{Derivation of the equation for $\protect\rho _{1}=\left\vert 
\protect\phi _{1}\right\vert ^{2}$}

The equation resulting from $E_{+}$.

This is the equation for $\phi _{1}$.%
\begin{eqnarray}
&&i\frac{\partial \phi _{1}}{\partial t}-2ib\phi _{1}  \label{57901} \\
&=&-\frac{1}{2}\frac{\partial ^{2}\phi _{1}}{\partial x^{2}}-\frac{1}{2}%
\left[ \frac{\partial \left( a-a^{\ast }\right) }{\partial x}\phi
_{1}+\left( a-a^{\ast }\right) \frac{\partial \phi _{1}}{\partial x}\right] 
\notag \\
&&-\frac{1}{2}\frac{\partial \phi _{1}}{\partial x}\left( a-a^{\ast }\right)
-\frac{1}{2}\left( a-a^{\ast }\right) ^{2}\phi _{1}  \notag \\
&&-\frac{1}{2}\frac{\partial ^{2}\phi _{1}}{\partial y^{2}}-\frac{i}{2}\left[
\frac{\partial \left( a+a^{\ast }\right) }{\partial y}\phi _{1}+\left(
a+a^{\ast }\right) \frac{\partial \phi _{1}}{\partial y}\right]  \notag \\
&&-\frac{i}{2}\frac{\partial \phi _{1}}{\partial y}\left( a+a^{\ast }\right)
+\frac{1}{2}\left( a+a^{\ast }\right) ^{2}\phi _{1}  \notag \\
&&-\frac{1}{m\kappa }\left( \rho _{1}-\rho _{2}\right) \phi _{1}  \notag
\end{eqnarray}

Now we write this equation for the complex conjugate, $\phi _{1}^{\ast }$.%
\begin{eqnarray}
&&-i\frac{\partial \phi _{1}^{\ast }}{\partial t}+2ib^{\ast }\phi _{1}^{\ast
}  \label{57902} \\
&=&-\frac{1}{2}\frac{\partial ^{2}\phi _{1}^{\ast }}{\partial x^{2}}-\frac{1%
}{2}\left[ \frac{\partial \left( a^{\ast }-a\right) }{\partial x}\phi
_{1}^{\ast }+\left( a^{\ast }-a\right) \frac{\partial \phi _{1}^{\ast }}{%
\partial x}\right]  \notag \\
&&-\frac{1}{2}\frac{\partial \phi _{1}^{\ast }}{\partial x}\left( a^{\ast
}-a\right) -\frac{1}{2}\left( a^{\ast }-a\right) ^{2}\phi _{1}^{\ast } 
\notag \\
&&-\frac{1}{2}\frac{\partial ^{2}\phi _{1}^{\ast }}{\partial y^{2}}+\frac{i}{%
2}\left[ \frac{\partial \left( a^{\ast }+a\right) }{\partial y}\phi
_{1}^{\ast }+\left( a^{\ast }+a\right) \frac{\partial \phi _{1}^{\ast }}{%
\partial y}\right]  \notag \\
&&+\frac{i}{2}\frac{\partial \phi _{1}^{\ast }}{\partial y}\left( a^{\ast
}+a\right) +\frac{1}{2}\left( a^{\ast }+a\right) ^{2}\phi _{1}^{\ast } 
\notag \\
&&-\frac{1}{m\kappa }\left( \rho _{1}-\rho _{2}\right) \phi _{1}^{\ast } 
\notag
\end{eqnarray}

Now we multiply the first equation (for $\phi _{1}$) with $\phi _{1}^{\ast }$%
.%
\begin{eqnarray}
&&i\phi _{1}^{\ast }\frac{\partial \phi _{1}}{\partial t}-2ib\phi _{1}^{\ast
}\phi _{1}  \label{57903} \\
&=&-\frac{1}{2}\phi _{1}^{\ast }\frac{\partial ^{2}\phi _{1}}{\partial x^{2}}%
-\frac{1}{2}\left[ \frac{\partial \left( a-a^{\ast }\right) }{\partial x}%
\phi _{1}^{\ast }\phi _{1}+\left( a-a^{\ast }\right) \phi _{1}^{\ast }\frac{%
\partial \phi _{1}}{\partial x}\right]  \notag \\
&&-\frac{1}{2}\phi _{1}^{\ast }\frac{\partial \phi _{1}}{\partial x}\left(
a-a^{\ast }\right) -\frac{1}{2}\left( a-a^{\ast }\right) ^{2}\phi _{1}^{\ast
}\phi _{1}  \notag \\
&&-\frac{1}{2}\phi _{1}^{\ast }\frac{\partial ^{2}\phi _{1}}{\partial y^{2}}-%
\frac{i}{2}\left[ \frac{\partial \left( a+a^{\ast }\right) }{\partial y}\phi
_{1}^{\ast }\phi _{1}+\left( a+a^{\ast }\right) \phi _{1}^{\ast }\frac{%
\partial \phi _{1}}{\partial y}\right]  \notag \\
&&-\frac{i}{2}\phi _{1}^{\ast }\frac{\partial \phi _{1}}{\partial y}\left(
a+a^{\ast }\right) +\frac{1}{2}\left( a+a^{\ast }\right) ^{2}\phi _{1}^{\ast
}\phi _{1}  \notag \\
&&-\frac{1}{\kappa }\left( \rho _{1}-\rho _{2}\right) \phi _{1}^{\ast }\phi
_{1}  \notag
\end{eqnarray}

\bigskip

Similarly, we multiply the second equation (for $\phi _{1}^{\ast }$) by $%
\phi _{1}$.%
\begin{eqnarray}
&&-i\phi _{1}\frac{\partial \phi _{1}^{\ast }}{\partial t}+2ib^{\ast }\phi
_{1}\phi _{1}^{\ast }  \label{57904} \\
&=&-\frac{1}{2}\phi _{1}\frac{\partial ^{2}\phi _{1}^{\ast }}{\partial x^{2}}%
-\frac{1}{2}\left[ \frac{\partial \left( a^{\ast }-a\right) }{\partial x}%
\phi _{1}\phi _{1}^{\ast }+\left( a^{\ast }-a\right) \phi _{1}\frac{\partial
\phi _{1}^{\ast }}{\partial x}\right]  \notag \\
&&-\frac{1}{2}\phi _{1}\frac{\partial \phi _{1}^{\ast }}{\partial x}\left(
a^{\ast }-a\right) -\frac{1}{2}\left( a^{\ast }-a\right) ^{2}\phi _{1}\phi
_{1}^{\ast }  \notag \\
&&-\frac{1}{2}\phi _{1}\frac{\partial ^{2}\phi _{1}^{\ast }}{\partial y^{2}}+%
\frac{i}{2}\left[ \frac{\partial \left( a^{\ast }+a\right) }{\partial y}\phi
_{1}\phi _{1}^{\ast }+\left( a^{\ast }+a\right) \phi _{1}\frac{\partial \phi
_{1}^{\ast }}{\partial y}\right]  \notag \\
&&+\frac{i}{2}\phi _{1}\frac{\partial \phi _{1}^{\ast }}{\partial y}\left(
a^{\ast }+a\right) +\frac{1}{2}\left( a^{\ast }+a\right) ^{2}\phi _{1}\phi
_{1}^{\ast }  \notag \\
&&-\frac{1}{m\kappa }\left( \rho _{1}-\rho _{2}\right) \phi _{1}\phi
_{1}^{\ast }  \notag
\end{eqnarray}

\bigskip

Here we begin the combination of the first two equations, one for $\phi _{1}$
and the second for $\phi _{1}^{\ast }$. We will work line by line. We \emph{%
substract} the two equations, with the intention of getting a time
derivative of the modulus $\phi _{1}\phi _{1}^{\ast }$.%
\begin{eqnarray}
&&\text{first line }\left( i\phi _{1}^{\ast }\frac{\partial \phi _{1}}{%
\partial t}-2ib\phi _{1}^{\ast }\phi _{1}\right) -\left( -i\phi _{1}\frac{%
\partial \phi _{1}^{\ast }}{\partial t}+2ib^{\ast }\phi _{1}\phi _{1}^{\ast
}\right)  \label{57905} \\
&=&i\frac{\partial }{\partial t}\left( \phi _{1}\phi _{1}^{\ast }\right)
-2i\left( b+b^{\ast }\right) \left\vert \phi _{1}\right\vert ^{2}  \notag
\end{eqnarray}%
\begin{equation}
\text{first term of the second line }\left( -\frac{1}{2}\phi _{1}^{\ast }%
\frac{\partial ^{2}\phi _{1}}{\partial x^{2}}\right) -\left( -\frac{1}{2}%
\phi _{1}\frac{\partial ^{2}\phi _{1}^{\ast }}{\partial x^{2}}\right)
\label{57906}
\end{equation}%
\begin{eqnarray}
&&\text{second term of the second line }  \label{57907} \\
&&\left( -\frac{1}{2}\left[ \frac{\partial \left( a-a^{\ast }\right) }{%
\partial x}\phi _{1}^{\ast }\phi _{1}+\left( a-a^{\ast }\right) \phi
_{1}^{\ast }\frac{\partial \phi _{1}}{\partial x}\right] \right) -  \notag \\
&&-\left( -\frac{1}{2}\left[ \frac{\partial \left( a^{\ast }-a\right) }{%
\partial x}\phi _{1}\phi _{1}^{\ast }+\left( a^{\ast }-a\right) \phi _{1}%
\frac{\partial \phi _{1}^{\ast }}{\partial x}\right] \right)  \notag \\
&=&-\frac{1}{2}\left[ \frac{\partial \left( a-a^{\ast }\right) }{\partial x}%
\left( \phi _{1}^{\ast }\phi _{1}+\phi _{1}\phi _{1}^{\ast }\right) \right. 
\notag \\
&&\left. +\left( a-a^{\ast }\right) \left( \phi _{1}^{\ast }\frac{\partial
\phi _{1}}{\partial x}+\phi _{1}\frac{\partial \phi _{1}^{\ast }}{\partial x}%
\right) \right]  \notag \\
&=&-\frac{\partial \left( a-a^{\ast }\right) }{\partial x}\left\vert \phi
_{1}\right\vert ^{2}-\frac{1}{2}\left( a-a^{\ast }\right) \frac{\partial }{%
\partial x}\left( \left\vert \phi _{1}\right\vert ^{2}\right)  \notag
\end{eqnarray}%
\begin{eqnarray}
&&\text{first term of the third line}  \label{57908} \\
&&\left( -\frac{1}{2}\phi _{1}^{\ast }\frac{\partial \phi _{1}}{\partial x}%
\left( a-a^{\ast }\right) \right) -\left( -\frac{1}{2}\phi _{1}\frac{%
\partial \phi _{1}^{\ast }}{\partial x}\left( a^{\ast }-a\right) \right) 
\notag \\
&=&-\frac{1}{2}\left( a-a^{\ast }\right) \left( \phi _{1}^{\ast }\frac{%
\partial \phi _{1}}{\partial x}+\phi _{1}\frac{\partial \phi _{1}^{\ast }}{%
\partial x}\right)  \notag \\
&=&-\frac{1}{2}\left( a-a^{\ast }\right) \frac{\partial }{\partial x}%
\left\vert \phi _{1}\right\vert ^{2}  \notag
\end{eqnarray}%
\begin{eqnarray*}
&&\text{second term of the third line} \\
&&\left( -\frac{1}{2}\left( a-a^{\ast }\right) ^{2}\phi _{1}^{\ast }\phi
_{1}\right) -\left( -\frac{1}{2}\left( a^{\ast }-a\right) ^{2}\phi _{1}\phi
_{1}^{\ast }\right) \\
&=&0
\end{eqnarray*}%
\begin{eqnarray}
&&\text{first term of the fourth line}  \label{57909} \\
&&\left( -\frac{1}{2}\phi _{1}^{\ast }\frac{\partial ^{2}\phi _{1}}{\partial
y^{2}}\right) -\left( -\frac{1}{2}\phi _{1}\frac{\partial ^{2}\phi
_{1}^{\ast }}{\partial y^{2}}\right)  \notag \\
&=&-\frac{1}{2}\left( \phi _{1}^{\ast }\frac{\partial ^{2}\phi _{1}}{%
\partial y^{2}}-\phi _{1}\frac{\partial ^{2}\phi _{1}^{\ast }}{\partial y^{2}%
}\right)  \notag
\end{eqnarray}%
\begin{eqnarray}
&&\text{second term of the fourth line}  \label{57910} \\
&&\left( -\frac{i}{2}\left[ \frac{\partial \left( a+a^{\ast }\right) }{%
\partial y}\phi _{1}^{\ast }\phi _{1}+\left( a+a^{\ast }\right) \phi
_{1}^{\ast }\frac{\partial \phi _{1}}{\partial y}\right] \right)  \notag \\
&&-\left( \frac{i}{2}\left[ \frac{\partial \left( a^{\ast }+a\right) }{%
\partial y}\phi _{1}\phi _{1}^{\ast }+\left( a^{\ast }+a\right) \phi _{1}%
\frac{\partial \phi _{1}^{\ast }}{\partial y}\right] \right)  \notag \\
&=&-i\frac{\partial \left( a+a^{\ast }\right) }{\partial y}\left\vert \phi
_{1}\right\vert ^{2}-\frac{i}{2}\left( a+a^{\ast }\right) \frac{\partial }{%
\partial y}\left\vert \phi _{1}\right\vert ^{2}  \notag
\end{eqnarray}%
\begin{eqnarray}
&&\text{first term of the fifth line}  \label{57911} \\
&&\left( -\frac{i}{2}\phi _{1}^{\ast }\frac{\partial \phi _{1}}{\partial y}%
\left( a+a^{\ast }\right) \right) -\left( \frac{i}{2}\phi _{1}\frac{\partial
\phi _{1}^{\ast }}{\partial y}\left( a^{\ast }+a\right) \right)  \notag \\
&=&-\frac{i}{2}\left( a+a^{\ast }\right) \frac{\partial }{\partial y}%
\left\vert \phi _{1}\right\vert ^{2}  \notag
\end{eqnarray}%
\begin{eqnarray}
&&\text{second term of the fifth line}  \label{57912} \\
&&\left( \frac{1}{2}\left( a+a^{\ast }\right) ^{2}\phi _{1}^{\ast }\phi
_{1}\right) -\left( \frac{1}{2}\left( a^{\ast }+a\right) ^{2}\phi _{1}\phi
_{1}^{\ast }\right)  \notag \\
&=&0  \notag
\end{eqnarray}%
\begin{eqnarray}
&&\text{term of the sixth line}  \label{57913} \\
&&\frac{1}{m\kappa }\left( -\left( \rho _{1}-\rho _{2}\right) \phi
_{1}^{\ast }\phi _{1}\right) -\left( -\left( \rho _{1}-\rho _{2}\right) \phi
_{1}\phi _{1}^{\ast }\right)  \notag \\
&=&0  \notag
\end{eqnarray}

\bigskip

What results:%
\begin{eqnarray}
&&i\frac{\partial }{\partial t}\left\vert \phi _{1}\right\vert ^{2}-2i\left(
b+b^{\ast }\right) \left\vert \phi _{1}\right\vert ^{2}\ \ \text{{\small %
first line}}  \label{57914} \\
&=&-\frac{1}{2}\left( \phi _{1}^{\ast }\frac{\partial ^{2}\phi _{1}}{%
\partial x^{2}}-\phi _{1}\frac{\partial ^{2}\phi _{1}^{\ast }}{\partial x^{2}%
}\right) \ \ \text{{\small first term of the second line}}  \notag \\
&&-\frac{\partial \left( a-a^{\ast }\right) }{\partial x}\left\vert \phi
_{1}\right\vert ^{2}-\frac{1}{2}\left( a-a^{\ast }\right) \frac{\partial }{%
\partial x}\left( \left\vert \phi _{1}\right\vert ^{2}\right) \ \text{%
{\small second term of the second line}}  \notag \\
&&-\frac{1}{2}\left( a-a^{\ast }\right) \frac{\partial }{\partial x}%
\left\vert \phi _{1}\right\vert ^{2}\ \ \text{{\small first term of the
third line}}  \notag \\
&&-\frac{1}{2}\left( \phi _{1}^{\ast }\frac{\partial ^{2}\phi _{1}}{\partial
y^{2}}-\phi _{1}\frac{\partial ^{2}\phi _{1}^{\ast }}{\partial y^{2}}\right)
\ \ \text{{\small first term of the fourth line}}  \notag \\
&&-i\frac{\partial \left( a+a^{\ast }\right) }{\partial y}\left\vert \phi
_{1}\right\vert ^{2}-\frac{i}{2}\left( a+a^{\ast }\right) \frac{\partial }{%
\partial y}\left\vert \phi _{1}\right\vert ^{2}\ \ \text{{\small second term
of the fourth line}}  \notag \\
&&-\frac{i}{2}\left( a+a^{\ast }\right) \frac{\partial }{\partial y}%
\left\vert \phi _{1}\right\vert ^{2}\ \ \text{{\small first term of the
fifth line}}  \notag
\end{eqnarray}

\bigskip

\subsubsection{Derivation of the equation for $\protect\rho _{2}=\left\vert 
\protect\phi _{2}\right\vert ^{2}$}

The equation resulting from $E_{-}$. We use Eqs.(\ref{55139}), (\ref{574})
and (\ref{578})%
\begin{eqnarray}
&&i\frac{\partial \phi _{2}}{\partial t}+2ib\phi _{2}  \label{57915} \\
&=&-\frac{1}{2}\frac{\partial ^{2}\phi _{2}}{\partial x^{2}}+\frac{1}{2}%
\left[ \frac{\partial \left( a-a^{\ast }\right) }{\partial x}\phi
_{2}+\left( a-a^{\ast }\right) \frac{\partial \phi _{2}}{\partial x}\right] 
\notag \\
&&+\frac{1}{2}\frac{\partial \phi _{2}}{\partial x}\left( a-a^{\ast }\right)
-\frac{1}{2}\left( a-a^{\ast }\right) ^{2}\phi _{2}  \notag \\
&&-\frac{1}{2}\frac{\partial ^{2}\phi _{2}}{\partial y^{2}}+\frac{i}{2}\left[
\frac{\partial \left( a+a^{\ast }\right) }{\partial y}\phi _{2}+\left(
a+a^{\ast }\right) \frac{\partial \phi _{2}}{\partial y}\right]  \notag \\
&&+\frac{i}{2}\frac{\partial \phi _{2}}{\partial y}\left( a+a^{\ast }\right)
+\frac{1}{2}\left( a+a^{\ast }\right) ^{2}\phi _{2}  \notag \\
&&+\frac{1}{m\kappa }\left( \rho _{1}-\rho _{2}\right) \phi _{2}  \notag
\end{eqnarray}

Now we write this equation after taking the complex cojugate%
\begin{eqnarray}
&&-i\frac{\partial \phi _{2}^{\ast }}{\partial t}-2ib^{\ast }\phi _{2}^{\ast
}  \label{57916} \\
&=&-\frac{1}{2}\frac{\partial ^{2}\phi _{2}^{\ast }}{\partial x^{2}}+\frac{1%
}{2}\left[ \frac{\partial \left( a^{\ast }-a\right) }{\partial x}\phi
_{2}^{\ast }+\left( a^{\ast }-a\right) \frac{\partial \phi _{2}^{\ast }}{%
\partial x}\right]  \notag \\
&&+\frac{1}{2}\frac{\partial \phi _{2}^{\ast }}{\partial x}\left( a^{\ast
}-a\right) -\frac{1}{2}\left( a^{\ast }-a\right) ^{2}\phi _{2}^{\ast } 
\notag \\
&&-\frac{1}{2}\frac{\partial ^{2}\phi _{2}^{\ast }}{\partial y^{2}}-\frac{i}{%
2}\left[ \frac{\partial \left( a^{\ast }+a\right) }{\partial y}\phi
_{2}^{\ast }+\left( a^{\ast }+a\right) \frac{\partial \phi _{2}^{\ast }}{%
\partial y}\right]  \notag \\
&&-\frac{i}{2}\frac{\partial \phi _{2}^{\ast }}{\partial y}\left( a^{\ast
}+a\right) +\frac{1}{2}\left( a^{\ast }+a\right) ^{2}\phi _{2}^{\ast } 
\notag \\
&&+\frac{1}{m\kappa }\left( \rho _{1}-\rho _{2}\right) \phi _{2}^{\ast } 
\notag
\end{eqnarray}

\bigskip

The first equation is multiplied with $\phi _{2}^{\ast }$ and the result is%
\begin{eqnarray}
&&i\phi _{2}^{\ast }\frac{\partial \phi _{2}}{\partial t}+2ib\phi _{2}^{\ast
}\phi _{2}  \label{57917} \\
&=&-\frac{1}{2}\phi _{2}^{\ast }\frac{\partial ^{2}\phi _{2}}{\partial x^{2}}%
+\frac{1}{2}\left[ \frac{\partial \left( a-a^{\ast }\right) }{\partial x}%
\phi _{2}^{\ast }\phi _{2}+\left( a-a^{\ast }\right) \phi _{2}^{\ast }\frac{%
\partial \phi _{2}}{\partial x}\right]  \notag \\
&&+\frac{1}{2}\phi _{2}^{\ast }\frac{\partial \phi _{2}}{\partial x}\left(
a-a^{\ast }\right) -\frac{1}{2}\left( a-a^{\ast }\right) ^{2}\phi _{2}^{\ast
}\phi _{2}  \notag \\
&&-\frac{1}{2}\phi _{2}^{\ast }\frac{\partial ^{2}\phi _{2}}{\partial y^{2}}+%
\frac{i}{2}\left[ \frac{\partial \left( a+a^{\ast }\right) }{\partial y}\phi
_{2}^{\ast }\phi _{2}+\left( a+a^{\ast }\right) \phi _{2}^{\ast }\frac{%
\partial \phi _{2}}{\partial y}\right]  \notag \\
&&+\frac{i}{2}\phi _{2}^{\ast }\frac{\partial \phi _{2}}{\partial y}\left(
a+a^{\ast }\right) +\frac{1}{2}\left( a+a^{\ast }\right) ^{2}\phi _{2}^{\ast
}\phi _{2}  \notag \\
&&+\frac{1}{m\kappa }\left( \rho _{1}-\rho _{2}\right) \phi _{2}^{\ast }\phi
_{2}  \notag
\end{eqnarray}%
and the equation for $\phi _{2}^{\ast }$ is multiplied by $\phi _{2}$ with
the result%
\begin{eqnarray}
&&-i\phi _{2}\frac{\partial \phi _{2}^{\ast }}{\partial t}-2ib^{\ast }\phi
_{2}\phi _{2}^{\ast }  \label{57918} \\
&=&-\frac{1}{2}\phi _{2}\frac{\partial ^{2}\phi _{2}^{\ast }}{\partial x^{2}}%
+\frac{1}{2}\left[ \frac{\partial \left( a^{\ast }-a\right) }{\partial x}%
\phi _{2}\phi _{2}^{\ast }+\left( a^{\ast }-a\right) \phi _{2}\frac{\partial
\phi _{2}^{\ast }}{\partial x}\right]  \notag \\
&&+\frac{1}{2}\phi _{2}\frac{\partial \phi _{2}^{\ast }}{\partial x}\left(
a^{\ast }-a\right) -\frac{1}{2}\left( a^{\ast }-a\right) ^{2}\phi _{2}\phi
_{2}^{\ast }  \notag \\
&&-\frac{1}{2}\phi _{2}\frac{\partial ^{2}\phi _{2}^{\ast }}{\partial y^{2}}-%
\frac{i}{2}\left[ \frac{\partial \left( a^{\ast }+a\right) }{\partial y}\phi
_{2}\phi _{2}^{\ast }+\left( a^{\ast }+a\right) \phi _{2}\frac{\partial \phi
_{2}^{\ast }}{\partial y}\right]  \notag \\
&&-\frac{i}{2}\phi _{2}\frac{\partial \phi _{2}^{\ast }}{\partial y}\left(
a^{\ast }+a\right) +\frac{1}{2}\left( a^{\ast }+a\right) ^{2}\phi _{2}\phi
_{2}^{\ast }  \notag \\
&&+\frac{1}{m\kappa }\left( \rho _{1}-\rho _{2}\right) \phi _{2}\phi
_{2}^{\ast }  \notag
\end{eqnarray}

Now we will substract the two equations, in order to obtain the time
derivative $\partial /\partial t$ of the product $\phi _{2}^{\ast }\phi _{2}$%
. The terms are written one by one%
\begin{eqnarray}
&&\text{first term on the first line}  \label{57919} \\
i\phi _{2}^{\ast }\frac{\partial \phi _{2}}{\partial t}+i\phi _{2}\frac{%
\partial \phi _{2}^{\ast }}{\partial t} &=&i\frac{\partial }{\partial t}%
\left\vert \phi _{2}\right\vert ^{2}  \notag
\end{eqnarray}%
\begin{eqnarray}
&&\text{the second term of the first line}  \label{57920} \\
&&2ib\phi _{2}^{\ast }\phi _{2}+2ib^{\ast }\phi _{2}\phi _{2}^{\ast }  \notag
\\
&=&2i\left( b+b^{\ast }\right) \left\vert \phi _{2}\right\vert ^{2}  \notag
\end{eqnarray}%
\begin{eqnarray}
&&\text{ the first term of the second line}  \label{57921} \\
&&-\frac{1}{2}\phi _{2}^{\ast }\frac{\partial ^{2}\phi _{2}}{\partial x^{2}}+%
\frac{1}{2}\phi _{2}\frac{\partial ^{2}\phi _{2}^{\ast }}{\partial x^{2}} 
\notag
\end{eqnarray}%
\begin{eqnarray}
&&\text{the second term of the second line}  \label{57922} \\
&&\frac{1}{2}\left[ \frac{\partial \left( a-a^{\ast }\right) }{\partial x}%
\phi _{2}^{\ast }\phi _{2}+\left( a-a^{\ast }\right) \phi _{2}^{\ast }\frac{%
\partial \phi _{2}}{\partial x}\right]  \notag \\
&&-\frac{1}{2}\left[ \frac{\partial \left( a^{\ast }-a\right) }{\partial x}%
\phi _{2}\phi _{2}^{\ast }+\left( a^{\ast }-a\right) \phi _{2}\frac{\partial
\phi _{2}^{\ast }}{\partial x}\right]  \notag \\
&=&\frac{\partial \left( a-a^{\ast }\right) }{\partial x}\left\vert \phi
_{2}\right\vert ^{2}+\frac{1}{2}\left( a-a^{\ast }\right) \frac{\partial }{%
\partial x}\left( \left\vert \phi _{2}\right\vert ^{2}\right)  \notag
\end{eqnarray}%
\begin{eqnarray}
&&\text{the first term of the third line}  \label{57923} \\
&&\frac{1}{2}\phi _{2}^{\ast }\frac{\partial \phi _{2}}{\partial x}\left(
a-a^{\ast }\right) -\frac{1}{2}\phi _{2}\frac{\partial \phi _{2}^{\ast }}{%
\partial x}\left( a^{\ast }-a\right)  \notag \\
&=&\frac{1}{2}\left( a-a^{\ast }\right) \frac{\partial }{\partial x}\left(
\left\vert \phi _{2}\right\vert ^{2}\right)  \notag
\end{eqnarray}%
\begin{eqnarray}
&&\text{the second term of the third line}  \label{57924} \\
&&-\frac{1}{2}\left( a-a^{\ast }\right) ^{2}\phi _{2}^{\ast }\phi _{2}+\frac{%
1}{2}\left( a^{\ast }-a\right) ^{2}\phi _{2}\phi _{2}^{\ast }  \notag \\
&=&0  \notag
\end{eqnarray}%
\begin{eqnarray}
&&\text{the first term of the fourth line}  \label{57925} \\
&&-\frac{1}{2}\phi _{2}^{\ast }\frac{\partial ^{2}\phi _{2}}{\partial y^{2}}+%
\frac{1}{2}\phi _{2}\frac{\partial ^{2}\phi _{2}^{\ast }}{\partial y^{2}} 
\notag
\end{eqnarray}%
\begin{eqnarray}
&&\text{the second term of the fourth line}  \label{57926} \\
&&\frac{i}{2}\left[ \frac{\partial \left( a+a^{\ast }\right) }{\partial y}%
\phi _{2}^{\ast }\phi _{2}+\left( a+a^{\ast }\right) \phi _{2}^{\ast }\frac{%
\partial \phi _{2}}{\partial y}\right]  \notag \\
&&+\frac{i}{2}\left[ \frac{\partial \left( a^{\ast }+a\right) }{\partial y}%
\phi _{2}\phi _{2}^{\ast }+\left( a^{\ast }+a\right) \phi _{2}\frac{\partial
\phi _{2}^{\ast }}{\partial y}\right]  \notag \\
&=&i\frac{\partial \left( a+a^{\ast }\right) }{\partial y}\left\vert \phi
_{2}\right\vert ^{2}+\frac{i}{2}\left( a+a^{\ast }\right) \frac{\partial }{%
\partial y}\left( \left\vert \phi _{2}\right\vert ^{2}\right)  \notag
\end{eqnarray}%
\begin{eqnarray}
&&\text{the first term in the fifth line}  \label{57927} \\
&&\frac{i}{2}\phi _{2}^{\ast }\frac{\partial \phi _{2}}{\partial y}\left(
a+a^{\ast }\right) +\frac{i}{2}\phi _{2}\frac{\partial \phi _{2}^{\ast }}{%
\partial y}\left( a^{\ast }+a\right)  \notag \\
&=&\frac{i}{2}\left( a+a^{\ast }\right) \frac{\partial }{\partial y}\left(
\left\vert \phi _{2}\right\vert ^{2}\right)  \notag
\end{eqnarray}%
\begin{eqnarray}
&&\text{the second term in the fifth line}  \label{57928} \\
&&\frac{1}{2}\left( a+a^{\ast }\right) ^{2}\phi _{2}^{\ast }\phi _{2}-\frac{1%
}{2}\left( a^{\ast }+a\right) ^{2}\phi _{2}\phi _{2}^{\ast }  \notag \\
&=&0  \notag
\end{eqnarray}%
\begin{eqnarray}
&&\text{the term of the sixth line}  \label{57929} \\
&&\frac{1}{m\kappa }\left( \rho _{1}-\rho _{2}\right) \phi _{2}^{\ast }\phi
_{2}-\frac{1}{m\kappa }\left( \rho _{1}-\rho _{2}\right) \phi _{2}\phi
_{2}^{\ast }  \notag \\
&=&0  \notag
\end{eqnarray}

\bigskip

What results%
\begin{eqnarray}
&&i\frac{\partial }{\partial t}\left\vert \phi _{2}\right\vert ^{2}+2i\left(
b+b^{\ast }\right) \left\vert \phi _{2}\right\vert ^{2}\ {\small \ }\text{%
{\small first line}}  \label{57930} \\
&=&-\frac{1}{2}\phi _{2}^{\ast }\frac{\partial ^{2}\phi _{2}}{\partial x^{2}}%
+\frac{1}{2}\phi _{2}\frac{\partial ^{2}\phi _{2}^{\ast }}{\partial x^{2}}\
\ \text{{\small first term of the second line}}  \notag \\
&&+\frac{\partial \left( a-a^{\ast }\right) }{\partial x}\left\vert \phi
_{2}\right\vert ^{2}+\frac{1}{2}\left( a-a^{\ast }\right) \frac{\partial }{%
\partial x}\left( \left\vert \phi _{2}\right\vert ^{2}\right) \ \ \text{%
{\small the second term of the second line}}  \notag \\
&&+\frac{1}{2}\left( a-a^{\ast }\right) \frac{\partial }{\partial x}\left(
\left\vert \phi _{2}\right\vert ^{2}\right) \ \ \text{{\small the first term
of the third line}}  \notag \\
&&-\frac{1}{2}\phi _{2}^{\ast }\frac{\partial ^{2}\phi _{2}}{\partial y^{2}}+%
\frac{1}{2}\phi _{2}\frac{\partial ^{2}\phi _{2}^{\ast }}{\partial y^{2}}\ \ 
\text{{\small the first term of the fourth line}}  \notag \\
&&+i\frac{\partial \left( a+a^{\ast }\right) }{\partial y}\left\vert \phi
_{2}\right\vert ^{2}+\frac{i}{2}\left( a+a^{\ast }\right) \frac{\partial }{%
\partial y}\left( \left\vert \phi _{2}\right\vert ^{2}\right) \ \ \text{%
{\small the second term of the fourth line}}  \notag \\
&&+\frac{i}{2}\left( a+a^{\ast }\right) \frac{\partial }{\partial y}\left(
\left\vert \phi _{2}\right\vert ^{2}\right) \text{\ \ {\small the first term
of the fifth line}}  \notag
\end{eqnarray}

\bigskip

\subsubsection{Derivation of the equation for the difference $\Omega \equiv
\left\vert \protect\phi _{1}\right\vert ^{2}-\left\vert \protect\phi %
_{2}\right\vert ^{2}$}

Now let us substract the two equations such as to obtain the combination 
\begin{equation}
\Omega \equiv \left\vert \phi _{1}\right\vert ^{2}-\left\vert \phi
_{2}\right\vert ^{2}  \label{57931}
\end{equation}%
and%
\begin{equation}
\Xi \equiv \left\vert \phi _{1}\right\vert ^{2}+\left\vert \phi
_{2}\right\vert ^{2}  \label{57932}
\end{equation}%
. 
\begin{equation}
i\frac{\partial }{\partial t}\left\vert \phi _{1}\right\vert ^{2}-i\frac{%
\partial }{\partial t}\left\vert \phi _{2}\right\vert ^{2}=i\frac{\partial }{%
\partial t}\Omega \ \text{{\small first terms on the first lines}}
\label{57933}
\end{equation}%
\begin{equation}
-2i\left( b+b^{\ast }\right) \left\vert \phi _{1}\right\vert ^{2}-2i\left(
b+b^{\ast }\right) \left\vert \phi _{2}\right\vert ^{2}\ =-2i\left(
b+b^{\ast }\right) \Xi \ \text{{\small second terms of the first lines}}
\label{57934}
\end{equation}%
\begin{eqnarray}
&&-\frac{1}{2}\phi _{1}^{\ast }\frac{\partial ^{2}\phi _{1}}{\partial x^{2}}+%
\frac{1}{2}\phi _{1}\frac{\partial ^{2}\phi _{1}^{\ast }}{\partial x^{2}}+%
\frac{1}{2}\phi _{2}^{\ast }\frac{\partial ^{2}\phi _{2}}{\partial x^{2}}-%
\frac{1}{2}\phi _{2}\frac{\partial ^{2}\phi _{2}^{\ast }}{\partial x^{2}}
\label{57935} \\
&&-\frac{1}{2}\phi _{1}^{\ast }\frac{\partial ^{2}\phi _{1}}{\partial y^{2}}+%
\frac{1}{2}\phi _{1}\frac{\partial ^{2}\phi _{1}^{\ast }}{\partial y^{2}}+%
\frac{1}{2}\phi _{2}^{\ast }\frac{\partial ^{2}\phi _{2}}{\partial y^{2}}-%
\frac{1}{2}\phi _{2}\frac{\partial ^{2}\phi _{2}^{\ast }}{\partial y^{2}}\ 
\text{{\small terms with second order derivations}}  \notag
\end{eqnarray}%
\begin{eqnarray}
&&-\frac{\partial \left( a-a^{\ast }\right) }{\partial x}\left\vert \phi
_{1}\right\vert ^{2}-\frac{1}{2}\left( a-a^{\ast }\right) \frac{\partial }{%
\partial x}\left( \left\vert \phi _{1}\right\vert ^{2}\right) -\frac{%
\partial \left( a-a^{\ast }\right) }{\partial x}\left\vert \phi
_{2}\right\vert ^{2}-\frac{1}{2}\left( a-a^{\ast }\right) \frac{\partial }{%
\partial x}\left( \left\vert \phi _{2}\right\vert ^{2}\right)  \notag \\
&=&-\frac{\partial \left( a-a^{\ast }\right) }{\partial x}\Xi -\frac{1}{2}%
\left( a-a^{\ast }\right) \frac{\partial }{\partial x}\Xi \ \ \text{{\small %
the second terms of the second lines}}  \label{57936}
\end{eqnarray}%
\begin{eqnarray}
&&-\frac{1}{2}\left( a-a^{\ast }\right) \frac{\partial }{\partial x}\left(
\left\vert \phi _{1}\right\vert ^{2}\right) -\frac{1}{2}\left( a-a^{\ast
}\right) \frac{\partial }{\partial x}\left( \left\vert \phi _{2}\right\vert
^{2}\right)  \label{57937} \\
&=&-\frac{1}{2}\left( a-a^{\ast }\right) \frac{\partial }{\partial x}\Xi \ \ 
\text{{\small the first terms of the third lines}}  \notag
\end{eqnarray}%
\begin{eqnarray}
&&-i\frac{\partial \left( a+a^{\ast }\right) }{\partial y}\left\vert \phi
_{1}\right\vert ^{2}-\frac{i}{2}\left( a+a^{\ast }\right) \frac{\partial }{%
\partial y}\left( \left\vert \phi _{1}\right\vert ^{2}\right) -i\frac{%
\partial \left( a+a^{\ast }\right) }{\partial y}\left\vert \phi
_{2}\right\vert ^{2}-\frac{i}{2}\left( a+a^{\ast }\right) \frac{\partial }{%
\partial y}\left( \left\vert \phi _{2}\right\vert ^{2}\right)  \notag \\
&=&-i\frac{\partial \left( a+a^{\ast }\right) }{\partial y}\Xi -\frac{i}{2}%
\left( a+a^{\ast }\right) \frac{\partial }{\partial y}\Xi \ \ \text{{\small %
the second terms of the fourth lines}}  \label{57938}
\end{eqnarray}%
\begin{eqnarray}
&&-\frac{i}{2}\left( a+a^{\ast }\right) \frac{\partial }{\partial y}%
\left\vert \phi _{1}\right\vert ^{2}-\frac{i}{2}\left( a+a^{\ast }\right) 
\frac{\partial }{\partial y}\left( \left\vert \phi _{2}\right\vert
^{2}\right)  \label{57939} \\
&=&-\frac{i}{2}\left( a+a^{\ast }\right) \frac{\partial }{\partial y}\Xi \ 
\text{{\small the first term of the fifth line}}  \notag
\end{eqnarray}

\bigskip

We now collect the results%
\begin{eqnarray}
&&i\frac{\partial }{\partial t}\Omega -2i\left( b+b^{\ast }\right) \Xi
\label{57940} \\
&=&-\frac{1}{2}\phi _{1}^{\ast }\frac{\partial ^{2}\phi _{1}}{\partial x^{2}}%
+\frac{1}{2}\phi _{1}\frac{\partial ^{2}\phi _{1}^{\ast }}{\partial x^{2}}+%
\frac{1}{2}\phi _{2}^{\ast }\frac{\partial ^{2}\phi _{2}}{\partial x^{2}}-%
\frac{1}{2}\phi _{2}\frac{\partial ^{2}\phi _{2}^{\ast }}{\partial x^{2}} 
\notag \\
&&-\frac{1}{2}\phi _{1}^{\ast }\frac{\partial ^{2}\phi _{1}}{\partial y^{2}}+%
\frac{1}{2}\phi _{1}\frac{\partial ^{2}\phi _{1}^{\ast }}{\partial y^{2}}+%
\frac{1}{2}\phi _{2}^{\ast }\frac{\partial ^{2}\phi _{2}}{\partial y^{2}}-%
\frac{1}{2}\phi _{2}\frac{\partial ^{2}\phi _{2}^{\ast }}{\partial y^{2}} 
\notag \\
&&-\frac{\partial \left( a-a^{\ast }\right) }{\partial x}\Xi -\frac{1}{2}%
\left( a-a^{\ast }\right) \frac{\partial }{\partial x}\Xi  \notag \\
&&\ \ \ \ \ \ \ \ \ \ \ \ \ \ \ \ \ \ \ \ \ \ \ \ \ -\frac{1}{2}\left(
a-a^{\ast }\right) \frac{\partial }{\partial x}\Xi  \notag \\
&&-i\frac{\partial \left( a+a^{\ast }\right) }{\partial y}\Xi -\frac{i}{2}%
\left( a+a^{\ast }\right) \frac{\partial }{\partial y}\Xi  \notag \\
&&\ \ \ \ \ \ \ \ \ \ \ \ \ \ \ \ \ \ \ \ \ \ \ \ \ \ \ -\frac{i}{2}\left(
a+a^{\ast }\right) \frac{\partial }{\partial y}\Xi  \notag
\end{eqnarray}

\bigskip

The result can still be transformed%
\begin{eqnarray}
&&i\frac{\partial }{\partial t}\Omega  \label{57941} \\
&=&2i\left( b+b^{\ast }\right) \Xi  \notag \\
&&-\frac{1}{2}\phi _{1}^{\ast }\frac{\partial ^{2}\phi _{1}}{\partial x^{2}}+%
\frac{1}{2}\phi _{1}\frac{\partial ^{2}\phi _{1}^{\ast }}{\partial x^{2}}+%
\frac{1}{2}\phi _{2}^{\ast }\frac{\partial ^{2}\phi _{2}}{\partial x^{2}}-%
\frac{1}{2}\phi _{2}\frac{\partial ^{2}\phi _{2}^{\ast }}{\partial x^{2}} 
\notag \\
&&-\frac{1}{2}\phi _{1}^{\ast }\frac{\partial ^{2}\phi _{1}}{\partial y^{2}}+%
\frac{1}{2}\phi _{1}\frac{\partial ^{2}\phi _{1}^{\ast }}{\partial y^{2}}+%
\frac{1}{2}\phi _{2}^{\ast }\frac{\partial ^{2}\phi _{2}}{\partial y^{2}}-%
\frac{1}{2}\phi _{2}\frac{\partial ^{2}\phi _{2}^{\ast }}{\partial y^{2}} 
\notag \\
&&-\frac{\partial }{\partial x}\left[ \left( a-a^{\ast }\right) \Xi \right] 
\notag \\
&&-i\frac{\partial }{\partial y}\left[ \left( a+a^{\ast }\right) \Xi \right]
\notag
\end{eqnarray}

\bigskip

\bigskip

Now, if we re-insert the components of the potential%
\begin{eqnarray}
a-a^{\ast } &=&2A_{x}/H\equiv 2\overline{A}_{x}  \label{57942} \\
i\left( a+a^{\ast }\right) &=&2A_{y}/H\equiv 2\overline{A}_{y}  \notag
\end{eqnarray}%
and keep the complex coefficients $b$ of the zero-component potential $A_{0}$%
\begin{eqnarray}
&&i\frac{\partial }{\partial t}\Omega -2i\left( b+b^{\ast }\right) \Xi
\label{57944} \\
&=&F\left( \Delta ;\phi _{1},\phi _{2}\right)  \notag \\
&&-\frac{\partial }{\partial x}\left( 2\overline{A}_{x}\Xi \right) -\frac{%
\partial }{\partial y}\left( 2\overline{A}_{y}\Xi \right)  \notag
\end{eqnarray}%
where we have introduced the notation $F\left( \Delta ;\phi _{1},\phi
_{2}\right) $ for the terms that contain second order derivatives.%
\begin{eqnarray}
&&i\frac{\partial }{\partial t}\left( \rho _{1}-\rho _{2}\right) -2i\left(
b+b^{\ast }\right) \left( \rho _{1}+\rho _{2}\right) +\frac{\partial }{%
\partial x}\left[ 2\overline{A}_{x}\left( \rho _{1}+\rho _{2}\right) \right]
+\frac{\partial }{\partial y}\left[ 2\overline{A}_{y}\left( \rho _{1}+\rho
_{2}\right) \right]  \notag \\
&=&F\left( \Delta ;\phi _{1},\phi _{2}\right)  \label{57945}
\end{eqnarray}

\bigskip

We transform the first two terms of the second-order differential terms $%
F\left( \Delta ;\phi _{1},\phi _{2}\right) $%
\begin{eqnarray}
&&-\frac{1}{2}\phi _{1}^{\ast }\frac{\partial ^{2}\phi _{1}}{\partial x^{2}}+%
\frac{1}{2}\phi _{1}\frac{\partial ^{2}\phi _{1}^{\ast }}{\partial x^{2}}
\label{57946} \\
&=&-\frac{1}{2}\frac{\partial }{\partial x}\left[ \phi _{1}^{\ast }\frac{%
\partial \phi _{1}}{\partial x}\right] +\frac{1}{2}\left[ \left( \frac{%
\partial \phi _{1}^{\ast }}{\partial x}\right) \left( \frac{\partial \phi
_{1}}{\partial x}\right) \right]  \notag \\
&&+\frac{1}{2}\frac{\partial }{\partial x}\left[ \phi _{1}\frac{\partial
\phi _{1}^{\ast }}{\partial x}\right] -\frac{1}{2}\left[ \left( \frac{%
\partial \phi _{1}}{\partial x}\right) \left( \frac{\partial \phi _{1}^{\ast
}}{\partial x}\right) \right]  \notag \\
&=&-\frac{1}{2}\frac{\partial }{\partial x}\left[ \phi _{1}^{\ast }\frac{%
\partial \phi _{1}}{\partial x}-\phi _{1}\frac{\partial \phi _{1}^{\ast }}{%
\partial x}\right]  \notag \\
&=&-\frac{1}{2}\frac{\partial }{\partial x}\left[ \left( \phi _{1}^{\ast
}\right) ^{2}\frac{\partial }{\partial x}\left( \frac{\phi _{1}}{\phi
_{1}^{\ast }}\right) \right]  \notag
\end{eqnarray}%
and take also the other pairs%
\begin{eqnarray}
&&-\frac{1}{2}\phi _{1}^{\ast }\frac{\partial ^{2}\phi _{1}}{\partial y^{2}}+%
\frac{1}{2}\phi _{1}\frac{\partial ^{2}\phi _{1}^{\ast }}{\partial y^{2}}
\label{57947} \\
&=&-\frac{1}{2}\frac{\partial }{\partial y}\left[ \left( \phi _{1}^{\ast
}\right) ^{2}\frac{\partial }{\partial y}\left( \frac{\phi _{1}}{\phi
_{1}^{\ast }}\right) \right]  \notag
\end{eqnarray}%
\begin{eqnarray}
&&\frac{1}{2}\phi _{2}^{\ast }\frac{\partial ^{2}\phi _{2}}{\partial x^{2}}-%
\frac{1}{2}\phi _{2}\frac{\partial ^{2}\phi _{2}^{\ast }}{\partial x^{2}}
\label{57948} \\
&=&\frac{1}{2}\frac{\partial }{\partial x}\left[ \left( \phi _{2}^{\ast
}\right) ^{2}\frac{\partial }{\partial x}\left( \frac{\phi _{2}}{\phi
_{2}^{\ast }}\right) \right]  \notag
\end{eqnarray}%
\begin{eqnarray}
&&\frac{1}{2}\phi _{2}^{\ast }\frac{\partial ^{2}\phi _{2}}{\partial y^{2}}-%
\frac{1}{2}\phi _{2}\frac{\partial ^{2}\phi _{2}^{\ast }}{\partial y^{2}}
\label{57949} \\
&=&\frac{1}{2}\frac{\partial }{\partial y}\left[ \left( \phi _{2}^{\ast
}\right) ^{2}\frac{\partial }{\partial y}\left( \frac{\phi _{2}}{\phi
_{2}^{\ast }}\right) \right]  \notag
\end{eqnarray}%
Then%
\begin{eqnarray}
&&F\left( \Delta ;\phi _{1},\phi _{2}\right)  \label{57950} \\
&=&-\frac{1}{2}\phi _{1}^{\ast }\frac{\partial ^{2}\phi _{1}}{\partial x^{2}}%
+\frac{1}{2}\phi _{1}\frac{\partial ^{2}\phi _{1}^{\ast }}{\partial x^{2}}+%
\frac{1}{2}\phi _{2}^{\ast }\frac{\partial ^{2}\phi _{2}}{\partial x^{2}}-%
\frac{1}{2}\phi _{2}\frac{\partial ^{2}\phi _{2}^{\ast }}{\partial x^{2}} 
\notag \\
&&-\frac{1}{2}\phi _{1}^{\ast }\frac{\partial ^{2}\phi _{1}}{\partial y^{2}}+%
\frac{1}{2}\phi _{1}\frac{\partial ^{2}\phi _{1}^{\ast }}{\partial y^{2}}+%
\frac{1}{2}\phi _{2}^{\ast }\frac{\partial ^{2}\phi _{2}}{\partial y^{2}}-%
\frac{1}{2}\phi _{2}\frac{\partial ^{2}\phi _{2}^{\ast }}{\partial y^{2}} 
\notag \\
&=&-\frac{1}{2}\frac{\partial }{\partial x}\left[ \left( \phi _{1}^{\ast
}\right) ^{2}\frac{\partial }{\partial x}\left( \frac{\phi _{1}}{\phi
_{1}^{\ast }}\right) \right] -\frac{1}{2}\frac{\partial }{\partial y}\left[
\left( \phi _{1}^{\ast }\right) ^{2}\frac{\partial }{\partial y}\left( \frac{%
\phi _{1}}{\phi _{1}^{\ast }}\right) \right]  \notag \\
&&+\frac{1}{2}\frac{\partial }{\partial x}\left[ \left( \phi _{2}^{\ast
}\right) ^{2}\frac{\partial }{\partial x}\left( \frac{\phi _{2}}{\phi
_{2}^{\ast }}\right) \right] +\frac{1}{2}\frac{\partial }{\partial y}\left[
\left( \phi _{2}^{\ast }\right) ^{2}\frac{\partial }{\partial y}\left( \frac{%
\phi _{2}}{\phi _{2}^{\ast }}\right) \right]  \notag
\end{eqnarray}%
We replace the functions $\phi _{1}$ , $\phi _{2}$ and their conjugates with 
\begin{eqnarray}
\phi _{1} &=&\rho _{1}^{1/2}\exp \left( i\chi \right)  \label{57951} \\
\phi _{2} &=&\rho _{2}^{1/2}\exp \left( i\eta \right)  \notag
\end{eqnarray}

Then we obtain%
\begin{eqnarray}
&&-\frac{1}{2}\frac{\partial }{\partial x}\left[ \left( \phi _{1}^{\ast
}\right) ^{2}\frac{\partial }{\partial x}\left( \frac{\phi _{1}}{\phi
_{1}^{\ast }}\right) \right]  \label{57953} \\
&=&-\frac{1}{2}\frac{\partial }{\partial x}\left[ \rho _{1}\exp \left(
-2i\chi \right) \frac{\partial }{\partial x}\exp \left( 2i\chi \right) %
\right]  \notag \\
&=&-\frac{1}{2}2i\frac{\partial }{\partial x}\left[ \rho _{1}\frac{\partial
\chi }{\partial x}\right] =-i\frac{\partial }{\partial x}\left[ \rho _{1}%
\frac{\partial \chi }{\partial x}\right]  \notag
\end{eqnarray}%
\begin{eqnarray}
&&-\frac{1}{2}\frac{\partial }{\partial y}\left[ \left( \phi _{1}^{\ast
}\right) ^{2}\frac{\partial }{\partial y}\left( \frac{\phi _{1}}{\phi
_{1}^{\ast }}\right) \right]  \label{57954} \\
&=&-i\frac{\partial }{\partial y}\left[ \rho _{1}\frac{\partial \chi }{%
\partial y}\right]  \notag
\end{eqnarray}%
\begin{eqnarray}
&&\frac{1}{2}\frac{\partial }{\partial x}\left[ \left( \phi _{2}^{\ast
}\right) ^{2}\frac{\partial }{\partial x}\left( \frac{\phi _{2}}{\phi
_{2}^{\ast }}\right) \right]  \label{57955} \\
&=&i\frac{\partial }{\partial x}\left[ \rho _{2}\frac{\partial \eta }{%
\partial x}\right]  \notag
\end{eqnarray}%
\begin{eqnarray}
&&\frac{1}{2}\frac{\partial }{\partial y}\left[ \left( \phi _{2}^{\ast
}\right) ^{2}\frac{\partial }{\partial y}\left( \frac{\phi _{2}}{\phi
_{2}^{\ast }}\right) \right]  \label{57956} \\
&=&i\frac{\partial }{\partial y}\left[ \rho _{2}\frac{\partial \eta }{%
\partial y}\right]  \notag
\end{eqnarray}%
Then%
\begin{eqnarray}
&&F\left( \Delta ;\phi _{1},\phi _{2}\right)  \label{57957} \\
&=&-i\frac{\partial }{\partial x}\left[ \rho _{1}\frac{\partial \chi }{%
\partial x}\right] -i\frac{\partial }{\partial y}\left[ \rho _{1}\frac{%
\partial \chi }{\partial y}\right]  \notag \\
&&+i\frac{\partial }{\partial x}\left[ \rho _{2}\frac{\partial \eta }{%
\partial x}\right] +i\frac{\partial }{\partial y}\left[ \rho _{2}\frac{%
\partial \eta }{\partial y}\right]  \notag
\end{eqnarray}

We simply introduce this expression for $F$ in the equation derived before
for the difference $\rho _{1}-\rho _{2}$ and write 
\begin{eqnarray}
&&i\frac{\partial }{\partial t}\left( \rho _{1}-\rho _{2}\right) -2i\left(
b+b^{\ast }\right) \left( \rho _{1}+\rho _{2}\right) +\frac{\partial }{%
\partial x}\left[ 2\overline{A}_{x}\left( \rho _{1}+\rho _{2}\right) \right]
+\frac{\partial }{\partial y}\left[ 2\overline{A}_{y}\left( \rho _{1}+\rho
_{2}\right) \right]  \notag \\
&=&-i\frac{\partial }{\partial x}\left[ \rho _{1}\frac{\partial \chi }{%
\partial x}\right] -i\frac{\partial }{\partial y}\left[ \rho _{1}\frac{%
\partial \chi }{\partial y}\right] +i\frac{\partial }{\partial x}\left[ \rho
_{2}\frac{\partial \eta }{\partial x}\right] +i\frac{\partial }{\partial y}%
\left[ \rho _{2}\frac{\partial \eta }{\partial y}\right]  \label{e110}
\end{eqnarray}%
or%
\begin{eqnarray}
&&\frac{\partial }{\partial t}\left( \rho _{1}-\rho _{2}\right) -2\left(
b+b^{\ast }\right) \left( \rho _{1}+\rho _{2}\right)  \label{e120} \\
&&+\frac{\partial }{\partial x}\left[ \left( \frac{2\overline{A}_{x}}{i}+%
\frac{\partial \chi }{\partial x}\right) \rho _{1}+\left( \frac{2\overline{A}%
_{x}}{i}-\frac{\partial \eta }{\partial x}\right) \rho _{2}\right]  \notag \\
&&+\frac{\partial }{\partial y}\left[ \left( \frac{2\overline{A}_{y}}{i}+%
\frac{\partial \chi }{\partial y}\right) \rho _{1}+\left( \frac{2\overline{A}%
_{y}}{i}-\frac{\partial \eta }{\partial y}\right) \rho _{2}\right]  \notag \\
&=&0  \notag
\end{eqnarray}

This equation is derived from the equations of motion under the \emph{%
algebraic ansatz}.

There is no other approximation.

\bigskip

Here we can introduce definitions 
\begin{equation}
v_{x}^{\left( 1\right) }\equiv \frac{2\overline{A}_{x}}{i}+\frac{\partial
\chi }{\partial x}\ ,\ v_{y}^{\left( 1\right) }=\frac{2\overline{A}_{y}}{i}+%
\frac{\partial \chi }{\partial y}  \label{e121}
\end{equation}%
\begin{equation}
v_{x}^{\left( 2\right) }\equiv -\frac{2\overline{A}_{x}}{i}+\frac{\partial
\eta }{\partial x}\ ,\ v_{y}^{\left( 2\right) }=-\frac{2\overline{A}_{y}}{i}+%
\frac{\partial \eta }{\partial y}  \label{e122}
\end{equation}%
and we can write%
\begin{eqnarray}
&&\frac{\partial }{\partial t}\left( \rho _{1}-\rho _{2}\right) -2\left(
b+b^{\ast }\right) \left( \rho _{1}+\rho _{2}\right)  \label{e131} \\
&&+\frac{\partial }{\partial x}\left[ v_{x}^{\left( 1\right) }\rho
_{1}-v_{x}^{\left( 2\right) }\rho _{2}\right] +\frac{\partial }{\partial y}%
\left[ v_{y}^{\left( 1\right) }\rho _{1}-v_{y}^{\left( 2\right) }\rho _{2}%
\right]  \notag \\
&=&0  \notag
\end{eqnarray}

\bigskip

The equations derived until now, for $\rho _{1}$, $\rho _{2}$ and $\left(
\rho _{1}-\rho _{2}\right) $ have involved ONLY the second equation of motion%
\begin{equation}
iD_{0}\phi =-\frac{1}{2m}D_{k}D^{k}\phi -\frac{1}{2m\kappa }\left[ \left[
\phi ,\phi ^{\dagger }\right] ,\phi \right]  \label{e134}
\end{equation}%
and the potentials $A_{x,y}$, which under algebraic ansatz, are given in
terms of $a$ and $a^{\ast }$. In addition we use the expression of $A_{0}$
and its algebraic ansatz, which is imaginary, $b\in \mathit{Im}\mathbf{R}$.

Nothing else, in particular the \emph{second equation of motion}, or the
Gauss constraint. This has not been yet invoked.

\subsubsection{Approximate form of the equation for $\Omega =\protect\rho %
_{1}-\protect\rho _{2}$ close to self-duality}

When we are close to the SD state, we can approximate: $A_{0}$ is purely
imaginar close to SD, and%
\begin{equation}
b+b^{\ast }\approx 0  \label{57943}
\end{equation}%
\begin{equation}
\rho _{1}=\rho _{2}^{-1}=\rho =\exp \left( \psi \right)  \label{57952}
\end{equation}%
\textbf{\ }\ 
\begin{equation}
\chi \approx -\eta  \label{57958}
\end{equation}%
and we will keep however the two functions $\rho _{1}$ and $\rho _{2}$. The
approximation will only consists of taking the two phases as almost equal
and opposed.

The terms in the expression of $F\left( \Delta ;\phi _{1},\phi _{2}\right) $
become%
\begin{equation}
-i\frac{\partial }{\partial x}\left[ \rho _{1}\frac{\partial \chi }{\partial
x}\right] -i\frac{\partial }{\partial x}\left[ \rho _{2}\frac{\partial \chi 
}{\partial x}\right] =-i\frac{\partial }{\partial x}\left[ \left( \rho
_{1}+\rho _{2}\right) \frac{\partial \chi }{\partial x}\right]  \label{57959}
\end{equation}%
and%
\begin{equation}
-i\frac{\partial }{\partial y}\left[ \rho _{1}\frac{\partial \chi }{\partial
y}\right] -i\frac{\partial }{\partial y}\left[ \rho _{2}\frac{\partial \chi 
}{\partial y}\right] =-i\frac{\partial }{\partial y}\left[ \left( \rho
_{1}+\rho _{2}\right) \frac{\partial \chi }{\partial y}\right]  \label{57960}
\end{equation}%
which gives%
\begin{eqnarray}
&&F\left( \Delta ;\phi _{1},\phi _{2}\right)  \label{57961} \\
&=&-i\frac{\partial }{\partial x}\left[ \left( \rho _{1}+\rho _{2}\right) 
\frac{\partial \chi }{\partial x}\right] -i\frac{\partial }{\partial y}\left[
\left( \rho _{1}+\rho _{2}\right) \frac{\partial \chi }{\partial y}\right] 
\notag
\end{eqnarray}

At this point, the \emph{approximative} (due to the assumption $\chi \approx
-\eta $) form of the equation for the time-variation of 
\begin{equation}
\Omega \equiv \rho _{1}-\rho _{2}  \label{57962}
\end{equation}%
is%
\begin{eqnarray}
&&i\frac{\partial }{\partial t}\left( \rho _{1}-\rho _{2}\right) +\frac{%
\partial }{\partial x}\left[ 2\overline{A}_{x}\left( \rho _{1}+\rho
_{2}\right) \right] +\frac{\partial }{\partial y}\left[ 2\overline{A}%
_{y}\left( \rho _{1}+\rho _{2}\right) \right]  \label{57963} \\
&&\ \ \ \ \ \ \ \ \ \ \ \ \ \ \ \approx -i\frac{\partial }{\partial x}\left[
\left( \rho _{1}+\rho _{2}\right) \frac{\partial \chi }{\partial x}\right] -i%
\frac{\partial }{\partial y}\left[ \left( \rho _{1}+\rho _{2}\right) \frac{%
\partial \chi }{\partial y}\right]  \notag
\end{eqnarray}%
or%
\begin{eqnarray}
&&\frac{\partial }{\partial t}\left( \rho _{1}-\rho _{2}\right)
\label{57964} \\
&&+\frac{\partial }{\partial x}\left[ \left( \rho _{1}+\rho _{2}\right)
\left( \frac{2\overline{A}_{x}}{i}+\frac{\partial \chi }{\partial x}\right) %
\right]  \notag \\
&&+\frac{\partial }{\partial y}\left[ \left( \rho _{1}+\rho _{2}\right)
\left( \frac{2\overline{A}_{y}}{i}+\frac{\partial \chi }{\partial y}\right) %
\right]  \notag \\
&\approx &0\ \ \ \ \text{close to SD}  \notag
\end{eqnarray}

\textbf{NOTE. }The expression for the potential in the simpler problem of
the Liouville equation is%
\begin{equation}
A_{\mu }=\partial _{\mu }\chi +\widehat{\mathbf{e}}_{z}\times \mathbf{\nabla 
}\ln \rho  \label{57965}
\end{equation}%
where we note that in our case the components of the potential are \emph{%
imaginary}. Then $2\overline{A}_{x}$ and the term $i\partial _{x}\chi $ may
lead to the \emph{physical }part of the velocity%
\begin{equation}
v^{phys}\equiv \widehat{\mathbf{e}}_{z}\times \mathbf{\nabla }\ln \rho \ \ 
\text{at SD}  \label{57966}
\end{equation}%
And (still a problem with the factors $2$) we have%
\begin{equation}
\frac{\partial }{\partial t}\left( \rho _{1}-\rho _{2}\right) +\frac{%
\partial }{\partial x}\left[ v_{x}^{phys}\left( \rho _{1}+\rho _{2}\right) %
\right] +\frac{\partial }{\partial y}\left[ v_{y}^{phys}\left( \rho
_{1}+\rho _{2}\right) \right] =0  \label{57967}
\end{equation}%
This is \emph{NOT} the equation of continuity. \textbf{END.}

\bigskip

\subsection{Derivation of the equation for the sum $\Xi =\protect\rho _{1}+%
\protect\rho _{2}$}

Another operation that we can make with the two equations (for $\left\vert
\phi _{1}\right\vert ^{2}$ and respectively $\left\vert \phi _{2}\right\vert
^{2}$) consists of adding them. This will obtain in the left hand side the
time derivative of the \emph{sum} of the two functions, \emph{i.e.} $\Xi $.

The sum of the Eqs.(\ref{57914}) and (\ref{57930}) is made term by term%
\begin{eqnarray}
&&i\frac{\partial }{\partial t}\left\vert \phi _{1}\right\vert ^{2}-2i\left(
b+b^{\ast }\right) \left\vert \phi _{1}\right\vert ^{2}  \label{57968} \\
&&+i\frac{\partial }{\partial t}\left\vert \phi _{2}\right\vert
^{2}+2i\left( b+b^{\ast }\right) \left\vert \phi _{2}\right\vert ^{2}  \notag
\\
&=&i\frac{\partial }{\partial t}\Xi -2i\left( b+b^{\ast }\right) \Omega \ \
\ \text{{\small first line}}  \notag
\end{eqnarray}%
The terms with second order derivatives%
\begin{eqnarray}
&&-\frac{1}{2}\phi _{1}^{\ast }\frac{\partial ^{2}\phi _{1}}{\partial x^{2}}+%
\frac{1}{2}\phi _{1}\frac{\partial ^{2}\phi _{1}^{\ast }}{\partial x^{2}}-%
\frac{1}{2}\phi _{1}^{\ast }\frac{\partial ^{2}\phi _{1}}{\partial y^{2}}+%
\frac{1}{2}\phi _{1}\frac{\partial ^{2}\phi _{1}^{\ast }}{\partial y^{2}}
\label{57969} \\
&&-\frac{1}{2}\phi _{2}^{\ast }\frac{\partial ^{2}\phi _{2}}{\partial x^{2}}+%
\frac{1}{2}\phi _{2}\frac{\partial ^{2}\phi _{2}^{\ast }}{\partial x^{2}}-%
\frac{1}{2}\phi _{2}^{\ast }\frac{\partial ^{2}\phi _{2}}{\partial y^{2}}+%
\frac{1}{2}\phi _{2}\frac{\partial ^{2}\phi _{2}^{\ast }}{\partial y^{2}}%
\text{{\small terms with second order derivations}}  \notag
\end{eqnarray}%
\begin{eqnarray}
&&-\frac{\partial \left( a-a^{\ast }\right) }{\partial x}\left\vert \phi
_{1}\right\vert ^{2}-\frac{1}{2}\left( a-a^{\ast }\right) \frac{\partial }{%
\partial x}\left( \left\vert \phi _{1}\right\vert ^{2}\right) +\frac{%
\partial \left( a-a^{\ast }\right) }{\partial x}\left\vert \phi
_{2}\right\vert ^{2}+\frac{1}{2}\left( a-a^{\ast }\right) \frac{\partial }{%
\partial x}\left( \left\vert \phi _{2}\right\vert ^{2}\right)  \notag \\
&=&-\frac{\partial \left( a-a^{\ast }\right) }{\partial x}\Omega -\frac{1}{2}%
\left( a-a^{\ast }\right) \frac{\partial }{\partial x}\Omega \ \text{{\small %
terms of the second lines}}  \label{57970}
\end{eqnarray}%
\begin{eqnarray}
&&-\frac{1}{2}\left( a-a^{\ast }\right) \frac{\partial }{\partial x}%
\left\vert \phi _{1}\right\vert ^{2}+\frac{1}{2}\left( a-a^{\ast }\right) 
\frac{\partial }{\partial x}\left( \left\vert \phi _{2}\right\vert
^{2}\right)  \label{57971} \\
&=&-\frac{1}{2}\left( a-a^{\ast }\right) \frac{\partial }{\partial x}\Omega
\ \text{{\small terms of the third lines}}  \notag
\end{eqnarray}%
\begin{eqnarray}
&&-i\frac{\partial \left( a+a^{\ast }\right) }{\partial y}\left\vert \phi
_{1}\right\vert ^{2}-\frac{i}{2}\left( a+a^{\ast }\right) \frac{\partial }{%
\partial y}\left\vert \phi _{1}\right\vert ^{2}+i\frac{\partial \left(
a+a^{\ast }\right) }{\partial y}\left\vert \phi _{2}\right\vert ^{2}+\frac{i%
}{2}\left( a+a^{\ast }\right) \frac{\partial }{\partial y}\left( \left\vert
\phi _{2}\right\vert ^{2}\right)  \notag \\
&=&-i\frac{\partial \left( a+a^{\ast }\right) }{\partial y}\Omega -\frac{i}{2%
}\left( a+a^{\ast }\right) \frac{\partial }{\partial y}\Omega \ \text{%
{\small terms of the fourth lines}}  \label{57972}
\end{eqnarray}%
\begin{eqnarray}
&&-\frac{i}{2}\left( a+a^{\ast }\right) \frac{\partial }{\partial y}%
\left\vert \phi _{1}\right\vert ^{2}+\frac{i}{2}\left( a+a^{\ast }\right) 
\frac{\partial }{\partial y}\left( \left\vert \phi _{2}\right\vert
^{2}\right)  \label{57973} \\
&=&-\frac{i}{2}\left( a+a^{\ast }\right) \frac{\partial }{\partial y}\Omega
\ \text{{\small terms of the fifth lines}}  \notag
\end{eqnarray}

Let consider what results%
\begin{eqnarray}
&&i\frac{\partial }{\partial t}\Xi -2i\left( b+b^{\ast }\right) \Omega
\label{57974} \\
&=&-\frac{1}{2}\phi _{1}^{\ast }\frac{\partial ^{2}\phi _{1}}{\partial x^{2}}%
+\frac{1}{2}\phi _{1}\frac{\partial ^{2}\phi _{1}^{\ast }}{\partial x^{2}}-%
\frac{1}{2}\phi _{1}^{\ast }\frac{\partial ^{2}\phi _{1}}{\partial y^{2}}+%
\frac{1}{2}\phi _{1}\frac{\partial ^{2}\phi _{1}^{\ast }}{\partial y^{2}} 
\notag \\
&&-\frac{1}{2}\phi _{2}^{\ast }\frac{\partial ^{2}\phi _{2}}{\partial x^{2}}+%
\frac{1}{2}\phi _{2}\frac{\partial ^{2}\phi _{2}^{\ast }}{\partial x^{2}}-%
\frac{1}{2}\phi _{2}^{\ast }\frac{\partial ^{2}\phi _{2}}{\partial y^{2}}+%
\frac{1}{2}\phi _{2}\frac{\partial ^{2}\phi _{2}^{\ast }}{\partial y^{2}} 
\notag \\
&&-\frac{\partial \left( a-a^{\ast }\right) }{\partial x}\Omega -\frac{1}{2}%
\left( a-a^{\ast }\right) \frac{\partial }{\partial x}\Omega -\frac{1}{2}%
\left( a-a^{\ast }\right) \frac{\partial }{\partial x}\Omega  \notag \\
&&-i\frac{\partial \left( a+a^{\ast }\right) }{\partial y}\Omega -\frac{i}{2}%
\left( a+a^{\ast }\right) \frac{\partial }{\partial y}\Omega -\frac{i}{2}%
\left( a+a^{\ast }\right) \frac{\partial }{\partial y}\Omega  \notag
\end{eqnarray}%
\begin{eqnarray}
&&i\frac{\partial }{\partial t}\Xi -2i\left( b+b^{\ast }\right) \Omega
\label{57975} \\
&=&G\left( \Delta ;\phi _{1},\phi _{2}\right)  \notag \\
&&-\frac{\partial }{\partial x}\left[ \left( a-a^{\ast }\right) \Omega %
\right] -i\frac{\partial }{\partial y}\left[ \left( a+a^{\ast }\right)
\Omega \right]  \notag
\end{eqnarray}

We will have to work on the function $G$ as for the previous case for $F$.

The treatment of the pairs of terms is identical 
\begin{eqnarray}
&&-\frac{1}{2}\phi _{1}^{\ast }\frac{\partial ^{2}\phi _{1}}{\partial x^{2}}+%
\frac{1}{2}\phi _{1}\frac{\partial ^{2}\phi _{1}^{\ast }}{\partial x^{2}}
\label{57976} \\
&=&-\frac{1}{2}\frac{\partial }{\partial x}\left[ \phi _{1}^{\ast }\frac{%
\partial \phi _{1}}{\partial x}\right] +\frac{1}{2}\left[ \left( \frac{%
\partial \phi _{1}^{\ast }}{\partial x}\right) \left( \frac{\partial \phi
_{1}}{\partial x}\right) \right]  \notag \\
&&+\frac{1}{2}\frac{\partial }{\partial x}\left[ \phi _{1}\frac{\partial
\phi _{1}^{\ast }}{\partial x}\right] -\frac{1}{2}\left[ \left( \frac{%
\partial \phi _{1}}{\partial x}\right) \left( \frac{\partial \phi _{1}^{\ast
}}{\partial x}\right) \right]  \notag \\
&=&-\frac{1}{2}\frac{\partial }{\partial x}\left[ \phi _{1}^{\ast }\frac{%
\partial \phi _{1}}{\partial x}-\phi _{1}\frac{\partial \phi _{1}^{\ast }}{%
\partial x}\right]  \notag \\
&=&-\frac{1}{2}\frac{\partial }{\partial x}\left[ \left( \phi _{1}^{\ast
}\right) ^{2}\frac{\partial }{\partial x}\left( \frac{\phi _{1}}{\phi
_{1}^{\ast }}\right) \right]  \notag
\end{eqnarray}%
\begin{eqnarray}
&&-\frac{1}{2}\phi _{1}^{\ast }\frac{\partial ^{2}\phi _{1}}{\partial y^{2}}+%
\frac{1}{2}\phi _{1}\frac{\partial ^{2}\phi _{1}^{\ast }}{\partial y^{2}}
\label{57977} \\
&=&-\frac{1}{2}\frac{\partial }{\partial y}\left[ \left( \phi _{1}^{\ast
}\right) ^{2}\frac{\partial }{\partial y}\left( \frac{\phi _{1}}{\phi
_{1}^{\ast }}\right) \right]  \notag
\end{eqnarray}%
\begin{eqnarray}
&&-\frac{1}{2}\phi _{2}^{\ast }\frac{\partial ^{2}\phi _{2}}{\partial x^{2}}+%
\frac{1}{2}\phi _{2}\frac{\partial ^{2}\phi _{2}^{\ast }}{\partial x^{2}}
\label{57978} \\
&=&-\frac{1}{2}\frac{\partial }{\partial x}\left[ \left( \phi _{2}^{\ast
}\right) ^{2}\frac{\partial }{\partial x}\left( \frac{\phi _{2}}{\phi
_{2}^{\ast }}\right) \right]  \notag
\end{eqnarray}%
\begin{eqnarray}
&&-\frac{1}{2}\phi _{2}^{\ast }\frac{\partial ^{2}\phi _{2}}{\partial y^{2}}+%
\frac{1}{2}\phi _{2}\frac{\partial ^{2}\phi _{2}^{\ast }}{\partial y^{2}}
\label{57979} \\
&=&-\frac{1}{2}\frac{\partial }{\partial y}\left[ \left( \phi _{2}^{\ast
}\right) ^{2}\frac{\partial }{\partial y}\left( \frac{\phi _{2}}{\phi
_{2}^{\ast }}\right) \right]  \notag
\end{eqnarray}%
The function $G$ becomes%
\begin{eqnarray}
&&G\left( \Delta ;\phi _{1},\phi _{2}\right)  \label{57980} \\
&=&-\frac{1}{2}\frac{\partial }{\partial x}\left[ \left( \phi _{1}^{\ast
}\right) ^{2}\frac{\partial }{\partial x}\left( \frac{\phi _{1}}{\phi
_{1}^{\ast }}\right) \right] -\frac{1}{2}\frac{\partial }{\partial y}\left[
\left( \phi _{1}^{\ast }\right) ^{2}\frac{\partial }{\partial y}\left( \frac{%
\phi _{1}}{\phi _{1}^{\ast }}\right) \right]  \notag \\
&&-\frac{1}{2}\frac{\partial }{\partial x}\left[ \left( \phi _{2}^{\ast
}\right) ^{2}\frac{\partial }{\partial x}\left( \frac{\phi _{2}}{\phi
_{2}^{\ast }}\right) \right] -\frac{1}{2}\frac{\partial }{\partial y}\left[
\left( \phi _{2}^{\ast }\right) ^{2}\frac{\partial }{\partial y}\left( \frac{%
\phi _{2}}{\phi _{2}^{\ast }}\right) \right]  \notag
\end{eqnarray}%
We note the difference relative to the expression of $F$, that the two terms
involving $\phi _{2}$ are now with the opposite sign.%
\begin{eqnarray}
&&G\left( \Delta ;\phi _{1},\phi _{2}\right)  \label{57981} \\
&=&-i\frac{\partial }{\partial x}\left[ \rho _{1}\frac{\partial \chi }{%
\partial x}\right] -i\frac{\partial }{\partial y}\left[ \rho _{1}\frac{%
\partial \chi }{\partial y}\right]  \notag \\
&&-i\frac{\partial }{\partial x}\left[ \rho _{2}\frac{\partial \eta }{%
\partial x}\right] -i\frac{\partial }{\partial y}\left[ \rho _{2}\frac{%
\partial \eta }{\partial y}\right]  \notag
\end{eqnarray}%
We insert this in the equation for the sum $\Xi $%
\begin{eqnarray}
&&i\frac{\partial }{\partial t}\Xi -2i\left( b+b^{\ast }\right) \left( \rho
_{1}-\rho _{2}\right) +\frac{\partial }{\partial x}\left[ \left( a-a^{\ast
}\right) \left( \rho _{1}-\rho _{2}\right) \right] +i\frac{\partial }{%
\partial y}\left[ \left( a+a^{\ast }\right) \left( \rho _{1}-\rho
_{2}\right) \right]  \notag \\
&=&-i\frac{\partial }{\partial x}\left[ \rho _{1}\frac{\partial \chi }{%
\partial x}\right] -i\frac{\partial }{\partial y}\left[ \rho _{1}\frac{%
\partial \chi }{\partial y}\right] -i\frac{\partial }{\partial x}\left[ \rho
_{2}\frac{\partial \eta }{\partial x}\right] -i\frac{\partial }{\partial y}%
\left[ \rho _{2}\frac{\partial \eta }{\partial y}\right]  \label{e150}
\end{eqnarray}%
and replace the potentials%
\begin{eqnarray}
&&i\frac{\partial }{\partial t}\left( \rho _{1}+\rho _{2}\right) -2i\left(
b+b^{\ast }\right) \left( \rho _{1}-\rho _{2}\right) +\frac{\partial }{%
\partial x}\left[ 2\overline{A}_{x}\left( \rho _{1}-\rho _{2}\right) \right]
+\frac{\partial }{\partial y}\left[ 2\overline{A}_{y}\left( \rho _{1}-\rho
_{2}\right) \right]  \notag \\
&=&-i\frac{\partial }{\partial x}\left[ \rho _{1}\frac{\partial \chi }{%
\partial x}\right] -i\frac{\partial }{\partial y}\left[ \rho _{1}\frac{%
\partial \chi }{\partial y}\right] -i\frac{\partial }{\partial x}\left[ \rho
_{2}\frac{\partial \eta }{\partial x}\right] -i\frac{\partial }{\partial y}%
\left[ \rho _{2}\frac{\partial \eta }{\partial y}\right]  \label{e151}
\end{eqnarray}%
\begin{eqnarray}
&&\frac{\partial }{\partial t}\left( \rho _{1}+\rho _{2}\right) -2\left(
b+b^{\ast }\right) \left( \rho _{1}-\rho _{2}\right)  \label{e153} \\
&&+\frac{\partial }{\partial x}\left[ \left( \frac{2\overline{A}_{x}}{i}+%
\frac{\partial \chi }{\partial x}\right) \rho _{1}+\left( -\frac{2\overline{A%
}_{x}}{i}+\frac{\partial \eta }{\partial x}\right) \rho _{2}\right]  \notag
\\
&&+\frac{\partial }{\partial y}\left[ \left( \frac{2\overline{A}_{y}}{i}+%
\frac{\partial \chi }{\partial y}\right) \rho _{1}+\left( -\frac{2\overline{A%
}_{y}}{i}+\frac{\partial \eta }{\partial y}\right) \rho _{2}\right]  \notag
\\
&=&0  \notag
\end{eqnarray}

There is \emph{no} approximation of the type \ "close to SD ".

Using the notations introducing so-called velocity fields $\mathbf{v}%
^{\left( 1\right) }$ and $\mathbf{v}^{\left( 2\right) }$ we have%
\begin{eqnarray}
&&\frac{\partial }{\partial t}\left( \rho _{1}+\rho _{2}\right) -2\left(
b+b^{\ast }\right) \left( \rho _{1}-\rho _{2}\right)  \label{e154} \\
&&+\frac{\partial }{\partial x}\left[ v_{x}^{\left( 1\right) }\rho
_{1}+v_{x}^{\left( 2\right) }\rho _{2}\right] +\frac{\partial }{\partial y}%
\left[ v_{y}^{\left( 1\right) }\rho _{1}+v_{y}^{\left( 2\right) }\rho _{2}%
\right]  \notag \\
&=&0  \notag
\end{eqnarray}

Only the \emph{algebraic ansatz} is used.

\bigskip

\subsubsection{Approximative form of the equation for $\Xi =\protect\rho %
_{1}+\protect\rho _{2}$ close to self-duality}

We assume that close to the SD we can approximate%
\begin{equation}
\chi \approx -\eta  \label{57982}
\end{equation}%
Then%
\begin{eqnarray}
&&-i\frac{\partial }{\partial x}\left[ \rho _{1}\frac{\partial \chi }{%
\partial x}\right] -i\frac{\partial }{\partial x}\left[ \rho _{2}\frac{%
\partial \eta }{\partial x}\right]  \label{57983} \\
&\approx &-i\frac{\partial }{\partial x}\left[ \left( \rho _{1}-\rho
_{2}\right) \frac{\partial \chi }{\partial x}\right]  \notag
\end{eqnarray}%
and%
\begin{eqnarray}
&&-i\frac{\partial }{\partial y}\left[ \rho _{1}\frac{\partial \chi }{%
\partial y}\right] -i\frac{\partial }{\partial y}\left[ \rho _{2}\frac{%
\partial \eta }{\partial y}\right]  \label{57984} \\
&\approx &-i\frac{\partial }{\partial y}\left[ \left( \rho _{1}-\rho
_{2}\right) \frac{\partial \chi }{\partial y}\right]  \notag
\end{eqnarray}%
and $G$ becomes%
\begin{eqnarray}
&&G\left( \Delta ;\phi _{1},\phi _{2}\right)  \label{57985} \\
&\approx &-i\frac{\partial }{\partial x}\left[ \left( \rho _{1}-\rho
_{2}\right) \frac{\partial \chi }{\partial x}\right] -i\frac{\partial }{%
\partial y}\left[ \left( \rho _{1}-\rho _{2}\right) \frac{\partial \chi }{%
\partial y}\right]  \notag
\end{eqnarray}%
\begin{eqnarray}
&&i\frac{\partial }{\partial t}\Xi -2i\left( b+b^{\ast }\right) \Omega
\label{57986} \\
&=&-i\frac{\partial }{\partial x}\left[ \left( \rho _{1}-\rho _{2}\right) 
\frac{\partial \chi }{\partial x}\right] -i\frac{\partial }{\partial y}\left[
\left( \rho _{1}-\rho _{2}\right) \frac{\partial \chi }{\partial y}\right] 
\notag \\
&&-\frac{\partial }{\partial x}\left[ \left( a-a^{\ast }\right) \Omega %
\right] -i\frac{\partial }{\partial y}\left[ \left( a+a^{\ast }\right)
\Omega \right]  \notag
\end{eqnarray}%
In addition we consider that close to SD%
\begin{equation}
b+b^{\ast }\approx 0  \label{57987}
\end{equation}%
\begin{eqnarray}
&&i\frac{\partial }{\partial t}\Xi +i\frac{\partial }{\partial x}\left[
\Omega \frac{\partial \chi }{\partial x}\right] +\frac{\partial }{\partial x}%
\left[ \left( a-a^{\ast }\right) \Omega \right]  \label{57988} \\
&&+i\frac{\partial }{\partial y}\left[ \Omega \frac{\partial \chi }{\partial
y}\right] +i\frac{\partial }{\partial y}\left[ \left( a+a^{\ast }\right)
\Omega \right]  \notag \\
&=&0  \notag
\end{eqnarray}%
\begin{equation}
i\frac{\partial }{\partial t}\Xi +\frac{\partial }{\partial x}\left[ \left( 2%
\overline{A}_{x}+i\frac{\partial \chi }{\partial x}\right) \Omega \right] +%
\frac{\partial }{\partial y}\left[ \left( 2\overline{A}_{y}+i\frac{\partial
\chi }{\partial y}\right) \Omega \right] =0  \label{57989}
\end{equation}

For comparison we place together the two equations%
\begin{eqnarray}
&&i\frac{\partial }{\partial t}\left( \rho _{1}-\rho _{2}\right)
\label{57990} \\
&&+\frac{\partial }{\partial x}\left[ \left( \rho _{1}+\rho _{2}\right)
\left( 2\overline{A}_{x}+i\frac{\partial \chi }{\partial x}\right) \right] 
\notag \\
&&+\frac{\partial }{\partial y}\left[ \left( \rho _{1}+\rho _{2}\right)
\left( 2\overline{A}_{y}+i\frac{\partial \chi }{\partial y}\right) \right] 
\notag \\
&\approx &0\ \ \ \text{close to SD}  \notag
\end{eqnarray}%
and%
\begin{eqnarray}
&&i\frac{\partial }{\partial t}\left( \rho _{1}+\rho _{2}\right)
\label{57991} \\
&&+\frac{\partial }{\partial x}\left[ \left( \rho _{1}-\rho _{2}\right)
\left( 2\overline{A}_{x}+i\frac{\partial \chi }{\partial x}\right) \right] 
\notag \\
&&+\frac{\partial }{\partial y}\left[ \left( \rho _{1}-\rho _{2}\right)
\left( 2\overline{A}_{y}+i\frac{\partial \chi }{\partial y}\right) \right] 
\notag \\
&\approx &0\ \ \ \text{close to SD}  \notag
\end{eqnarray}

The potential is actually imaginary. Schematically one can write, 
\begin{eqnarray}
\frac{\partial }{\partial t}\Omega +\mathit{div}\left( \mathbf{v}^{\left(
1\right) }\Xi \right) &\approx &0\ \ \text{close to SD}  \label{700} \\
\frac{\partial }{\partial t}\Xi +\mathit{div}\left( \mathbf{v}^{\left(
1\right) }\Omega \right) &\approx &0\ \ \text{close to SD}  \notag
\end{eqnarray}%
where%
\begin{eqnarray}
v_{x}^{\left( 1\right) } &\equiv &\frac{2\overline{A}_{x}}{i}+\frac{\partial
\chi }{\partial x}  \label{701} \\
v_{y}^{\left( 1\right) } &=&\frac{2\overline{A}_{y}}{i}+\frac{\partial \chi 
}{\partial y}  \notag
\end{eqnarray}

\bigskip

\subsection{Derivation of the equation for $\protect\rho _{1}$}

We have obtained equations for the functions%
\begin{eqnarray}
\Omega &\equiv &\rho _{1}-\rho _{2}  \label{e302} \\
\Xi &\equiv &\rho _{1}+\rho _{2}  \notag
\end{eqnarray}%
These are%
\begin{eqnarray}
&&\frac{\partial }{\partial t}\left( \rho _{1}-\rho _{2}\right) -2\left(
b+b^{\ast }\right) \left( \rho _{1}+\rho _{2}\right)  \label{e303} \\
&&+\frac{\partial }{\partial x}\left[ \left( \frac{2\overline{A}_{x}}{i}+%
\frac{\partial \chi }{\partial x}\right) \rho _{1}+\left( \frac{2\overline{A}%
_{x}}{i}-\frac{\partial \eta }{\partial x}\right) \rho _{2}\right]  \notag \\
&&+\frac{\partial }{\partial y}\left[ \left( \frac{2\overline{A}_{y}}{i}+%
\frac{\partial \chi }{\partial y}\right) \rho _{1}+\left( \frac{2\overline{A}%
_{y}}{i}-\frac{\partial \eta }{\partial y}\right) \rho _{2}\right]  \notag \\
&=&0  \notag
\end{eqnarray}%
and 
\begin{eqnarray}
&&\frac{\partial }{\partial t}\left( \rho _{1}+\rho _{2}\right) -2\left(
b+b^{\ast }\right) \left( \rho _{1}-\rho _{2}\right)  \label{e304} \\
&&+\frac{\partial }{\partial x}\left[ \left( \frac{2\overline{A}_{x}}{i}+%
\frac{\partial \chi }{\partial x}\right) \rho _{1}+\left( -\frac{2\overline{A%
}_{x}}{i}+\frac{\partial \eta }{\partial x}\right) \rho _{2}\right]  \notag
\\
&&+\frac{\partial }{\partial y}\left[ \left( \frac{2\overline{A}_{y}}{i}+%
\frac{\partial \chi }{\partial y}\right) \rho _{1}+\left( -\frac{2\overline{A%
}_{y}}{i}+\frac{\partial \eta }{\partial y}\right) \rho _{2}\right]  \notag
\\
&=&0  \notag
\end{eqnarray}%
These equations are general, do not contain approximation close to SD.

We will combine them to obtain the equation for $\rho _{1}$.

\bigskip

\textbf{NOTE. }If we take as starting point forms of the equations that have
been obtained at previous levels, we will repeat some calculations.

We start from the equations for the difference $\Omega $ and for the sum $%
\Xi $.

For the difference $\rho _{1}-\rho _{2}$:%
\begin{eqnarray}
&&i\frac{\partial }{\partial t}\left( \rho _{1}-\rho _{2}\right) -2i\left(
b+b^{\ast }\right) \left( \rho _{1}+\rho _{2}\right) +\frac{\partial }{%
\partial x}\left[ 2\overline{A}_{x}\left( \rho _{1}+\rho _{2}\right) \right]
+\frac{\partial }{\partial y}\left[ 2\overline{A}_{y}\left( \rho _{1}+\rho
_{2}\right) \right]  \notag \\
&=&F\left( \Delta ;\phi _{1},\phi _{2}\right)  \label{1300}
\end{eqnarray}%
where%
\begin{eqnarray}
&&F\left( \Delta ;\phi _{1},\phi _{2}\right)  \label{1301} \\
&=&-i\frac{\partial }{\partial x}\left[ \rho _{1}\frac{\partial \chi }{%
\partial x}\right] -i\frac{\partial }{\partial y}\left[ \rho _{1}\frac{%
\partial \chi }{\partial y}\right] +i\frac{\partial }{\partial x}\left[ \rho
_{2}\frac{\partial \eta }{\partial x}\right] +i\frac{\partial }{\partial y}%
\left[ \rho _{2}\frac{\partial \eta }{\partial y}\right]  \notag
\end{eqnarray}

\bigskip

For the sum $\rho _{1}+\rho _{2}$:%
\begin{eqnarray}
&&i\frac{\partial }{\partial t}\left( \rho _{1}+\rho _{2}\right) -2i\left(
b+b^{\ast }\right) \left( \rho _{1}-\rho _{2}\right) +\frac{\partial }{%
\partial x}\left[ 2\overline{A}_{x}\left( \rho _{1}-\rho _{2}\right) \right]
+\frac{\partial }{\partial y}\left[ 2\overline{A}_{y}\left( \rho _{1}-\rho
_{2}\right) \right]  \notag \\
&=&G\left( \Delta ;\phi _{1},\phi _{2}\right)  \label{1302}
\end{eqnarray}%
where%
\begin{eqnarray}
&&G\left( \Delta ;\phi _{1},\phi _{2}\right)  \label{1303} \\
&=&-i\frac{\partial }{\partial x}\left[ \rho _{1}\frac{\partial \chi }{%
\partial x}\right] -i\frac{\partial }{\partial y}\left[ \rho _{1}\frac{%
\partial \chi }{\partial y}\right] -i\frac{\partial }{\partial x}\left[ \rho
_{2}\frac{\partial \eta }{\partial x}\right] -i\frac{\partial }{\partial y}%
\left[ \rho _{2}\frac{\partial \eta }{\partial y}\right]  \notag
\end{eqnarray}

\bigskip

These equations can be combined to become equations for only $\rho _{1}$ and
respectively $\rho _{2}$, which is not exact since the velocity field
depends on both variables and the separation is not possible. \textbf{END.}

Adding the two equations we obtain%
\begin{eqnarray}
&&2i\frac{\partial }{\partial t}\rho _{1}-4i\left( b+b^{\ast }\right) \rho
_{1}+\frac{\partial }{\partial x}\left[ 4\overline{A}_{x}\rho _{1}\right] +%
\frac{\partial }{\partial y}\left[ 4\overline{A}_{y}\rho _{1}\right]
\label{1304} \\
&=&-2i\frac{\partial }{\partial x}\left[ \rho _{1}\frac{\partial \chi }{%
\partial x}\right] -2i\frac{\partial }{\partial y}\left[ \rho _{1}\frac{%
\partial \chi }{\partial y}\right]  \notag
\end{eqnarray}%
and can be written as%
\begin{equation}
\frac{\partial }{\partial t}\rho _{1}-2\left( b+b^{\ast }\right) \rho _{1}+%
\frac{\partial }{\partial x}\left[ \left( \frac{2\overline{A}_{x}}{i}+\frac{%
\partial \chi }{\partial x}\right) \rho _{1}\right] +\frac{\partial }{%
\partial y}\left[ \left( \frac{2\overline{A}_{y}}{i}+\frac{\partial \chi }{%
\partial y}\right) \rho _{1}\right] =0  \label{1305}
\end{equation}%
There is no approximation of the type\ "close to SD ".

This can be written as%
\begin{equation}
\left[ \frac{\partial }{\partial t}-2\left( b+b^{\ast }\right) \right] \rho
_{1}+\frac{\partial }{\partial x}\left( v_{x}^{\left( 1\right) }\rho
_{1}\right) +\frac{\partial }{\partial y}\left( v_{y}^{\left( 1\right) }\rho
_{1}\right) =0  \label{e401}
\end{equation}%
If we define%
\begin{equation}
\frac{\partial }{\partial t^{\prime }}\equiv \frac{\partial }{\partial t}%
-2\left( b+b^{\ast }\right)  \label{e402}
\end{equation}%
and remember that we dispose of the definition%
\begin{equation}
v_{x}^{\left( 1\right) }\equiv \frac{2\overline{A}_{x}}{i}+\frac{\partial
\chi }{\partial x}\ ,\ v_{y}^{\left( 1\right) }=\frac{2\overline{A}_{y}}{i}+%
\frac{\partial \chi }{\partial y}  \label{e403}
\end{equation}%
we obtain%
\begin{equation}
\frac{\partial }{\partial t^{\prime }}\rho _{1}+\mathit{div}\left( \mathbf{v}%
^{\left( 1\right) }\rho _{1}\right) =0  \label{e404}
\end{equation}%
At SD, $\partial /\partial t^{\prime }\rightarrow \partial /\partial t$.

\subsection{Derivation of the equation for $\protect\rho _{2}$}

Now we substract the two equations%
\begin{eqnarray}
&&-2i\frac{\partial }{\partial t}\rho _{2}-4i\left( b+b^{\ast }\right) \rho
_{2}+\frac{\partial }{\partial x}\left[ 4\overline{A}_{x}\rho _{2}\right] +%
\frac{\partial }{\partial y}\left[ 4\overline{A}_{y}\rho _{2}\right]
\label{1306} \\
&=&2i\frac{\partial }{\partial x}\left[ \rho _{2}\frac{\partial \eta }{%
\partial x}\right] +2i\frac{\partial }{\partial y}\left[ \rho _{2}\frac{%
\partial \eta }{\partial y}\right]  \notag
\end{eqnarray}%
or%
\begin{equation}
\frac{\partial }{\partial t}\rho _{2}+2\left( b+b^{\ast }\right) \rho _{2}+%
\frac{\partial }{\partial x}\left[ \left( -\frac{2\overline{A}_{x}}{i}+\frac{%
\partial \eta }{\partial x}\right) \rho _{2}\right] +\frac{\partial }{%
\partial y}\left[ \left( -\frac{2\overline{A}_{y}}{i}+\frac{\partial \eta }{%
\partial y}\right) \rho _{2}\right] =0  \label{1307}
\end{equation}

\bigskip

Now, we can use the definition%
\begin{equation}
v_{x}^{\left( 2\right) }\equiv -\frac{2\overline{A}_{x}}{i}+\frac{\partial
\eta }{\partial x}\ ,\ v_{y}^{\left( 2\right) }=-\frac{2\overline{A}_{y}}{i}+%
\frac{\partial \eta }{\partial y}  \label{e501}
\end{equation}%
together with%
\begin{equation}
\frac{\partial }{\partial t^{\prime \prime }}\equiv \frac{\partial }{%
\partial t}+2\left( b+b^{\ast }\right)  \label{e503}
\end{equation}%
and write%
\begin{equation}
\frac{\partial }{\partial t^{\prime \prime }}\rho _{2}+\frac{\partial }{%
\partial x}\left( v_{x}^{\left( 2\right) }\rho _{2}\right) +\frac{\partial }{%
\partial y}\left( v_{y}^{\left( 2\right) }\rho _{2}\right) =0  \label{e504}
\end{equation}%
\begin{equation}
\frac{\partial }{\partial t^{\prime \prime }}\rho _{2}+\mathit{div}\left( 
\mathbf{v}^{\left( 2\right) }\rho _{2}\right) =0  \label{e505}
\end{equation}%
We know that $\partial /\partial t^{\prime \prime }\rightarrow \partial
/\partial t$ at SD, where $b+b^{\ast }=0$. Visibly, at SD, where $\eta
=-\chi $ the two velocity fields $\mathbf{v}^{\left( 1\right) }$ and $%
\mathbf{v}^{\left( 2\right) }$ are simply opposite.

\section{Appendix E. The current of the Euler FT} \label{App:AppendixE}

\renewcommand{\theequation}{E.\arabic{equation}} \setcounter{equation}{0}

\subsection{General expressions for the current's components}

The formula for the FT current in the Euler case is 
\begin{eqnarray}
J^{0} &=&\left[ \phi ,\phi ^{\dagger }\right]  \label{f10} \\
J^{i} &=&-\frac{i}{2m}\left( \left[ \phi ^{\dagger },D_{i}\phi \right] -%
\left[ \left( D_{i}\phi \right) ^{\dagger },\phi \right] \right)  \notag
\end{eqnarray}%
\begin{eqnarray}
J^{i} &=&-\frac{i}{2m}\left( \left[ \phi ^{\dagger },\frac{\partial \phi }{%
\partial x^{i}}+\left[ A_{i},\phi \right] \right] -\left[ \left( \frac{%
\partial \phi }{\partial x^{i}}+\left[ A_{i},\phi \right] \right) ^{\dagger
},\phi \right] \right)  \label{f11} \\
&=&-\frac{i}{2m}\left[ \phi ^{\dagger }\left( \frac{\partial \phi }{\partial
x^{i}}+\left[ A_{i},\phi \right] \right) -\left( \frac{\partial \phi }{%
\partial x^{i}}+\left[ A_{i},\phi \right] \right) \phi ^{\dagger }\right. 
\notag \\
&&\ \ \ \left. -\left( \frac{\partial \phi }{\partial x^{i}}+\left[
A_{i},\phi \right] \right) ^{\dagger }\phi +\phi \left( \frac{\partial \phi 
}{\partial x^{i}}+\left[ A_{i},\phi \right] \right) ^{\dagger }\right] 
\notag \\
&=&-\frac{i}{2m}\left\{ \phi ^{\dagger }\frac{\partial \phi }{\partial x^{i}}%
+\phi ^{\dagger }\left( A_{i}\phi -\phi A_{i}\right) -\frac{\partial \phi }{%
\partial x^{i}}\phi ^{\dagger }-\left( A_{i}\phi -\phi A_{i}\right) \phi
^{\dagger }\right.  \notag \\
&&\left. -\left( \frac{\partial \phi ^{\dagger }}{\partial x^{i}}+\left(
\phi ^{\dagger }A_{i}^{\dagger }-A_{i}^{\dagger }\phi ^{\dagger }\right)
\right) \phi +\phi \left( \frac{\partial \phi ^{\dagger }}{\partial x^{i}}%
+\left( \phi ^{\dagger }A_{i}^{\dagger }-A_{i}^{\dagger }\phi ^{\dagger
}\right) \right) \right\}  \notag
\end{eqnarray}%
Let us collect the part that depends only on $\phi $ and $\phi ^{\dagger }$
and separately the part that depends on $A_{i}$ and $A_{i}^{\dagger }$.%
\begin{eqnarray}
J^{i} &=&-\frac{i}{2m}\left\{ \phi ^{\dagger }\frac{\partial \phi }{\partial
x^{i}}-\frac{\partial \phi }{\partial x^{i}}\phi ^{\dagger }-\frac{\partial
\phi ^{\dagger }}{\partial x^{i}}\phi +\phi \frac{\partial \phi ^{\dagger }}{%
\partial x^{i}}\right.  \label{f103} \\
&&+\phi ^{\dagger }A_{i}\phi -\phi ^{\dagger }\phi A_{i}-A_{i}\phi \phi
^{\dagger }+\phi A_{i}\phi ^{\dagger }  \notag \\
&&\left. -\phi ^{\dagger }A_{i}^{\dagger }\phi +A_{i}^{\dagger }\phi
^{\dagger }\phi +\phi \phi ^{\dagger }A_{i}^{\dagger }-\phi A_{i}^{\dagger
}\phi ^{\dagger }\right\}  \notag
\end{eqnarray}%
This expression will be used later just as a check for the result of the
derivation presented below.

The current for $\mu \equiv k$ (space components) is 
\begin{eqnarray}
J^{k} &=&-\frac{i}{2m}\left\{ \phi ^{\dagger }\left( \partial ^{k}\phi
\right) -\left( \partial ^{k}\phi \right) \phi ^{\dagger }-\left( \partial
^{k}\phi ^{\dagger }\right) \phi +\phi \left( \partial ^{k}\phi ^{\dagger
}\right) \right.  \label{f104} \\
&&\left. +\left[ \phi ^{\dagger },\left[ A^{k},\phi \right] \right] +\left[
\phi ,\left[ \phi ^{\dagger },A^{k\dagger }\right] \right] \right\}  \notag
\\
&\equiv &\Lambda _{1}^{k}+\Lambda _{2}^{k}  \notag
\end{eqnarray}%
where 
\begin{eqnarray}
\Lambda _{1}^{k} &\equiv &-\frac{i}{2m}\left\{ \phi ^{\dagger }\left(
\partial ^{k}\phi \right) -\left( \partial ^{k}\phi \right) \phi ^{\dagger
}-\left( \partial ^{k}\phi ^{\dagger }\right) \phi +\phi \left( \partial
^{k}\phi ^{\dagger }\right) \right\}  \label{f105} \\
\Lambda _{2}^{k} &\equiv &-\frac{i}{2m}\left( \left[ \phi ^{\dagger },\left[
A^{k},\phi \right] \right] +\left[ \phi ,\left[ \phi ^{\dagger },A^{k\dagger
}\right] \right] \right)  \notag
\end{eqnarray}

\subsubsection{The expression of the first part of the current, $\Lambda
_{1} $}

The terms containing space and time derivatives (here the symbol $\Psi $ is
replaced by $\phi $) 
\begin{equation}
\Lambda _{1}^{k}=-\frac{i}{2m}\left[ \phi ^{\dagger }\left( \partial
^{k}\phi \right) -\left( \partial ^{k}\phi \right) \phi ^{\dagger }-\left(
\partial ^{k}\phi ^{\dagger }\right) \phi +\phi \left( \partial ^{k}\phi
^{\dagger }\right) \right]  \label{f106}
\end{equation}%
where we have to insert 
\begin{eqnarray}
\phi &=&\phi _{1}E_{+}+\phi _{2}E_{-}  \label{f107} \\
\phi ^{\dagger } &=&\phi _{1}^{\ast }E_{-}+\phi _{2}^{\ast }E_{+}  \notag
\end{eqnarray}%
This consists of two commutators.

The first commutator is 
\begin{eqnarray}
\left[ \phi ^{\dagger },\partial ^{k}\phi \right] &=&\phi ^{\dagger }\left(
\partial ^{k}\phi \right) -\left( \partial ^{k}\phi \right) \phi ^{\dagger }
\label{f108} \\
&=&\left( \phi _{1}^{\ast }E_{-}+\phi _{2}^{\ast }E_{+}\right) \left( \frac{%
\partial \phi _{1}}{\partial x_{k}}E_{+}+\frac{\partial \phi _{2}}{\partial
x_{k}}E_{-}\right)  \notag \\
&&-\left( \frac{\partial \phi _{1}}{\partial x_{k}}E_{+}+\frac{\partial \phi
_{2}}{\partial x_{k}}E_{-}\right) \left( \phi _{1}^{\ast }E_{-}+\phi
_{2}^{\ast }E_{+}\right)  \notag \\
&=&\phi _{1}^{\ast }\frac{\partial \phi _{1}}{\partial x_{k}}E_{-}E_{+}+%
\underline{\phi _{1}^{\ast }\frac{\partial \phi _{2}}{\partial x_{k}}%
E_{-}E_{-}}+\underbrace{\phi _{2}^{\ast }\frac{\partial \phi _{1}}{\partial
x_{k}}E_{+}E_{+}}+\phi _{2}^{\ast }\frac{\partial \phi _{2}}{\partial x_{k}}%
E_{+}E_{-}  \notag \\
&&-\phi _{1}^{\ast }\frac{\partial \phi _{1}}{\partial x_{k}}E_{+}E_{-}%
\underbrace{-\phi _{2}^{\ast }\frac{\partial \phi _{1}}{\partial x_{k}}%
E_{+}E_{+}}\underline{-\phi _{1}^{\ast }\frac{\partial \phi _{2}}{\partial
x_{k}}E_{-}E_{-}}-\phi _{2}^{\ast }\frac{\partial \phi _{2}}{\partial x_{k}}%
E_{-}E_{+}  \notag
\end{eqnarray}%
The coefficients of $E_{-}E_{-}$ and of $E_{+}E_{+}$ cancel. The result is 
\begin{eqnarray}
&&\left[ \phi ^{\dagger },\partial ^{k}\phi \right]  \label{f109} \\
&=&\phi _{1}^{\ast }\frac{\partial \phi _{1}}{\partial x_{k}}\left[
E_{-},E_{+}\right] +\phi _{2}^{\ast }\frac{\partial \phi _{2}}{\partial x_{k}%
}\left[ E_{+},E_{-}\right]  \notag
\end{eqnarray}%
Here we must use the commutators of the generators of the algebra and obtain 
\begin{equation}
\left[ \phi ^{\dagger },\partial ^{k}\phi \right] =-\left( \phi _{1}^{\ast }%
\frac{\partial \phi _{1}}{\partial x_{k}}-\phi _{2}^{\ast }\frac{\partial
\phi _{2}}{\partial x_{k}}\right) H  \label{f110}
\end{equation}

The second commutator in $\Lambda _{1}^{k}$ is 
\begin{eqnarray}
\left[ \phi ,\partial ^{k}\phi ^{\dagger }\right] &=&\phi \left( \partial
^{k}\phi ^{\dagger }\right) -\left( \partial ^{k}\phi ^{\dagger }\right) \phi
\label{f111} \\
&=&\left( \phi _{1}E_{+}+\phi _{2}E_{-}\right) \left( \frac{\partial \phi
_{1}^{\ast }}{\partial x_{k}}E_{-}+\frac{\partial \phi _{2}^{\ast }}{%
\partial x_{k}}E_{+}\right)  \notag \\
&&-\left( \frac{\partial \phi _{1}^{\ast }}{\partial x_{k}}E_{-}+\frac{%
\partial \phi _{2}^{\ast }}{\partial x_{k}}E_{+}\right) \left( \phi
_{1}E_{+}+\phi _{2}E_{-}\right)  \notag \\
&=&\phi _{1}\frac{\partial \phi _{1}^{\ast }}{\partial x_{k}}E_{+}E_{-}+%
\underline{\phi _{2}\frac{\partial \phi _{1}^{\ast }}{\partial x_{k}}%
E_{-}E_{-}}+\underbrace{\phi _{1}\frac{\partial \phi _{2}^{\ast }}{\partial
x_{k}}E_{+}E_{+}}+\phi _{2}\frac{\partial \phi _{2}^{\ast }}{\partial x_{k}}%
E_{-}E_{+}  \notag \\
&&-\phi _{1}\frac{\partial \phi _{1}^{\ast }}{\partial x_{k}}E_{-}E_{+}%
\underbrace{-\phi _{1}\frac{\partial \phi _{2}^{\ast }}{\partial x_{k}}%
E_{+}E_{+}}\underline{-\phi _{2}\frac{\partial \phi _{1}^{\ast }}{\partial
x_{k}}E_{-}E_{-}}-\phi _{2}\frac{\partial \phi _{2}^{\ast }}{\partial x_{k}}%
E_{+}E_{-}  \notag
\end{eqnarray}%
As above, the coefficients of the terms $E_{+}E_{+}$ and respectively $%
E_{-}E_{-}$ cancel. The other represent commutators that can be expressed by 
$H$: 
\begin{eqnarray}
&&\left[ \phi ,\partial ^{k}\phi ^{\dagger }\right]  \label{f112} \\
&=&\phi _{1}\frac{\partial \phi _{1}^{\ast }}{\partial x_{k}}\left[
E_{+},E_{-}\right] -\phi _{2}\frac{\partial \phi _{2}^{\ast }}{\partial x_{k}%
}\left[ E_{+},E_{-}\right]  \notag \\
&=&\left( \phi _{1}\frac{\partial \phi _{1}^{\ast }}{\partial x_{k}}-\phi
_{2}\frac{\partial \phi _{2}^{\ast }}{\partial x_{k}}\right) H  \notag
\end{eqnarray}

Putting together these results we have

\begin{eqnarray}
\Lambda _{1}^{k} &=&-\frac{i}{2m}\left[ \phi ^{\dagger }\left( \partial
^{k}\phi \right) -\left( \partial ^{k}\phi \right) \phi ^{\dagger }-\left(
\partial ^{k}\phi ^{\dagger }\right) \phi +\phi \left( \partial ^{k}\phi
^{\dagger }\right) \right]  \label{f113} \\
&=&-\frac{i}{2m}\left\{ \left[ \phi ^{\dagger },\partial ^{k}\phi \right] +%
\left[ \phi ,\partial ^{k}\phi ^{\dagger }\right] \right\}  \notag \\
&=&-\frac{i}{2m}\left[ -\left( \phi _{1}^{\ast }\frac{\partial \phi _{1}}{%
\partial x_{k}}-\phi _{2}^{\ast }\frac{\partial \phi _{2}}{\partial x_{k}}%
\right) H+\left( \phi _{1}\frac{\partial \phi _{1}^{\ast }}{\partial x_{k}}%
-\phi _{2}\frac{\partial \phi _{2}^{\ast }}{\partial x_{k}}\right) H\right] 
\notag \\
&=&-\frac{i}{2m}\left[ \phi _{1}\frac{\partial \phi _{1}^{\ast }}{\partial
x_{k}}-\phi _{1}^{\ast }\frac{\partial \phi _{1}}{\partial x_{k}}-\phi _{2}%
\frac{\partial \phi _{2}^{\ast }}{\partial x_{k}}+\phi _{2}^{\ast }\frac{%
\partial \phi _{2}}{\partial x_{k}}\right] H  \notag
\end{eqnarray}%
The derivatives look like the derivatives of ratios $\phi /\phi ^{\ast }$ if
we multiply by the adequet denominator.%
\begin{eqnarray}
\phi _{1}\frac{\partial \phi _{1}^{\ast }}{\partial x_{k}}-\phi _{1}^{\ast }%
\frac{\partial \phi _{1}}{\partial x_{k}} &=&-\left( \phi _{1}^{\ast
}\right) ^{2}\frac{\frac{\partial \phi _{1}}{\partial x_{k}}\phi _{1}^{\ast
}-\phi _{1}\frac{\partial \phi _{1}^{\ast }}{\partial x_{k}}}{\left( \phi
_{1}^{\ast }\right) ^{2}}=  \label{f114} \\
&=&-\left( \phi _{1}^{\ast }\right) ^{2}\frac{\partial }{\partial x_{k}}%
\left( \frac{\phi _{1}}{\phi _{1}^{\ast }}\right)  \notag
\end{eqnarray}%
\begin{eqnarray}
-\phi _{2}\frac{\partial \phi _{2}^{\ast }}{\partial x_{k}}+\phi _{2}^{\ast }%
\frac{\partial \phi _{2}}{\partial x_{k}} &=&\left( \phi _{2}^{\ast }\right)
^{2}\frac{\frac{\partial \phi _{2}}{\partial x_{k}}\phi _{2}^{\ast }-\phi
_{2}\frac{\partial \phi _{2}^{\ast }}{\partial x_{k}}}{\left( \phi
_{2}^{\ast }\right) ^{2}}  \label{f116} \\
&=&\left( \phi _{2}^{\ast }\right) ^{2}\frac{\partial }{\partial x_{k}}%
\left( \frac{\phi _{2}}{\phi _{2}^{\ast }}\right)  \notag
\end{eqnarray}%
Then this part is%
\begin{eqnarray}
\Lambda _{1}^{k} &=&-\frac{i}{2m}\left[ \phi ^{\dagger }\left( \partial
^{k}\phi \right) -\left( \partial ^{k}\phi \right) \phi ^{\dagger }-\left(
\partial ^{k}\phi ^{\dagger }\right) \phi +\phi \left( \partial ^{k}\phi
^{\dagger }\right) \right]  \label{f120} \\
&=&-\frac{i}{2m}\left[ -\left( \phi _{1}^{\ast }\right) ^{2}\frac{\partial }{%
\partial x_{k}}\left( \frac{\phi _{1}}{\phi _{1}^{\ast }}\right) +\left(
\phi _{2}^{\ast }\right) ^{2}\frac{\partial }{\partial x_{k}}\left( \frac{%
\phi _{2}}{\phi _{2}^{\ast }}\right) \right] H  \notag
\end{eqnarray}

\bigskip

Postponing a reformulation of this expression, we just represent here the
functions $\phi _{1}$ and $\phi _{2}$ as they are defined, we have%
\begin{eqnarray}
\rho _{1} &=&\left\vert \phi _{1}\right\vert ^{2}=\exp \left( \psi
_{1}\right)  \label{f130} \\
\rho _{2} &=&\left\vert \phi _{2}\right\vert ^{2}=\exp \left( \psi
_{2}\right)  \notag
\end{eqnarray}%
Then%
\begin{equation}
\phi _{1}=\exp \left( \frac{\psi _{1}}{2}\right) \exp \left( i\chi \right)
\label{f161}
\end{equation}%
\begin{equation}
\phi _{2}=\exp \left( \frac{\psi _{2}}{2}\right) \exp \left( i\eta \right)
\label{f162}
\end{equation}%
Then%
\begin{eqnarray}
\frac{\phi _{1}}{\phi _{1}^{\ast }} &=&\exp \left( 2i\chi \right)
\label{f163} \\
\frac{\phi _{2}}{\phi _{2}^{\ast }} &=&\exp \left( 2i\eta \right)  \notag
\end{eqnarray}%
\begin{eqnarray}
\left( \phi _{1}^{\ast }\right) ^{2} &=&\rho _{1}\exp \left( -2i\chi \right)
\label{f164} \\
\left( \phi _{2}^{\ast }\right) ^{2} &=&\rho _{2}\exp \left( -2i\eta \right)
\notag
\end{eqnarray}%
and%
\begin{eqnarray}
\Lambda _{1} &=&-\frac{i}{2m}\left[ -\left( \phi _{1}^{\ast }\right) ^{2}%
\frac{\partial }{\partial x_{k}}\left( \frac{\phi _{1}}{\phi _{1}^{\ast }}%
\right) +\left( \phi _{2}^{\ast }\right) ^{2}\frac{\partial }{\partial x_{k}}%
\left( \frac{\phi _{2}}{\phi _{2}^{\ast }}\right) \right] H  \label{f165} \\
&=&-\frac{i}{2m}\left[ -\rho _{1}\exp \left( -2i\chi \right) \frac{\partial 
}{\partial x_{k}}\exp \left( 2i\chi \right) +\rho _{2}\exp \left( -2i\eta
\right) \frac{\partial }{\partial x_{k}}\exp \left( 2i\eta \right) \right] H
\notag \\
&=&-\frac{i}{2m}\left[ -\rho _{1}2i\frac{\partial \chi }{\partial x_{k}}%
+\rho _{2}2i\frac{\partial \eta }{\partial x_{k}}\right] H  \notag \\
&=&\frac{1}{m}\left( -\rho _{1}\frac{\partial \chi }{\partial x_{k}}+\rho
_{2}\frac{\partial \eta }{\partial x_{k}}\right) H  \notag
\end{eqnarray}

\bigskip

\subsubsection{The expression of the second part of the current, $\Lambda
_{2}$}

According to the expansion done above we have to calculate%
\begin{eqnarray}
\Lambda _{2} &=&-\frac{i}{2m}\left\{ \phi ^{\dagger }A_{i}\phi -\phi
^{\dagger }\phi A_{i}-A_{i}\phi \phi ^{\dagger }+\phi A_{i}\phi ^{\dagger
}\right.  \label{f170} \\
&&\left. -\phi ^{\dagger }A_{i}^{\dagger }\phi +A_{i}^{\dagger }\phi
^{\dagger }\phi +\phi \phi ^{\dagger }A_{i}^{\dagger }-\phi A_{i}^{\dagger
}\phi ^{\dagger }\right\}  \notag
\end{eqnarray}%
Let us replace here%
\begin{eqnarray}
\phi &\equiv &\phi =\phi _{1}E_{+}+\phi _{2}E_{-}  \label{f172} \\
\phi ^{\dagger } &=&\phi ^{\dagger }=\phi _{1}^{\ast }E_{-}+\phi _{2}^{\ast
}E_{+}  \notag
\end{eqnarray}%
and the formulas 
\begin{eqnarray}
A_{x} &=&\frac{1}{2}\left( a-a^{\ast }\right) H  \label{f173} \\
A_{y} &=&\frac{i}{2}\left( a+a^{\ast }\right) H  \notag
\end{eqnarray}

\paragraph{Calculation of the $x$ component}

We ignore for the moment the coefficient $\left( -i/2\right) $.

First term on the first line%
\begin{eqnarray}
\phi ^{\dagger }A_{x}\phi &=&\left( \phi _{1}^{\ast }E_{-}+\phi _{2}^{\ast
}E_{+}\right) \frac{1}{2}\left( a-a^{\ast }\right) H\left( \phi
_{1}E_{+}+\phi _{2}E_{-}\right)  \label{f180} \\
&=&\phi _{1}^{\ast }\phi _{1}\frac{1}{2}\left( a-a^{\ast }\right) \ \
E_{-}HE_{+}  \notag \\
&&+\phi _{2}^{\ast }\phi _{1}\frac{1}{2}\left( a-a^{\ast }\right) \ \
E_{+}HE_{+}  \notag \\
&&+\phi _{1}^{\ast }\phi _{2}\frac{1}{2}\left( a-a^{\ast }\right) \ \
E_{-}HE_{-}  \notag \\
&&+\phi _{2}^{\ast }\phi _{2}\frac{1}{2}\left( a-a^{\ast }\right) \ \
E_{+}HE_{-}  \notag
\end{eqnarray}

The second term on the first line%
\begin{eqnarray}
-\phi ^{\dagger }\phi A_{i} &=&-\left( \phi _{1}^{\ast }E_{-}+\phi
_{2}^{\ast }E_{+}\right) \left( \phi _{1}E_{+}+\phi _{2}E_{-}\right) \frac{1%
}{2}\left( a-a^{\ast }\right) H  \label{f181} \\
&=&-\phi _{1}^{\ast }\phi _{1}\frac{1}{2}\left( a-a^{\ast }\right) \ \
E_{-}E_{+}H  \notag \\
&&-\phi _{1}^{\ast }\phi _{2}\frac{1}{2}\left( a-a^{\ast }\right) \ \
E_{-}E_{-}H  \notag \\
&&-\phi _{2}^{\ast }\phi _{1}\frac{1}{2}\left( a-a^{\ast }\right) \ \
E_{+}E_{+}H  \notag \\
&&-\phi _{2}^{\ast }\phi _{2}\frac{1}{2}\left( a-a^{\ast }\right) \ \
E_{+}E_{-}H  \notag
\end{eqnarray}

The third term on the first line%
\begin{eqnarray}
-A_{i}\phi \phi ^{\dagger } &=&-\frac{1}{2}\left( a-a^{\ast }\right) H\left(
\phi _{1}E_{+}+\phi _{2}E_{-}\right) \left( \phi _{1}^{\ast }E_{-}+\phi
_{2}^{\ast }E_{+}\right)  \label{f182} \\
&=&-\frac{1}{2}\left( a-a^{\ast }\right) \phi _{1}\phi _{1}^{\ast }\ \
HE_{+}E_{-}  \notag \\
&&-\frac{1}{2}\left( a-a^{\ast }\right) \phi _{1}\phi _{2}^{\ast }\ \
HE_{+}E_{+}  \notag \\
&&-\frac{1}{2}\left( a-a^{\ast }\right) \phi _{2}\phi _{1}^{\ast }\ \
HE_{-}E_{-}  \notag \\
&&-\frac{1}{2}\left( a-a^{\ast }\right) \phi _{2}\phi _{2}^{\ast }\ \
HE_{-}E_{+}  \notag
\end{eqnarray}

The fourth term on the first line%
\begin{eqnarray*}
\phi A_{i}\phi ^{\dagger } &=&\left( \phi _{1}E_{+}+\phi _{2}E_{-}\right) 
\frac{1}{2}\left( a-a^{\ast }\right) H\left( \phi _{1}^{\ast }E_{-}+\phi
_{2}^{\ast }E_{+}\right) \\
&=&\phi _{1}\phi _{1}^{\ast }\frac{1}{2}\left( a-a^{\ast }\right) \ \
E_{+}HE_{-} \\
&&+\phi _{1}\phi _{2}^{\ast }\frac{1}{2}\left( a-a^{\ast }\right) \ \
E_{+}HE_{+} \\
&&+\phi _{2}\phi _{1}^{\ast }\frac{1}{2}\left( a-a^{\ast }\right) \ \
E_{-}HE_{-} \\
&&+\phi _{2}\phi _{2}^{\ast }\frac{1}{2}\left( a-a^{\ast }\right) \ \
E_{-}HE_{+}
\end{eqnarray*}

\bigskip

Now we go to the second line in the detailed expression of $\Lambda _{2}$;

The first term is similar to the first term of the first line, but $A_{i}$
is now \emph{daggered}:%
\begin{eqnarray}
-\phi ^{\dagger }A_{i}^{\dagger }\phi &=&-\left( \phi _{1}^{\ast }E_{-}+\phi
_{2}^{\ast }E_{+}\right) \frac{1}{2}\left( a^{\ast }-a\right) H\left( \phi
_{1}E_{+}+\phi _{2}E_{-}\right)  \label{f184} \\
&=&-\phi _{1}^{\ast }\phi _{1}\frac{1}{2}\left( a^{\ast }-a\right) \ \
E_{-}HE_{+}  \notag \\
&&-\phi _{2}^{\ast }\phi _{1}\frac{1}{2}\left( a^{\ast }-a\right) \ \
E_{+}HE_{+}  \notag \\
&&-\phi _{1}^{\ast }\phi _{2}\frac{1}{2}\left( a^{\ast }-a\right) \ \
E_{-}HE_{-}  \notag \\
&&-\phi _{2}^{\ast }\phi _{2}\frac{1}{2}\left( a^{\ast }-a\right) \ \
E_{+}HE_{-}  \notag
\end{eqnarray}

the second term of the second line is $A_{i}^{\dagger }\phi ^{\dagger }\phi $%
, or%
\begin{eqnarray}
A_{i}^{\dagger }\phi ^{\dagger }\phi &=&\frac{1}{2}\left( a^{\ast }-a\right)
H\left( \phi _{1}^{\ast }E_{-}+\phi _{2}^{\ast }E_{+}\right) \left( \phi
_{1}E_{+}+\phi _{2}E_{-}\right)  \label{f185} \\
&=&\frac{1}{2}\left( a^{\ast }-a\right) \phi _{1}^{\ast }\phi _{1}\ \
HE_{-}E_{+}  \notag \\
&&+\frac{1}{2}\left( a^{\ast }-a\right) \phi _{1}^{\ast }\phi _{2}\ \
HE_{-}E_{-}  \notag \\
&&+\frac{1}{2}\left( a^{\ast }-a\right) \phi _{2}^{\ast }\phi _{1}\ \
HE_{+}E_{+}  \notag \\
&&+\frac{1}{2}\left( a^{\ast }-a\right) \phi _{2}^{\ast }\phi _{2}\ \
HE_{+}E_{-}  \notag
\end{eqnarray}

the third term in the second line%
\begin{eqnarray}
\phi \phi ^{\dagger }A_{i}^{\dagger } &=&\left( \phi _{1}E_{+}+\phi
_{2}E_{-}\right) \left( \phi _{1}^{\ast }E_{-}+\phi _{2}^{\ast }E_{+}\right) 
\frac{1}{2}\left( a^{\ast }-a\right) H  \label{f187} \\
&=&\frac{1}{2}\left( a^{\ast }-a\right) \phi _{1}\phi _{1}^{\ast }\ \
E_{+}E_{-}H  \notag \\
&&+\frac{1}{2}\left( a^{\ast }-a\right) \phi _{1}\phi _{2}^{\ast }\ \
E_{+}E_{+}H  \notag \\
&&+\frac{1}{2}\left( a^{\ast }-a\right) \phi _{2}\phi _{1}^{\ast }\ \
E_{-}E_{-}H  \notag \\
&&+\frac{1}{2}\left( a^{\ast }-a\right) \phi _{2}\phi _{2}^{\ast }\ \
E_{-}E_{+}H  \notag
\end{eqnarray}

the fourth term in the second line%
\begin{eqnarray}
-\phi A_{i}^{\dagger }\phi ^{\dagger } &=&-\left( \phi _{1}E_{+}+\phi
_{2}E_{-}\right) \frac{1}{2}\left( a^{\ast }-a\right) H\left( \phi
_{1}^{\ast }E_{-}+\phi _{2}^{\ast }E_{+}\right)  \label{f189} \\
&=&-\phi _{1}\phi _{1}^{\ast }\frac{1}{2}\left( a^{\ast }-a\right) \ \
E_{+}HE_{-}  \notag \\
&&-\phi _{1}\phi _{2}^{\ast }\frac{1}{2}\left( a^{\ast }-a\right) \ \
E_{+}HE_{+}  \notag \\
&&-\phi _{2}\phi _{1}^{\ast }\frac{1}{2}\left( a^{\ast }-a\right) \ \
E_{-}HE_{-}  \notag \\
&&-\phi _{2}\phi _{2}^{\ast }\frac{1}{2}\left( a^{\ast }-a\right) \ \
E_{-}HE_{+}  \notag
\end{eqnarray}

\bigskip

We now collect the coefficients of the terms%
\begin{eqnarray}
&&\text{for}\ \ \phi _{1}^{\ast }\phi _{1}\frac{1}{2}\left( a-a^{\ast
}\right) \ \ \text{these are}  \label{f190} \\
&&+E_{-}HE_{+}  \notag \\
&&-E_{-}E_{+}H  \notag \\
&&-HE_{+}E_{-}  \notag \\
&&+E_{+}HE_{-}  \notag \\
&&+E_{-}HE_{+}  \notag \\
&&-HE_{-}E_{+}  \notag \\
&&-E_{+}E_{-}H  \notag \\
&&+E_{+}HE_{-}  \notag
\end{eqnarray}%
We can combine these operator products%
\begin{eqnarray}
&&E_{-}\left( HE_{+}-E_{+}H\right) \ \ \ \ \ \text{this is}\ \ E_{-}\left(
2E_{+}\right)  \label{f191} \\
&&-\left( HE_{+}-E_{+}H\right) E_{-}\ \ \text{this is}\ \ -\left(
2E_{+}\right) E_{-}  \notag \\
&&+\left( E_{-}H-HE_{-}\right) E_{+}\ \ \text{this is }-\left(
-2E_{-}\right) E_{+}  \notag \\
&&-E_{+}\left( E_{-}H-HE_{-}\right) \ \ \text{this is}\ \ -E_{+}\left(
-\right) \left( -2E_{-}\right)  \notag
\end{eqnarray}%
or%
\begin{equation}
2\left[ E_{-}E_{+}-E_{+}E_{-}+E_{-}E_{+}-E_{+}E_{-}\right] =2\left[ -H-H%
\right] =-4H  \label{f193}
\end{equation}%
Finally from this term we obtain%
\begin{equation}
\left( -4\right) \phi _{1}^{\ast }\phi _{1}\frac{1}{2}\left( a-a^{\ast
}\right) \ \ H  \label{f194}
\end{equation}

\bigskip

The next term

\begin{eqnarray}
&&\text{for}\ \ \phi _{2}^{\ast }\phi _{1}\frac{1}{2}\left( a-a^{\ast
}\right) \text{\ \ these are}  \label{f195} \\
&&E_{+}HE_{+}  \notag \\
&&-E_{+}E_{+}H  \notag \\
&&-HE_{+}E_{+}  \notag \\
&&+E_{+}HE_{+}  \notag \\
&&+\ E_{+}HE_{+}  \notag \\
&&-HE_{+}E_{+}  \notag \\
&&-E_{+}E_{+}H  \notag \\
&&+E_{+}HE_{+}  \notag
\end{eqnarray}%
and we combine the product of operators%
\begin{eqnarray}
&&E_{+}\left( HE_{+}-E_{+}H\right) \ \ \ \text{this is }\ \ E_{+}\left(
2E_{+}\right)  \label{f199} \\
&&-\left( HE_{+}-E_{+}H\right) E_{+}\ \ \text{this is}\ \ -\left(
2E_{+}\right) E_{+}  \notag \\
&&-\left( HE_{+}-E_{+}H\right) E_{+}\text{\ \ this is}\ \ -\left(
2E_{+}\right) E_{+}  \notag \\
&&+E_{+}\left( HE_{+}-E_{+}H\right) \text{\ \ this is}\ \ E_{+}\left(
2E_{+}\right)  \notag
\end{eqnarray}%
and we find%
\begin{equation*}
2\left[ 0\right]
\end{equation*}%
which makes that the term contribute zith \emph{zero}%
\begin{equation}
\phi _{2}^{\ast }\phi _{1}\frac{1}{2}\left( a-a^{\ast }\right) \ \ \times 2%
\left[ 0\right] =0  \label{f201}
\end{equation}

\bigskip

The next term%
\begin{eqnarray}
&&\text{for}\ \phi _{1}^{\ast }\phi _{2}\frac{1}{2}\left( a-a^{\ast }\right) 
\text{\ \ these are}  \label{f202} \\
&&\ E_{-}HE_{-}  \notag \\
&&-E_{-}E_{-}H  \notag \\
&&-HE_{-}E_{-}  \notag \\
&&+E_{-}HE_{-}  \notag \\
&&+E_{-}HE_{-}  \notag \\
&&-HE_{-}E_{-}  \notag \\
&&-E_{-}E_{-}H  \notag \\
&&+E_{-}HE_{-}  \notag
\end{eqnarray}%
We combine the porducts of operators%
\begin{eqnarray}
&&E_{-}\left( HE_{-}-E_{-}H\right) \ \ \text{this is}\ \ E_{-}\left(
-2E_{-}\right)  \label{f204} \\
&&-\left( HE_{-}-E_{-}H\right) E_{-}\ \ \text{this is}\ \ -\left(
-2E_{-}\right) E_{-}  \notag \\
&&-\left( HE_{-}-E_{-}H\right) E_{-}\text{\ \ this is}\ \ -\left(
-2E_{-}\right) E_{-}  \notag \\
&&+E_{-}\left( HE_{-}-E_{-}H\right) \text{\ \ this is}\ \ E_{-}\left(
-2E_{-}\right)  \notag
\end{eqnarray}%
which gives finally%
\begin{equation*}
2\left[ 0\right]
\end{equation*}%
and this term does not contribute to the final expression%
\begin{equation}
\phi _{1}^{\ast }\phi _{2}\frac{1}{2}\left( a-a^{\ast }\right) \ \times \ 2%
\left[ 0\right] =0  \label{f205}
\end{equation}

\bigskip

The next term is%
\begin{eqnarray}
&&\text{for}\ \ \phi _{2}^{\ast }\phi _{2}\frac{1}{2}\left( a-a^{\ast
}\right) \text{\ \ these are}  \label{f206} \\
&&E_{+}HE_{-}  \notag \\
&&-E_{+}E_{-}H  \notag \\
&&-HE_{-}E_{+}  \notag \\
&&+E_{-}HE_{+}  \notag \\
&&+E_{+}HE_{-}  \notag \\
&&-HE_{+}E_{-}  \notag \\
&&-E_{-}E_{+}H  \notag \\
&&+E_{-}HE_{+}  \notag
\end{eqnarray}%
we combine the products of operators%
\begin{eqnarray}
&&E_{+}\left( HE_{-}-E_{-}H\right) \ \ \text{this is}\ \ E_{+}\left(
-2E_{-}\right)  \label{f211} \\
&&-\left( HE_{-}-E_{-}H\right) E_{+}\text{\ \ this is }-\left(
-2E_{-}\right) E_{+}  \notag \\
&&-\left( HE_{+}-E_{+}H\right) E_{-}\text{\ \ this is}\ \ -\left(
2E_{+}\right) E_{-}  \notag \\
&&+E_{-}\left( HE_{+}-E_{+}H\right) \text{\ \ this is}\ \ E_{-}\left(
2E_{+}\right)  \notag
\end{eqnarray}%
which gives%
\begin{equation}
2\left[ -E_{+}E_{-}+E_{-}E_{+}-E_{+}E_{-}+E_{-}E_{+}\right] =-4\left[
E_{+}E_{-}-E_{-}E_{+}\right] =-4H  \label{f212}
\end{equation}%
and it results that the contribution of this term is%
\begin{equation}
\phi _{2}^{\ast }\phi _{2}\frac{1}{2}\left( a-a^{\ast }\right) \left(
-4\right) \ \ H  \label{f214}
\end{equation}

\bigskip

We put together the two terms%
\begin{eqnarray}
\Lambda _{2}^{x} &=&-\frac{i}{2m}\left[ \left( -4\right) \phi _{1}^{\ast
}\phi _{1}\frac{1}{2}\left( a-a^{\ast }\right) \ \ H+\left( -4\right) \phi
_{2}^{\ast }\phi _{2}\frac{1}{2}\left( a-a^{\ast }\right) \ \ H\right] 
\notag \\
&=&\frac{i}{m}\left( a-a^{\ast }\right) \left( \rho _{1}+\rho _{2}\right) \ H
\label{f2015}
\end{eqnarray}%
This will be confirmed by a cross check below.

\bigskip

Now the $x$-component of the current is%
\begin{eqnarray}
J^{x}/H &=&\Lambda _{1}^{x}+\Lambda _{2}^{x}\ \ /H  \label{f216} \\
&=&\frac{1}{m}\left( -\rho _{1}\frac{\partial \chi }{\partial x_{k}}+\rho
_{2}\frac{\partial \eta }{\partial x_{k}}\right) +\frac{i}{m}\left(
a-a^{\ast }\right) \left( \rho _{1}+\rho _{2}\right)  \notag
\end{eqnarray}

\textbf{NOTE} that we use the symbolic writting $J^{x}/H$ and similar to
denote the coefficient of the $H$ generator in the alegbraic expression of $%
J^{x}$. In other situations we use the notation%
\begin{equation}
A_{x}=\overline{A}_{x}H  \label{f217}
\end{equation}%
to separate in $A_{x}$ the coefficient $\overline{A}_{x}$ from the algebraic
generator $H$.

\paragraph{Calculation of the $y$ component}

It differs from the $x$ term by the insertion of%
\begin{eqnarray}
A_{y} &=&\frac{i}{2}\left( a+a^{\ast }\right) H  \label{f301} \\
A_{y}^{\dagger } &=&-\frac{i}{2}\left( a^{\ast }+a\right) H=-A_{y}  \notag
\end{eqnarray}%
It has similar properties as $A_{x}$ and $A_{x}^{\dagger }$. We just need to
replace%
\begin{eqnarray}
a-a^{\ast } &\rightarrow &a+a^{\ast }  \label{f302} \\
\frac{1}{2} &\rightarrow &\frac{i}{2}  \notag
\end{eqnarray}%
in $\Lambda _{2}^{x}$ to obtain%
\begin{eqnarray}
\Lambda _{2}^{y} &=&-\frac{i}{2m}\left[ \left( -4\right) \phi _{1}^{\ast
}\phi _{1}\frac{i}{2}\left( a+a^{\ast }\right) \ \ H+\left( -4\right) \phi
_{2}^{\ast }\phi _{2}\frac{i}{2}\left( a+a^{\ast }\right) \ \ H\right] 
\notag \\
&=&-\frac{1}{m}\left( a+a^{\ast }\right) \left( \rho _{1}+\rho _{2}\right) \
H  \label{f305}
\end{eqnarray}

Now, for the current%
\begin{eqnarray}
J^{y}/H &=&\Lambda _{1}^{y}+\Lambda _{2}^{y}\ \ /H  \label{f307} \\
&=&\frac{1}{m}\left( -\rho _{1}\frac{\partial \chi }{\partial x_{k}}+\rho
_{2}\frac{\partial \eta }{\partial x_{k}}\right) -\frac{1}{m}\left(
a+a^{\ast }\right) \left( \rho _{1}+\rho _{2}\right)  \notag
\end{eqnarray}

\bigskip

\subsubsection{The time component of the Euler current}

This is given by 
\begin{eqnarray}
J^{0} &=&\left[ \phi ,\phi ^{\dagger }\right]  \label{f308} \\
&=&\left[ \phi _{1}E_{+}+\phi _{2}E_{-},\phi _{1}^{\ast }E_{-}+\phi
_{2}^{\ast }E_{+}\right]  \notag \\
&=&\phi _{1}\phi _{1}^{\ast }\left[ E_{+},E_{-}\right] +\phi _{2}\phi
_{2}^{\ast }\left[ E_{-},E_{+}\right]  \notag \\
&=&\left\vert \phi _{1}\right\vert ^{2}H-\left\vert \phi _{2}\right\vert
^{2}H  \notag
\end{eqnarray}%
or 
\begin{equation}
J^{0}=\left( \rho _{1}-\rho _{2}\right) H  \label{f315}
\end{equation}%
This is the \emph{charge} and we see that it is the vorticity, since%
\begin{equation}
\rho _{1}-\rho _{2}=-\frac{\kappa \omega }{2}  \label{f318}
\end{equation}

\subsection{The expression of the EULER current $J^{\protect\mu }$}

Finally 
\begin{equation}
J^{\mu }=\Lambda _{1}^{\mu }+\Lambda _{2}^{\mu }  \label{f402}
\end{equation}%
gives 
\begin{eqnarray}
J^{x} &=&\frac{1}{m}\left[ -\rho _{1}\frac{\partial \chi }{\partial x}+\rho
_{2}\frac{\partial \eta }{\partial x}+i(a-a^{\ast })\left( \rho _{1}+\rho
_{2}\right) \right] H  \label{f403} \\
J^{y} &=&\frac{1}{m}\left[ -\rho _{1}\frac{\partial \chi }{\partial y}+\rho
_{2}\frac{\partial \eta }{\partial y}-\left( a+a^{\ast }\right) \left( \rho
_{1}+\rho _{2}\right) \right] H  \notag \\
J^{0} &=&\left( \rho _{1}-\rho _{2}\right) H  \notag
\end{eqnarray}

\bigskip

We give a slightly different expression for the components of the current,
introducing the potentials $\overline{A}_{x,y}$.%
\begin{eqnarray}
J^{x}\ /H &=&\frac{1}{m}\left[ -\rho _{1}\frac{\partial \chi }{\partial x}%
+\rho _{2}\frac{\partial \eta }{\partial x}+i(a-a^{\ast })\left( \rho
_{1}+\rho _{2}\right) \right]  \label{f405} \\
&=&\frac{1}{m}\left[ -\rho _{1}\frac{\partial \chi }{\partial x}+\rho _{2}%
\frac{\partial \eta }{\partial x}-\frac{2\overline{A}_{x}}{i}\left( \rho
_{1}+\rho _{2}\right) \right]  \notag
\end{eqnarray}%
\begin{eqnarray}
J^{y}\ /H &=&\frac{1}{m}\left[ -\rho _{1}\frac{\partial \chi }{\partial y}%
+\rho _{2}\frac{\partial \eta }{\partial y}-\left( a+a^{\ast }\right) \left(
\rho _{1}+\rho _{2}\right) \right]  \label{f406} \\
&=&\frac{1}{m}\left[ -\rho _{1}\frac{\partial \chi }{\partial y}+\rho _{2}%
\frac{\partial \eta }{\partial y}-\frac{2\overline{A}_{y}}{i}\left( \rho
_{1}+\rho _{2}\right) \right]  \notag
\end{eqnarray}%
\begin{equation}
J^{0}\ /H=\rho _{1}-\rho _{2}  \label{f407}
\end{equation}

\subsection{Expression of the Euler current at self - duality}

At self-duality (and only at self-duality) we can replace the functions $a$
and $a^{\ast }$ that define the potentials $A_{\pm }$ with expressions of
the functions $\phi _{1,2}$ and $\phi _{1,2}^{\ast }$ coming from the first
equation of self-duality, $D_{-}\phi =0$.

We will replace the potentials $a$ and $a^{\ast }$ using%
\begin{equation}
a+a^{\ast }=-\frac{1}{2}\frac{\partial \psi }{\partial x}-\frac{\partial
\chi }{\partial y}\ \ \text{(pure real)}  \label{f603}
\end{equation}%
\begin{equation}
a-a^{\ast }=i\left( \frac{1}{2}\frac{\partial \psi }{\partial y}-\frac{%
\partial \chi }{\partial x}\right) \ \ \text{(pure imaginary)}  \label{f604}
\end{equation}

\subsubsection{The $x$ component of the current, $J^{x}$, at SD}

For the $x$-component we use Eq.(\ref{f604}). We have 
\begin{eqnarray}
\left[ mJ^{x}\right] \ /H &=&-\exp \left( \psi \right) \frac{\partial \chi }{%
\partial x}+\exp \left( -\psi \right) \frac{\partial \eta }{\partial x}%
+i(a-a^{\ast })\left( \rho _{1}+\rho _{2}\right)  \label{f605} \\
&=&-\exp \left( \psi \right) \frac{\partial \chi }{\partial x}+\exp \left(
-\psi \right) \frac{\partial \eta }{\partial x}+i\frac{i}{2}\left[ \frac{%
\partial \psi }{\partial y}-\frac{\partial \left( 2\chi \right) }{\partial x}%
\right] \left( \rho _{1}+\rho _{2}\right)  \notag \\
&=&-\left[ \frac{\partial \left( \psi /2\right) }{\partial y}-\frac{\partial
\chi }{\partial x}\right] \left( \rho _{1}+\rho _{2}\right) -\exp \left(
\psi \right) \frac{\partial \chi }{\partial x}+\exp \left( -\psi \right) 
\frac{\partial \eta }{\partial x}  \notag
\end{eqnarray}%
\textbf{NOTE}

Before going further we explore the possibilities of this equation. For this
we replace since we are already at SD%
\begin{eqnarray}
\rho _{1} &\rightarrow &\exp \left( \psi \right)  \label{f640} \\
\rho _{2} &\rightarrow &\exp \left( -\psi \right)  \notag
\end{eqnarray}%
\begin{eqnarray}
\left[ mJ^{x}\right] \ /H &=&-\left( \rho _{1}+\rho _{2}\right) \frac{%
\partial \left( \psi /2\right) }{\partial y}  \label{f641} \\
&&+\underline{\frac{\partial \chi }{\partial x}\exp \left( \psi \right) }+%
\frac{\partial \chi }{\partial x}\exp \left( -\psi \right)  \notag \\
&&-\underline{\exp \left( \psi \right) \frac{\partial \chi }{\partial x}}%
+\exp \left( -\psi \right) \frac{\partial \eta }{\partial x}  \notag
\end{eqnarray}

We find the expression%
\begin{equation}
\left[ mJ^{x}\right] \ /H=-\left( \rho _{1}+\rho _{2}\right) \frac{\partial
\left( \psi /2\right) }{\partial y}+\frac{\partial \chi }{\partial x}\exp
\left( -\psi \right) +\exp \left( -\psi \right) \frac{\partial \eta }{%
\partial x}  \label{f642}
\end{equation}%
where we can use%
\begin{equation}
\chi =-\eta  \label{f644}
\end{equation}%
and obtain%
\begin{equation}
\left[ mJ^{x}\right] \ /H=-\left( \rho _{1}+\rho _{2}\right) \frac{\partial
\left( \psi /2\right) }{\partial y}  \label{f645}
\end{equation}%
Finally%
\begin{eqnarray}
\left[ mJ^{x}\right] \ /H &=&-\frac{\partial }{\partial y}\frac{1}{2}\left(
\rho _{1}-\rho _{2}\right)  \label{f646} \\
&=&\frac{\kappa }{4}\frac{\partial }{\partial y}\omega \ \ \text{at SD} 
\notag
\end{eqnarray}

\bigskip

\subsubsection{The $y$ component of the current, $J^{y}$ at SD}

Now the $y$ component of the current. We will use Eq.(\ref{f603}) and obtain%
\begin{eqnarray}
\left[ mJ^{y}\right] \ /H &=&-\exp \left( \psi \right) \frac{\partial \chi }{%
\partial y}+\exp \left( -\psi \right) \frac{\partial \eta }{\partial y}%
-\left( a+a^{\ast }\right) \left( \rho _{1}+\rho _{2}\right)  \label{f670} \\
&=&-\exp \left( \psi \right) \frac{\partial \chi }{\partial y}+\exp \left(
-\psi \right) \frac{\partial \eta }{\partial y}+\left[ \frac{\partial \left(
\psi /2\right) }{\partial x}+\frac{\partial \chi }{\partial y}\right] \left(
\rho _{1}+\rho _{2}\right)  \notag \\
&=&-\exp \left( \psi \right) \frac{\partial \chi }{\partial y}+\exp \left(
-\psi \right) \frac{\partial \eta }{\partial y}+\left[ \frac{\partial \left(
\psi /2\right) }{\partial x}+\frac{\partial \chi }{\partial y}\right] \left(
\rho _{1}+\rho _{2}\right)  \notag
\end{eqnarray}%
Expanding%
\begin{eqnarray}
\left[ mJ^{y}\right] \ /H &=&\frac{\partial \left( \psi /2\right) }{\partial
x}\left( \rho _{1}+\rho _{2}\right) +\underline{\exp \left( \psi \right) 
\frac{\partial \chi }{\partial y}}+\exp \left( -\psi \right) \frac{\partial
\chi }{\partial y}  \label{f672} \\
&&-\underline{\exp \left( \psi \right) \frac{\partial \chi }{\partial y}}%
+\exp \left( -\psi \right) \frac{\partial \eta }{\partial y}  \notag
\end{eqnarray}%
and the two underlined terms cancel each other. The relation%
\begin{equation}
\chi =-\eta  \label{f673}
\end{equation}%
leads to%
\begin{equation}
\left[ mJ^{y}\right] \ /H=\frac{\partial \left( \psi /2\right) }{\partial x}%
\left( \rho _{1}+\rho _{2}\right)  \label{f674}
\end{equation}

Finally%
\begin{eqnarray}
\left[ mJ^{y}\right] \ /H &=&\frac{\partial }{\partial x}\frac{1}{2}\left(
\rho _{1}-\rho _{2}\right)  \label{f678} \\
&=&-\frac{\kappa }{4}\frac{\partial }{\partial x}\omega \ \ \text{at SD} 
\notag
\end{eqnarray}

\subsubsection{Summary, at SD}

We list them again%
\begin{equation}
\left[ mJ^{x}\right] \ /H=-\left[ \frac{\partial \left( \psi /2\right) }{%
\partial y}-\frac{\partial \chi }{\partial x}\right] \left( \rho _{1}+\rho
_{2}\right) -\exp \left( \psi \right) \frac{\partial \chi }{\partial x}+\exp
\left( -\psi \right) \frac{\partial \eta }{\partial x}  \label{f681}
\end{equation}%
\begin{equation}
\left[ mJ^{y}\right] \ /H=+\left[ \frac{\partial \left( \psi /2\right) }{%
\partial x}+\frac{\partial \chi }{\partial y}\right] \left( \rho _{1}+\rho
_{2}\right) -\exp \left( \psi \right) \frac{\partial \chi }{\partial y}+\exp
\left( -\psi \right) \frac{\partial \eta }{\partial y}  \label{f682}
\end{equation}

When the phases are replaced as $\chi =-\eta $ it is obtained%
\begin{eqnarray}
\left[ mJ^{x}\right] \ /H &=&-\frac{\partial }{\partial y}\frac{1}{2}\left(
\rho _{1}-\rho _{2}\right)  \label{f702} \\
&=&\frac{\kappa }{4}\frac{\partial }{\partial y}\omega \ \ \ \text{at SD} 
\notag
\end{eqnarray}%
and%
\begin{eqnarray}
\left[ mJ^{y}\right] \ /H &=&\frac{\partial }{\partial x}\frac{1}{2}\left(
\rho _{1}-\rho _{2}\right)  \label{f705} \\
&=&-\frac{\kappa }{4}\frac{\partial }{\partial x}\omega \ \ \ \text{at SD} 
\notag
\end{eqnarray}

To this we have to add%
\begin{equation}
J^{0}=\rho _{1}-\rho _{2}=-\frac{\kappa }{2}\omega \ \ \ \text{at SD}
\label{f708}
\end{equation}

\bigskip

Then the covariant conservation of the current results%
\begin{equation}
D_{\mu }J^{\mu }=0  \label{f710}
\end{equation}

\bigskip

We \textbf{NOTE} that the current appears as the rotational of the density
of vorticity.

\bigskip

\section{Appendix F. The current projected along the streamlines and the
perpendicular direction} \label{App:AppendixF}

\renewcommand{\theequation}{F.\arabic{equation}} \setcounter{equation}{0}

The expressions of the current components are%
\begin{equation}
\left[ mJ^{x}\right] \ /H=-\left[ \frac{\partial \left( \psi /2\right) }{%
\partial y}-\frac{\partial \chi }{\partial x}\right] \left( \rho _{1}+\rho
_{2}\right) -\exp \left( \psi \right) \frac{\partial \chi }{\partial x}+\exp
\left( -\psi \right) \frac{\partial \eta }{\partial x}  \label{g10}
\end{equation}%
\begin{equation}
\left[ mJ^{y}\right] \ /H=+\left[ \frac{\partial \left( \psi /2\right) }{%
\partial x}+\frac{\partial \chi }{\partial y}\right] \left( \rho _{1}+\rho
_{2}\right) -\exp \left( \psi \right) \frac{\partial \chi }{\partial y}+\exp
\left( -\psi \right) \frac{\partial \eta }{\partial y}  \label{g11}
\end{equation}%
and without the phases, taking into account that at SD $\chi =-\eta $.%
\begin{equation}
\left[ mJ^{x}\right] \ /H=-\frac{\partial }{\partial y}\frac{1}{2}\left(
\rho _{1}-\rho _{2}\right) \text{\ \ at SD}  \label{g12}
\end{equation}%
\begin{equation}
\left[ mJ^{y}\right] \ /H=\frac{\partial }{\partial x}\frac{1}{2}\left( \rho
_{1}-\rho _{2}\right) \text{\ \ at SD}  \label{g13}
\end{equation}

\subsection{Projection formulas}

We will make a change of the system of reference in plane%
\begin{equation}
\left( \widehat{\mathbf{e}}_{x},\widehat{\mathbf{e}}_{y}\right) \rightarrow
\left( \widehat{\mathbf{e}}_{\psi },\widehat{\mathbf{e}}_{\perp }\right)
\label{g101}
\end{equation}%
where we have to define the two versors.

The infinitesimal displacement along the streamline is represented by the
vector%
\begin{eqnarray}
\mathbf{dl}_{\parallel } &=&\left( \delta x,\delta y\right)  \label{g102} \\
&=&\delta x\widehat{\mathbf{e}}_{x}+\delta y\widehat{\mathbf{e}}_{y}  \notag
\end{eqnarray}%
with the length%
\begin{eqnarray}
\left\vert \mathbf{dl}_{\parallel }\right\vert &=&\sqrt{\left( \delta
x\right) ^{2}+\left( \delta y\right) ^{2}}  \label{g103} \\
&=&\delta x\sqrt{1+\left( \frac{\delta y}{\delta x}\right) ^{2}}  \notag
\end{eqnarray}%
and the versor is%
\begin{equation}
\widehat{\mathbf{e}}_{\psi }=\frac{\mathbf{dl}_{\parallel }}{\left\vert 
\mathbf{dl}_{\parallel }\right\vert }=\frac{1}{\sqrt{1+\left( \frac{\partial
y}{\partial x}\right) ^{2}}}\widehat{\mathbf{e}}_{x}+\frac{\frac{\partial y}{%
\partial x}}{\sqrt{1+\left( \frac{\partial y}{\partial x}\right) ^{2}}}%
\widehat{\mathbf{e}}_{y}  \label{g104}
\end{equation}%
where the streamline is represented in tow forms%
\begin{eqnarray}
\psi \left( x,y\right) &=&\text{const}  \label{g105} \\
y &=&y\left( x\right)  \notag
\end{eqnarray}%
From the theorem of implicit functions we get%
\begin{eqnarray}
\frac{\partial \psi }{\partial x}+\frac{\partial \psi }{\partial y}\frac{%
\partial y}{\partial x} &=&0  \label{g106} \\
\frac{\partial y}{\partial x} &=&-\frac{\frac{\partial \psi }{\partial x}}{%
\frac{\partial \psi }{\partial y}}  \notag
\end{eqnarray}%
and the versor along the streamline is%
\begin{equation}
\widehat{\mathbf{e}}_{\psi }=\frac{\frac{\partial \psi }{\partial y}}{\sqrt{%
\left( \frac{\partial \psi }{\partial y}\right) ^{2}+\left( \frac{\partial
\psi }{\partial x}\right) ^{2}}}\widehat{\mathbf{e}}_{x}+\left( -\frac{\frac{%
\partial \psi }{\partial x}}{\frac{\partial \psi }{\partial y}}\right) \frac{%
\frac{\partial \psi }{\partial y}}{\sqrt{\left( \frac{\partial \psi }{%
\partial y}\right) ^{2}+\left( \frac{\partial \psi }{\partial x}\right) ^{2}}%
}\widehat{\mathbf{e}}_{y}  \label{g112}
\end{equation}%
We replace%
\begin{equation}
\sqrt{\left( \frac{\partial \psi }{\partial y}\right) ^{2}+\left( \frac{%
\partial \psi }{\partial x}\right) ^{2}}=\left\vert \mathbf{\nabla }\psi
\right\vert  \label{g114}
\end{equation}%
and we have%
\begin{equation}
\widehat{\mathbf{e}}_{\psi }=\frac{1}{\left\vert \mathbf{\nabla }\psi
\right\vert }\frac{\partial \psi }{\partial y}\widehat{\mathbf{e}}_{x}-\frac{%
1}{\left\vert \mathbf{\nabla }\psi \right\vert }\frac{\partial \psi }{%
\partial x}\widehat{\mathbf{e}}_{y}  \label{g115}
\end{equation}%
The other versor, perpendicular on the streamline, is defined by a vector
product%
\begin{eqnarray}
\widehat{\mathbf{e}}_{\perp } &=&\widehat{\mathbf{e}}_{z}\times \widehat{%
\mathbf{e}}_{\psi }=\left( 
\begin{array}{ccc}
\widehat{\mathbf{e}}_{x} & \widehat{\mathbf{e}}_{y} & \widehat{\mathbf{e}}%
_{z} \\ 
0 & 0 & 1 \\ 
\frac{1}{\left\vert \mathbf{\nabla }\psi \right\vert }\frac{\partial \psi }{%
\partial y} & -\frac{1}{\left\vert \mathbf{\nabla }\psi \right\vert }\frac{%
\partial \psi }{\partial x} & 0%
\end{array}%
\right)  \label{g116} \\
&=&\widehat{\mathbf{e}}_{x}\left( +\frac{1}{\left\vert \mathbf{\nabla }\psi
\right\vert }\frac{\partial \psi }{\partial x}\right) +\widehat{\mathbf{e}}%
_{y}\left( +\frac{1}{\left\vert \mathbf{\nabla }\psi \right\vert }\frac{%
\partial \psi }{\partial y}\right)  \notag
\end{eqnarray}%
We have the transformation%
\begin{equation}
\left( 
\begin{array}{c}
\widehat{\mathbf{e}}_{\psi } \\ 
\widehat{\mathbf{e}}_{\perp }%
\end{array}%
\right) =\left( 
\begin{array}{cc}
\frac{1}{\left\vert \mathbf{\nabla }\psi \right\vert }\frac{\partial \psi }{%
\partial y} & -\frac{1}{\left\vert \mathbf{\nabla }\psi \right\vert }\frac{%
\partial \psi }{\partial x} \\ 
\frac{1}{\left\vert \mathbf{\nabla }\psi \right\vert }\frac{\partial \psi }{%
\partial x} & \frac{1}{\left\vert \mathbf{\nabla }\psi \right\vert }\frac{%
\partial \psi }{\partial y}%
\end{array}%
\right) \left( 
\begin{array}{c}
\widehat{\mathbf{e}}_{x} \\ 
\widehat{e}_{y}%
\end{array}%
\right)  \label{g117}
\end{equation}%
The determinant of this matrix is%
\begin{equation}
\det \left( 
\begin{array}{cc}
\frac{1}{\left\vert \mathbf{\nabla }\psi \right\vert }\frac{\partial \psi }{%
\partial y} & -\frac{1}{\left\vert \mathbf{\nabla }\psi \right\vert }\frac{%
\partial \psi }{\partial x} \\ 
\frac{1}{\left\vert \mathbf{\nabla }\psi \right\vert }\frac{\partial \psi }{%
\partial x} & \frac{1}{\left\vert \mathbf{\nabla }\psi \right\vert }\frac{%
\partial \psi }{\partial y}%
\end{array}%
\right) =\frac{1}{\left\vert \mathbf{\nabla }\psi \right\vert ^{2}}\left( 
\frac{\partial \psi }{\partial y}\right) ^{2}+\frac{1}{\left\vert \mathbf{%
\nabla }\psi \right\vert ^{2}}\left( \frac{\partial \psi }{\partial x}%
\right) ^{2}=1  \label{g119}
\end{equation}%
and the inverse transformation%
\begin{equation}
\left( 
\begin{array}{c}
\widehat{\mathbf{e}}_{x} \\ 
\widehat{e}_{y}%
\end{array}%
\right) =\left( 
\begin{array}{cc}
\frac{1}{\left\vert \mathbf{\nabla }\psi \right\vert }\frac{\partial \psi }{%
\partial y} & \frac{1}{\left\vert \mathbf{\nabla }\psi \right\vert }\frac{%
\partial \psi }{\partial x} \\ 
-\frac{1}{\left\vert \mathbf{\nabla }\psi \right\vert }\frac{\partial \psi }{%
\partial x} & \frac{1}{\left\vert \mathbf{\nabla }\psi \right\vert }\frac{%
\partial \psi }{\partial y}%
\end{array}%
\right) \left( 
\begin{array}{c}
\widehat{\mathbf{e}}_{\psi } \\ 
\widehat{\mathbf{e}}_{\perp }%
\end{array}%
\right)  \label{g121}
\end{equation}%
or%
\begin{eqnarray}
\widehat{\mathbf{e}}_{x} &=&\frac{1}{\left\vert \mathbf{\nabla }\psi
\right\vert }\frac{\partial \psi }{\partial y}\widehat{\mathbf{e}}_{\psi }+%
\frac{1}{\left\vert \mathbf{\nabla }\psi \right\vert }\frac{\partial \psi }{%
\partial x}\widehat{\mathbf{e}}_{\perp }  \label{g124} \\
\widehat{\mathbf{e}}_{y} &=&-\frac{1}{\left\vert \mathbf{\nabla }\psi
\right\vert }\frac{\partial \psi }{\partial x}\widehat{\mathbf{e}}_{\psi }+%
\frac{1}{\left\vert \mathbf{\nabla }\psi \right\vert }\frac{\partial \psi }{%
\partial y}\widehat{\mathbf{e}}_{\perp }  \notag
\end{eqnarray}

Now we rotate the vector 
\begin{eqnarray}
\mathbf{J} &=&\widehat{\mathbf{e}}_{x}J_{x}+\widehat{\mathbf{e}}_{y}J_{y}
\label{g128} \\
&=&\widehat{\mathbf{e}}_{\psi }J_{\psi }+\widehat{\mathbf{e}}_{\perp
}J_{\perp }  \notag
\end{eqnarray}%
\begin{eqnarray}
\mathbf{J} &=&J_{x}\left( \frac{1}{\left\vert \mathbf{\nabla }\psi
\right\vert }\frac{\partial \psi }{\partial y}\widehat{\mathbf{e}}_{\psi }+%
\frac{1}{\left\vert \mathbf{\nabla }\psi \right\vert }\frac{\partial \psi }{%
\partial x}\widehat{\mathbf{e}}_{\perp }\right) +J_{y}\left( -\frac{1}{%
\left\vert \mathbf{\nabla }\psi \right\vert }\frac{\partial \psi }{\partial x%
}\widehat{\mathbf{e}}_{\psi }+\frac{1}{\left\vert \mathbf{\nabla }\psi
\right\vert }\frac{\partial \psi }{\partial y}\widehat{\mathbf{e}}_{\perp
}\right)  \notag \\
&=&\frac{1}{\left\vert \mathbf{\nabla }\psi \right\vert }\left( J_{x}\frac{%
\partial \psi }{\partial y}-J_{y}\frac{\partial \psi }{\partial x}\right) 
\widehat{\mathbf{e}}_{\psi }  \notag \\
&&+\frac{1}{\left\vert \mathbf{\nabla }\psi \right\vert }\left( J_{x}\frac{%
\partial \psi }{\partial x}+J_{y}\frac{\partial \psi }{\partial y}\right) 
\widehat{\mathbf{e}}_{\perp }  \label{g131}
\end{eqnarray}

Now we calculate the two components using the expressions of $J_{x,y}$,%
\begin{eqnarray}
J_{\psi } &=&\frac{1}{\left\vert \mathbf{\nabla }\psi \right\vert }\left(
J_{x}\frac{\partial \psi }{\partial y}-J_{y}\frac{\partial \psi }{\partial x}%
\right)  \label{g132} \\
&=&\frac{1}{\left\vert \mathbf{\nabla }\psi \right\vert m}\left\{ \frac{%
\partial \psi }{\partial y}\left( -\left[ \frac{\partial \left( \psi
/2\right) }{\partial y}-\frac{\partial \chi }{\partial x}\right] \left( \rho
_{1}+\rho _{2}\right) -\exp \left( \psi \right) \frac{\partial \chi }{%
\partial x}+\exp \left( -\psi \right) \frac{\partial \eta }{\partial x}%
\right) \right.  \notag \\
&&\left. -\ \ \ \ \ \frac{\partial \psi }{\partial x}\ \left( -\left[ \frac{%
\partial \left( \psi /2\right) }{\partial x}+\frac{\partial \chi }{\partial y%
}\right] \left( \rho _{1}+\rho _{2}\right) -\exp \left( \psi \right) \frac{%
\partial \chi }{\partial y}+\exp \left( -\psi \right) \frac{\partial \eta }{%
\partial y}\right) \right\}  \notag
\end{eqnarray}%
and%
\begin{eqnarray}
J_{\perp } &=&\frac{1}{\left\vert \mathbf{\nabla }\psi \right\vert }\left(
J_{x}\frac{\partial \psi }{\partial x}+J_{y}\frac{\partial \psi }{\partial y}%
\right)  \label{g133} \\
&=&\frac{1}{\left\vert \mathbf{\nabla }\psi \right\vert m}\left\{ \frac{%
\partial \psi }{\partial x}\left( -\left[ \frac{\partial \left( \psi
/2\right) }{\partial y}-\frac{\partial \chi }{\partial x}\right] \left( \rho
_{1}+\rho _{2}\right) -\exp \left( \psi \right) \frac{\partial \chi }{%
\partial x}+\exp \left( -\psi \right) \frac{\partial \eta }{\partial x}%
\right) \right.  \notag \\
&&\left. +\ \ \ \ \ \frac{\partial \psi }{\partial y}\ \left( -\left[ \frac{%
\partial \left( \psi /2\right) }{\partial x}+\frac{\partial \chi }{\partial y%
}\right] \left( \rho _{1}+\rho _{2}\right) -\exp \left( \psi \right) \frac{%
\partial \chi }{\partial y}+\exp \left( -\psi \right) \frac{\partial \eta }{%
\partial y}\right) \right\}  \notag \\
&=&  \notag
\end{eqnarray}

\subsection{Using the final formulas for the current components}

We can use%
\begin{equation}
\left[ mJ^{x}\right] \ /H=-\frac{\partial }{\partial y}\frac{1}{2}\left(
\rho _{1}-\rho _{2}\right) \text{\ \ at SD}  \label{g141}
\end{equation}%
\begin{equation}
\left[ mJ^{y}\right] \ /H=\frac{\partial }{\partial x}\frac{1}{2}\left( \rho
_{1}-\rho _{2}\right) \text{\ \ at SD}  \label{g142}
\end{equation}%
or the equivalent forms%
\begin{equation}
\left[ mJ^{x}\right] \ /H=-\frac{1}{2}\left( \rho _{1}+\rho _{2}\right) 
\frac{\partial \psi }{\partial y}\text{\ \ at SD}  \label{g143}
\end{equation}%
\begin{equation}
\left[ mJ^{y}\right] \ /H=\frac{1}{2}\left( \rho _{1}+\rho _{2}\right) \frac{%
\partial \psi }{\partial x}\text{\ \ at SD}  \label{g150}
\end{equation}%
and obtain%
\begin{eqnarray}
J_{\psi } &=&\frac{1}{\left\vert \mathbf{\nabla }\psi \right\vert }\left(
J_{x}\frac{\partial \psi }{\partial y}-J_{y}\frac{\partial \psi }{\partial x}%
\right)  \label{g151} \\
&=&\frac{1}{\left\vert \mathbf{\nabla }\psi \right\vert m}\frac{1}{2}\left(
\rho _{1}+\rho _{2}\right) \left[ -\left( \frac{\partial \psi }{\partial y}%
\right) ^{2}-\left( \frac{\partial \psi }{\partial x}\right) ^{2}\right] 
\notag \\
&=&-\frac{1}{2m}\left( \rho _{1}+\rho _{2}\right) \left\vert \mathbf{\nabla }%
\psi \right\vert \text{\ \ at SD}  \notag
\end{eqnarray}%
and%
\begin{eqnarray}
J_{\perp } &=&\frac{1}{\left\vert \mathbf{\nabla }\psi \right\vert }\left(
J_{x}\frac{\partial \psi }{\partial x}+J_{y}\frac{\partial \psi }{\partial y}%
\right)  \label{g156} \\
&=&\frac{1}{\left\vert \mathbf{\nabla }\psi \right\vert m}\frac{1}{2}\left(
\rho _{1}+\rho _{2}\right) \left( -\frac{\partial \psi }{\partial y}\frac{%
\partial \psi }{\partial x}+\frac{\partial \psi }{\partial x}\frac{\partial
\psi }{\partial y}\right)  \notag \\
&=&0\text{\ \ at SD}  \notag
\end{eqnarray}

This indeed confirms that the only current is along the streamlines and the
current transversal to them vanishes.

\end{appendices}


\begin{thebibliography}{99}

\bibitem{Montgomery1991}
W.H. Matthaeus, W.T. Stribling, D.~Martinez, S.~Oughton, and D.~Montgomery.
\newblock Decaying, two-dimensional, navier-stokes turbulence at very long
  times.
\newblock {\em Physica D}, 51:531--538, 1991.

\bibitem{Montgomery1992}
D.~Montgomery, W.H. Matthaeus, W.T. Stribling, D.~Martinez, and S.~Oughton.
\newblock Relaxation in two-dimensions and the "sinh-poisson" equation.
\newblock {\em Phys. Fluids A}, 4:3--6, 1992.

\bibitem{TingChenLee}
A.C. Ting, H.H. Chen, and Y.C. Lee.
\newblock Exact solutions of a nonlinear boundary value problem: The vortices
  of the two-dimensional sinh-poisson equation.
\newblock {\em Physica D: Nonlinear Phenomena}, 26(1–3):37 -- 66, 1987.

\bibitem{MasonWoodhouse}
L.~J. Mason and N.~M.~J. Woodhouse.
\newblock {\em Integrability, self - duality and twistor theory}.
\newblock London Mathematical Society Monographs New Series. Clarendon Press,
  Oxford, 1996.

\bibitem{FlorinMadi2003}
F.~Spineanu and M.~Vlad.
\newblock Self-duality of the asymptotic relaxation states of fluids and
  plasmas.
\newblock {\em Phys. Rev. E}, 67:046309--1--4, 2003.

\bibitem{JackiwPi}
R.~Jackiew and So-Y. Pi.
\newblock Classical and quantal nonrelativistic chern-simons theory.
\newblock {\em Phys. Rev. Lett.}, 64:2969--2979, 1990.

\bibitem{KraichnanMontgomery}
R~H Kraichnan and D~Montgomery.
\newblock Two-dimensional turbulence.
\newblock {\em Reports on Progress in Physics}, 43(5):547--619, 1980.

\bibitem{RobertSommeria}
R.~Robert and J.~Sommeria.
\newblock Statistical equilibrium states for two dimensional flows.
\newblock {\em J. Fluid Mech}, 229:291, 1991.

\bibitem{RobertSommeria1992}
R.~Robert and J.~Sommeria.
\newblock Relaxation towards a statistical equilibrium state in two-dimensional
  perfect fluid dynamics.
\newblock {\em Phys. Rev. Lett.}, 69:2776--2779, 1992.

\bibitem{Chavanis1}
Pierre-Henri Chavanis.
\newblock Statistical mechanics of two-dimensional vortices and stellar
  systems.
\newblock In Thierry Dauxois, Stefano Ruffo, Ennio Arimondo, and Martin
  Wilkens, editors, {\em Dynamics and Thermodynamics of Systems with Long-Range
  Interactions}, volume 602 of {\em Lecture Notes in Physics}, pages 208--289.
  Springer Berlin Heidelberg, 2002.

\bibitem{Onsager}
L.~Onsager.
\newblock Statistical hydrodynamics.
\newblock {\em Il Nuovo Cimento Series 9}, 6(2):279--287, 1949.

\bibitem{MajdaWang}
A.J. Majda and X.~Wang.
\newblock {\em Non-linear dynamics and statistical theories for basic
  geophysical flows}.
\newblock Cambridge University Press, 2006.

\bibitem{DritschelLuciaPoje}
D.G Dritschel, M.~Lucia, and A.C. Poje.
\newblock Equilibrium statistics and dynamics of point vortex flows on the
  sphere.
\newblock {\em http://arxiv.org/pdf/1407.3151.pdf}, 2014.

\bibitem{EdwardsTaylor}
S.F. Edwards and J.B. Taylor.
\newblock Negative temperatures states of two-dimensional plasmas and vortex
  fluids.
\newblock {\em Proc. R. Soc. Lond. A}, 336:257--271, 1974.

\bibitem{JoyceMontgomery}
G.~Joyce and D.C. Montgomery.
\newblock Negative temperature states of the two-dimensional guiding-centre
  plasma.
\newblock {\em J. Plasma Phys.}, 10:107--121, 1973.

\bibitem{Taylor1993}
J.B. Taylor.
\newblock Filamentation, current profile and transport in a tokamak.
\newblock {\em Phys. Fluids B5}, pages 4378--4383, December 1993.

\bibitem{Bethuel}
F.~Bethuel, G.~Orlandi, and D.~Smets.
\newblock Collisions and phase-vortex interactions in dissipative
  ginzburg-landau dynamics.
\newblock {\em Duke Math. J.}, 130:523--614, 2005.

\bibitem{DunneJackiwTrugenberg}
Gerald~V. Dunne, R.~Jackiw, So-Young Pi, and Carlo~A. Trugenberger.
\newblock Self-dual chern-simons solitons and two-dimensional nonlinear
  equations.
\newblock {\em Phys. Rev. D}, 43:1332--1345, Feb 1991.

\bibitem{DunneBook}
G.~Dunne.
\newblock {\em Self - dual Chern - Simons theories}, volume~36 of {\em Lecture
  Notes in Physics}.
\newblock Springer Verlag, Berlin Heidelberg, 1995.

\bibitem{MartinSiggiaRose}
P.C. Martin, E.D. Siggia, and H.A. Rose.
\newblock Statistical dynamics of classical systems.
\newblock {\em Phys. Rev. A}, 8:423--437, 1973.

\bibitem{RVJensen}
R.V. Jensen.
\newblock Functional integral approach to classical statistical dynamics.
\newblock {\em J. Stat. Phys.}, 5:183--210, 1981.

\bibitem{JacobsRebbi}
L.~Jacobs and C.~Rebbi.
\newblock Interaction energy of superconducting vortices.
\newblock {\em Phys. Rev. B}, 19:4486--4494, 1979.

\bibitem{Arthur}
K.~Arthur.
\newblock Interaction energy of chern-simons vortices.
\newblock {\em Phys. Lett. B}, 356:509--515, 1995.

\bibitem{9812103FuertesGuilarte}
W.G. Fuertes and J.M. Guilarte.
\newblock {Low energy vortex dynamics in Abelian Higgs systems}.
\newblock {\em arxiv.org}, hep-th:9812103, 1998.

\bibitem{Manton}
N.S. Manton.
\newblock Statistical mechanics of vortices.
\newblock {\em Nucl. Phys. B}, 400[FS]:624--632, 1993.

\bibitem{Nauta}
B.J. Nauta and A.~Arrizabalaga.
\newblock Asymmetric chern-simons number diffusion from cp-violation.
\newblock {\em Nucl. Phys. B}, 635:255--285, 2002.

\bibitem{ArnoldSonYaffe}
P.~Arnold, D.~Son, and L.G. Yaffe.
\newblock The hot baryon violation rate is $o\left(\alpha_{W}^{5}t^{4}\right)$.
\newblock {\em Phys. Rev. D}, 55:6264--6273, 1997.

\bibitem{JackiwPiSymm}
R.~Jackiw and S.Y. Pi.
\newblock Finite and infinite symmetries in $2+1$ dimensional field theory.
\newblock {\em Nucl. Phys. B (Proc. Suppl)}, 33C:104--113, 1993.

\bibitem{collapse}
L.~Berge, A.~de~Bouard, and J.C. Saut.
\newblock Collapse of chern-simons - gauged matter fields.
\newblock {\em Phys. Rev. Lett}, 74(20), 1995.

\bibitem{Levich1987}
E.~Levich.
\newblock Certain problems in the theory of developed hydrodynamical
  turbulence.
\newblock {\em Physics Reports}, 151(3--4):129 -- 238, 1987.

\bibitem{Taylor1972}
J.B. Taylor.
\newblock Negative temperatures in two dimensional vortex motion.
\newblock {\em Phys. Letters}, 40A:1--2, 1972.

\bibitem{HopfingerVanHeijst}
E.J. Hopfinger and G.J.F. van Heijst.
\newblock Vortices in rotating fluids.
\newblock {\em Annu. Rev. Fluid Mech.}, 25:241--289, 1993.

\bibitem{CorcosSherman}
G.~M. Corcos and F.~S. Sherman.
\newblock Vorticity concentration and the dynamics of unstable free shear
  layers.
\newblock {\em Journal of Fluid Mechanics}, 73:241--264, 1 1976.

\bibitem{Mcwilliams}
James~C. McWilliams.
\newblock The emergence of isolated coherent vortices in turbulent flow.
\newblock {\em Journal of Fluid Mechanics}, 146:21--43, 9 1984.

\bibitem{FlorinMadi2005}
F.~Spineanu and M.~Vlad.
\newblock Stationary vortical flows in two-dimensional plasma and in planetary
  atmospheres.
\newblock {\em Phys. Rev. Lett.}, 94:235003--1--4, 2005.

\bibitem{FlorinMadiGAFD}
F.~Spineanu and M.~Vlad.
\newblock Relationships between the main parameters of the stationary
  two-dimensional vortical flows in the planetary atmosphere.
\newblock {\em Geophys. Astro. Fluid. Dyn.}, 103:223--244, 2009.

\bibitem{FlorinmadiEqualRadiixxx}
F.~Spineanu and M.~Vlad.
\newblock {A field theoretical prediction of the tropical cyclone properties }.
\newblock {\em arxiv.org}, physics:1310.2750, 2013.

\bibitem{FlorinMadiXXX} F.~Spineanu and M.~Vlad. \newblock The asymptotic
quasi-stationary states of the two-dimensional  magnetically confined plasma
and of the planetary atmosphere. \newblock {\em arxiv.org}, physics:0501020,
2005.

\end{thebibliography}

\end{document}